\documentclass[12pt]{elsarticle}
\makeatletter
\def\ps@pprintTitle{%
 \let\@oddhead\@empty
 \let\@evenhead\@empty
 \def\@oddfoot{\centerline{\thepage}}%
 \def\@evenfoot{\thepage}
 \let\@evenfoot\@oddfoot}
\makeatother

\usepackage[authormarkup=none]{changes} %% UNCOMMENT THIS LINE TO SHOW CHANGES
\definecolor{C0}{HTML}{1F77B4} 
\definecolor{C1}{HTML}{FF7F0E}
\definecolor{C2}{HTML}{008000}
\definechangesauthor[color=C0]{R1}
\definechangesauthor[color=C1]{R2}
\definechangesauthor[color=C2]{RA}

\usepackage{hyperref}

\usepackage[T1]{fontenc}
\usepackage{lmodern}
\usepackage{microtype}
\usepackage[margin=1in]{geometry}
\usepackage[]{algorithm2e}
\usepackage{amsmath,amssymb,amsthm,bm}
\usepackage{mathtools}
\usepackage{graphicx}
 \usepackage{algpseudocode} 
\usepackage{subcaption}
\biboptions{sort&compress}
\usepackage{epstopdf}
\usepackage{xcolor}
\AppendGraphicsExtensions{.tif}

\begin{document}

%\title{Graph neural networks for \replaced[id=R1, comment=C1]{simulating }{emulating} crack coalescence and propagation in brittle materials}
%\title{ACCURATE: ACCelerated Universal fRAcTure Emulator using Transfer Learning and Graph Neural Networks}
\title{A generalized machine learning framework for brittle crack problems using transfer learning and graph neural networks}
\author[auburn]{Roberto Perera}
\author[auburn]{Vinamra Agrawal\corref{cor1}}
\cortext[cor1]{Corresponding author: vinagr@auburn.edu}
\address[auburn]{Department of Aerospace Engineering, Auburn University, Auburn, AL, USA}

\begin{abstract}

    Despite their recent success, machine learning (ML) models such as graph neural networks (GNNs), suffer from drawbacks such as the need for large training datasets and poor performance for unseen cases. 
    In this work, we use transfer learning (TL) approaches to circumvent the need for retraining with large datasets.
    We apply TL to an existing ML framework, trained to predict multiple crack propagation and stress evolution in brittle materials under Mode-I loading.
    The new framework, ACCelerated Universal fRAcTure Emulator (ACCURATE), is generalized to a variety of crack problems by using a sequence of TL update steps including (i) arbitrary crack lengths, (ii) arbitrary crack orientations, (iii) square domains, (iv) horizontal domains, and (v) shear loadings.
    We show that using small training datasets of 20 simulations for each TL update step, ACCURATE achieved high prediction accuracy in Mode-I and Mode-II stress intensity factors, and crack paths for these problems. %case studies (i) - (iv).
    We demonstrate ACCURATE's ability to predict crack growth and stress evolution with high accuracy for unseen cases involving the combination of new boundary dimensions with arbitrary crack lengths and crack orientations in both tensile and shear loading. 
    We also demonstrate significantly accelerated simulation times of up to 2 orders of magnitude faster (200x) compared to an XFEM-based fracture model. 
    The ACCURATE framework provides a universal computational fracture mechanics model that can be easily modified or extended in future work.
   
\end{abstract}


\maketitle

\section{Introduction} \label{Introduction}

    Material defects such as microcracks play a critical role in the performance and reliability of materials and structures. 
    To better understand and predict microcrack propagation and coalescence, computational fracture mechanics models have significant advantages over costly experiments. 
    Despite their success, for higher complexity cases involving large systems of microcracks, these modeling techniques quickly become computationally expensive and time-consuming.
    The development and integration of Machine Learning (ML) methods have received considerable attention in the material science and solid mechanics community \cite{ZhangFinite2020,SETTGAST20191, DANOUN2022104436, Yangeabd7416, WENG2020103522,TENG2022104175,park4164581generalizable,YANG2020108509,WiltAccelerating2020,ZHANG2022115233,PERERA2021Optimized,C8MH00653A,GU201819,BESSA2017633,Vlassis2021Component,FUHG2022101446,Mozaffa2019Deep,Chandrashekar2022Quantifying,Kadeethum2021framework,KADEETHUM2022Continuous,Lizama2021framework}.
    Specific to fracture mechanics problems, prediction of crack growth \cite{Lew2021DeepLM, HSU2020197,SHARMA2021104071,ZHANG2022104309,OULADBRAHIM2022104200,CAO2020103631}, crack coalescence \cite{MOORE201846, HUNTER201987}, stress \cite{osti_1765066}, fracture toughness \cite{LIU2020105,Vahid2020Machine}, fatigue strength \cite{GEBHARDT2020103625}, and displacements \cite{IM2021114030} have been performed with high accuracy using various ML techniques.
    Recently, graph neural networks (GNNs) have shown high accuracy and speedup when simulating complex solid mechanics problems.
    GNNs work by integrating graph theory with deep neural networks.
    The graph-based implementation of GNNs (connecting nodes and edges) makes them a suitable candidate for many engineering problems \cite{VLASSIS2020113299,HEIDER2020112875,Xuanjie20223D-structure-attention,TSOPANIDIS2022110272,9022162,Zhang2022Predicting,CHOUDHARY2022111388,Perera2022Dynamic}.

    While ML methods have shown promise, their accuracy suffers for problems when problem-specific inputs (e.g., loading type, domain size, etc.,) are unknown to the training dataset.
    This requires generating new large training datasets and retraining for each possible problem-specific input making the approach computationally expensive. % when using a high-fidelity fracture mechanics model.
    A possible solution to circumvent large data drawbacks is the use of Transfer Learning (TL) methods \cite{yosinski2014advances}. 
    TL methods allow the transfer of learned information of the underlying physics across other problem-specific inputs.
    One way in which this information is transferred is through the pre-trained weights from a baseline ML model \cite{torrey2010transfer,Liu2021Knowledge}. 
    A classic demonstration of the advantage of using TL methods is object detection/recognition tasks using Convolutional Neural Networks (CNNs) \cite{Shin2016Deep,Imad2021Transfer,Shao2015Transfer,Talukdar2018Transfer,Bamne2020Transfer,tan2018survey}.
    For object detection/recognition, the initial layers are shown to capture basic shapes such as vertical and horizontal straight lines, and round edges, while the final layers capture more problem-specific features.
    For instance,  \cite{saeed2019automatic,GONG2022115136,zheng2021scalable} made use of TL strategies to develop object detection models for the detection of defects in CFRP, composite structures, and graphene, respectively.
    
    TL has also shown promising results towards material design \cite{kim2021deep}, prediction of material properties from molecular dynamic simulations \cite{Vlassis2022Molecular}, capture hierarchical microstructure representations \cite{decost2017exploring}, and predictions of structure-property from reconstructed microstructures \cite{li2018transfer}.
    Additionally, TL strategies have also been integrated to GNNs. 
    Similar to CNNs, the embeddings from the early stages of the GNN capture global features of the problem, while the final stages learn more problem-specific features \cite{kolluru2022transfer,chen2021atomsets}.
    For instance, Lee \textit{et al.} \cite{lee2021transfer} applied TL to the pre-trained Crystal Graph Convolutional Neural Network (CGCNN) introduced in \cite{xie2018crystal}, to improve predictions of target properties in crystal structures with scarce training datasets.
    The pre-trained CGCNN model was also used in recent work along with TL techniques to predict methane adsorption in metal-organic frameworks \cite{wang2022combining}.

    In this work, we leverage TL methods for predicting crack propagation, coalescence, and stress evolution in brittle materials. 
    In an earlier study, a Microcrack-GNN was introduced \cite{Perera20221Graph} capable of simulating crack growth, crack coalescence, and stress evolution in brittle materials with multiple microcracks under tension using GNNs.
    While MicroCrack-GNN showed high prediction accuracy and simulation speedup of 25x compared to an XFEM model, there were several limitations of the framework.
    For instance, the framework was only trained (with 864 simulations) for problem-specific cases involving tensile loads without accounting for shear-loading cases.
    The framework only considered cases involving a fixed domain size ($2000mm \times 3000mm$), three crack orientations ($0^{o}$, $60^{o}$, $120^{o}$), and fixed crack length ($300mm$).
    The accuracy of the framework was not tested for cases involving shear loads or different domain sizes.
    Finally, although the graph representation (i.e., nodes' and edges' features) included information to account for the crack length and crack orientation, it did not include information for different domain sizes {or loading types}.
    
\begin{figure}
        \centering 
        \includegraphics[width=\linewidth]{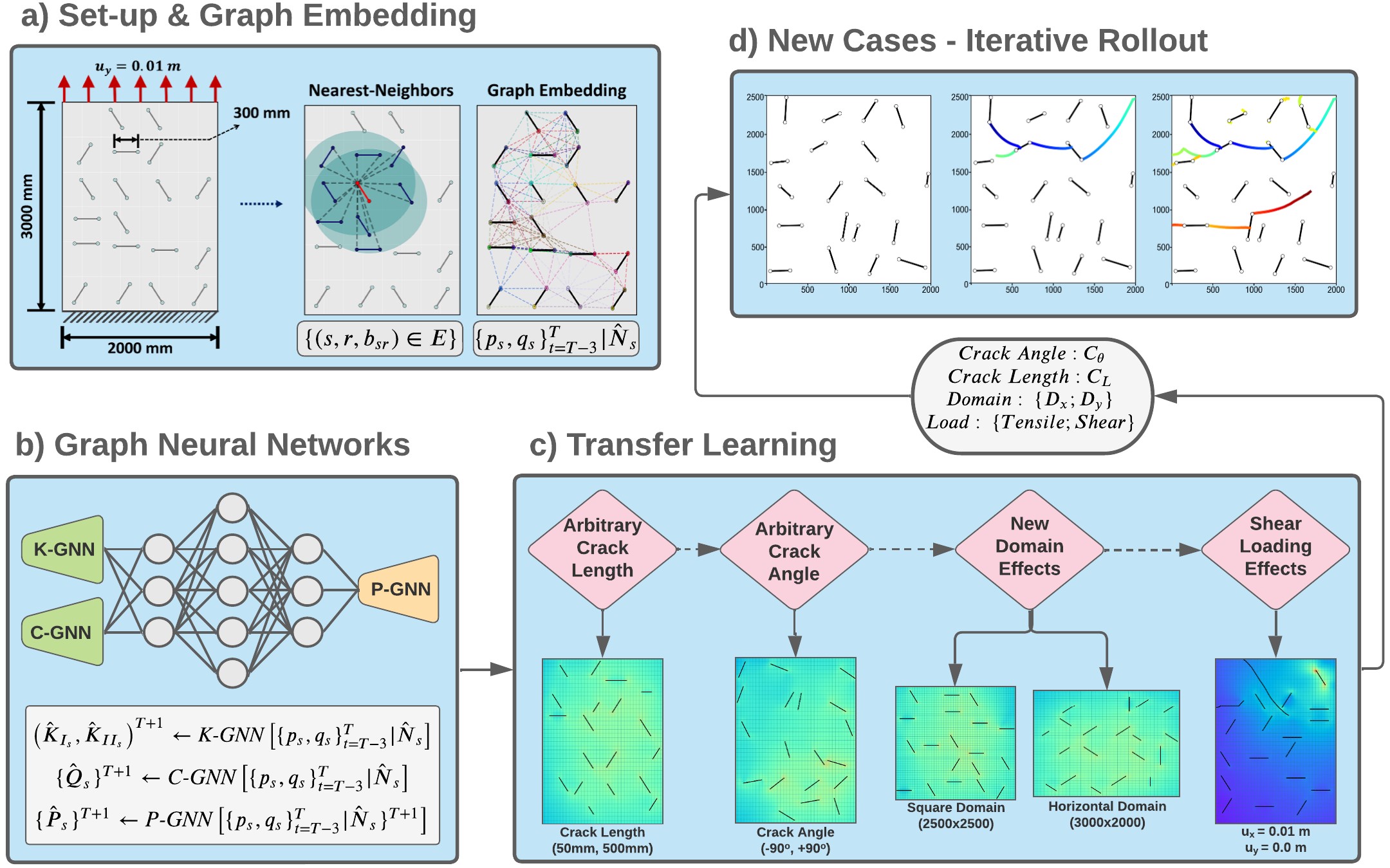}
        \centering
        \caption{Flowchart of the {ACCURATE} framework: a) Initial problem set-up, nearest-neighbor formulation and resulting TL graph embedding. b) Architecture of GNN models, \textit{K-GNN}, \textit{C-GNN}, and \textit{P-GNN}. c) Order of TL applications: arbitrary crack length, (ii) arbitrary crack angle, (iii) new domain dimensions (square and horizontal), and (iv) shear loading effects. d) Resulting iterative rollout for an unseen problem configuration.}
        \label{fig:MicroCrack-GNN_structure}
    \end{figure}

    Leveraging TL, we develop a generalized GNN framework, ACCelerated Universal fRAcTure Emulator (ACCURATE), shown in Figure \ref{fig:MicroCrack-GNN_structure}, capable of predicting crack propagation and stress evolution for a variety of fracture mechanics problems.
    We applied a sequence of 5 TL update steps involving new problem-specific inputs of: (i) arbitrary crack length, (ii) arbitrary crack orientation, (iii) new domain effects (square and horizontal), and (iv) shear loading effects (shown in Figure \ref{fig:MicroCrack-GNN_structure}c).    
    Through the use of TL in ACCURATE, we demonstrate the efficacy of TL approaches in using significantly smaller training samples, thus, reducing training time.
    The ACCURATE framework is able to predict unseen cases involving new {boundary dimensions} with both arbitrary angles and crack lengths subjected to either tensile loads or shear loads with high accuracy without the need for TL.
    The generalized graph representation of ACCURATE provides an easily-modifiable approach that can be extended to new problems in future works. 
    Lastly, ACCURATE achieves accelerated emulation times with 200x speedup compared to the XFEM-based surrogate model.
    
    The paper is organized as follows. 
    In Section \ref{sect:Methods}, we describe the extended finite element method (XFEM) based model used for gathering the training dataset, the set-up of the TL case studies, and the previous structure of MicroCrack-GNN. 
    {In Section \ref{sect:Framework}}, we introduce the structure of the ACCURATE framework.
    {In Section \ref{subsect:TL_training_samples_error}, we } analyze the effects of the number of training simulations used in TL, on the resulting prediction errors. 
    In Sections \ref{subsect:stress_evolution} and \ref{subsect:crack_growth}, we show ACCURATE's ability to emulate stress evolution by means of the predicted Mode-I and Mode-II stress intensity factors, and its ability to emulate crack propagation and coalescence for the test cases used for the TL update steps.
    In Section \ref{subsec:results_Keff_CProp}, we present the error analysis for the ACCURATE framework corresponding to each test case used in the TL sequential updates.
    In Section \ref{subsec:Unseen_Cases}, we demonstrate ACCURATE's ability to predict unseen cases involving new domain sizes subjected to tensile or shear loads with cracks of both arbitrary lengths and orientations.
    Finally, in Section \ref{subsec:simulation_time} we compare the required emulation times until failure for ACCURATE versus the XFEM fracture model.

\section{Methods}\label{sect:Methods}

    \subsection{XFEM-based surrogate model and TL simulation set-up}\label{subsect:Matlab}
        
        We use the XFEM-based model in \cite{SUTULA2018205,SUTULA2018225, SUTULA2018257} to generate the training, validation, and test datasets for the implementation of TL. 
        The XFEM-based model is capable of modeling the two-dimensional propagation of multiple cracks in brittle materials with arbitrary positions and orientations subjected to both tensile or shear loadings.
        The problem set-up follows a similar approach as \cite{Perera20221Graph}, where a domain with a maximum of 19 microcracks with random positions was used for each simulation. % as shown in Figure \ref{fig:setup}.
        Each simulation involves an isotropic, homogeneous, and perfectly brittle material with Young's Modulus of $E = 22.6 \ GPa$, Poisson's ratio of $\nu = 0.242$, and material toughness of $K_{crt} = 1.08 \ MPa \cdot \sqrt{m}$.
        Next, we fixed the bottom edge of the domain and apply a constant amplitude of $0.01\ m$ at the top edge {towards the positive y-direction (tensile load)}
        
        To investigate the use of TL we generated 6 new case studies and apply TL update steps sequentially as follows.
        \begin{itemize}
            \item Case 1: Vertical domain ($2000mm\times3000mm$) with fixed crack lengths ($300mm$) and crack orientations ($0^{o}$, $60^{o}$, and $120^{o}$) in Mode-I loading.
            \item Case 2: Vertical domain with arbitrary crack lengths (from $50mm$ to $500mm$) in Mode-I loading.
            \item Case 3: Vertical domain with arbitrary crack orientations (from $-90^{o}$ to $+90^{o}$) in Mode-I loading.
            \item Case 4: Square domain of $2500mm\times2500mm$ in Mode-I loading.
            \item Case 5: Horizontal domain of $3000mm\times2000mm$ in Mode-I loading.
            \item Case 6: Shear loading with the fixed bottom edge and constant displacement of $0.01\ m$ at the top edge towards the positive x-direction (shear load).
        \end{itemize}
        Using this, we generate a dataset of $35$ simulations with up to $101$ time-steps each, for each case study (Case 1 - Case 6).
        We split each dataset using $20$ simulations for the training set, $5$ simulations for the validation dataset, and $10$ simulations for the test dataset.
        A training data point consists of a sequence of $N_{seq}+1$ time steps (chosen randomly across all training simulations) where $1,\ldots, N_{seq}$ time steps are the input, and the prediction at $N_{seq}+1$ time step is the output. 
        The purpose of this method is to feed random cases and random time sequences to the GNN at each training iteration. 
        This approach ensures that the model learns to predict the future, given any random instance in time of any simulation.
        Following \cite{Perera20221Graph,sanchezgonzalez2020learning, battaglia2018relational,pfaff2021learning}, we use a sequence length of $N_{seq} = 4$, (i.e., {$\mathbf{\hat{T}} := \{T-3, \ T-2, \ T-1, \ T\}$ }).
        The total number of inputs for each dataset is computed as $N_{input} = N_{sim}\times (N_{steps}-N_{seq})$, resulting in $1,940$, $485$, and $970$  inputs for the training, validation, and test datasets, respectively.
        We note that MicroCrack-GNN used a total of $36,288$ inputs for training the framework (approximately 20x larger than the TL datasets).

    \subsection{MicroCrack-GNN}\label{subsect:GraphTheory}

        Here we briefly present the details of the Microcrack-GNN framework, detailed in \cite{Perera20221Graph}.
        In the MicroCrack-GNN model, the graph representation is described as {$\langle \mathbf{V},\mathbf{E} \rangle$}, where $\mathbf{V}$ represents all crack-tips as vertices, and $\mathbf{E}$ represents all the edges in the graph.
        Each vertex (i.e., crack-tip $v_{s}^{t}$) for a sequence of previous time-steps, {$\hat{\mathbf{T}}:= \{T-3, \ T-2, \ T-1, \ T\}$}, was defined by its position in Cartesian coordinates, $\hat{P}_{s}=(x_{s}, y_{s})$, its nearest-neighboring crack-tips $\hat{N}_{s}$ (i.e., neighboring cracks that lie within a zone of influence of {$r_{c} = 750 \ mm$ }), as well as its initial orientation in radians $\hat{O}_{s}=\theta_{s}$ (i.e., {$\{ \theta_{s} \} = $} $0^{o}$, $60^{o}$, {or} $120^{o}$).
        Additionally, the edges in the system ($e_{sr}^{t}$) were defined by $(s, r, b_{sr}) \in \mathbf{E}$, where $s$ is the index for the ``sender'' vertex, $r$ is the index for the ``receiver'' vertex, and $b_{sr}$ is a binary value specifying whether the ``sender'' vertex and the ``receiver'' vertex form part of the same pairwise neighbors.
        The resulting vertex and edges representation for MicroCrack-GNN is shown in equation (\ref{eq:vertex_edges})
        \begin{flalign}
        && v_{s}^{t} = \left(  \hat{P_{s}}^{t}, \hat{N_{s}}^{t}, \hat{O_{s}}^{t} \right) && \{t \in \mathbf{\hat{T}}\} \ \ ; \{ s \in \mathbf{V} \}, \ \nonumber\\
        && e_{sr}^{t}  = \left(v_{s}^{t}, v_{r}^{t}, b_{sr}^{t}\right) &&  \{t \in \mathbf{\hat{T}}\} \ \ ; \{(s,r,b_{sr}) \in \mathbf{E}\}. \label{eq:vertex_edges}
        \end{flalign}
        Next, the spatial message-passing process was applied to learn the vertices and edges relations in the latent space \cite{klicpera2020directional, li2017gated, zhang2020dynamic,gilmer2017neural}.
        The message-passing process was first applied to the vertices feature vector and the edges feature vector using two encoder Multi-Layer Perceptron (MLP) networks.
        The output embeddings, {$\{v_{s}^{'}\}^{\hat{T}}$} and {$\{e_{sr}^{'}\}^{\hat{T}}$}, are the encoded vertices' and edges' embedding in the latent space for time sequence $\hat{T}$.
        As shown in equation (\ref{eq:one-hot-encoded}) both the vertices' and edges' embedding were then concatenated in groups defined by the nearest-neighbors of microcrack $v_{s}$ (i.e., at $b_{sr} = 1$). 
        Then, the concatenated embeddings were used as input to an additional MLP network, $\mu_{G}$, for obtaining a one-hot encoded feature vector, $\{p_{s}\}^{\hat{T}}$, describing the vertex-edge-vertex relations in the latent space. 
        \begin{flalign}
        && \{p_{s}\}^{\hat{T}}  \longleftarrow \mu_{G}\left(\{v_{s}^{'}\}^{\hat{T}}, \sum_{r \in \mathcal{N}_{s}} \{e_{rs}^{'}\}^{\hat{T}}\right) &&  \{s \in \mathbf{V}\}. \label{eq:one-hot-encoded}
        \end{flalign}
        This procedure was then repeated for a series of update steps, $M$ (i.e., the spatial message-passing steps).
        For MicroCrack-GNN, the optimal number of message-passing steps was found using the 10-fold Cross-Validation method as $M=6$.
        In this work, we apply TL to ACCURATE using the pre-trained weights of MicroCrack-GNN's message-passing network shown in equation (\ref{eq:one-hot-encoded}).
        
\section{Description of ACCURATE}\label{sect:Framework}

    \subsection{Graph representation and spatial message-passing}\label{subsubsec:XFEM_GraphTheory}

        As the first step towards the development of ACCURATE (Figure \ref{fig:MicroCrack-GNN_structure}a), we define a new graph representation.
        The graph representation of ACCURATE {$\langle \mathbf{V},\mathbf{E} \rangle$} resembles that of the Microcrack-GNN, with critical additions to the node and edge feature vectors.
        {For the vertices, defined by $\xi_{s}$, we included four additional features to account for: (i) load type $\hat{\mathcal{F}_{s}} = (u_{x_{s}},u_{y_{s}})$ (i.e., Tension: $\{ u_{x} = 0.0, u_{y} = 0.01m \}$, Shear: $\{ u_{x} = 0.01m, u_{y} = 0.0 \}$), and (ii) effects in height and width of the domain; the horizontal and vertical distances to the right and top edges of the domain, $\hat{\mathcal{B}}_{s}=\left( d_{W_{s}}, d_{H_{s}} \right)$.}
        For the edges, defined by $\mathcal{E}_{sr}$, we introduced new spatial features and physics-informed features.
        The spatial features include the horizontal distance $\Delta \hat{\mathcal{X}}_{sr}=\left(x_{r} - x_{s} \right)$, the vertical distance $\Delta \hat{\mathcal{Y}}_{sr}=\left(y_{r} - y_{s} \right)$, the equivalent distance $\hat{\mathcal{L}}_{sr}= \sqrt{\Delta \hat{\mathcal{X}}_{sr}^{2} + \Delta \hat{\mathcal{Y}}_{sr}^{2} } $, and the relative crack orientation $\Delta \hat{O}_{sr}=\left(\theta_{r} - \theta_{s} \right)$ between the crack-tips.
        The physics-informed features $\Delta \hat{\Pi}_{sr}$, include the change in Mode-I stress intensity factor $\Delta \hat{K}_{I_{sr}}$, the change in Mode-II stress intensity factor $\Delta \hat{K}_{II_{sr}}$, and the change in effective stress intensity factor $\Delta \hat{K}_{eff_{sr}} = \left( \sqrt{\Delta \hat{K}_{I_{sr}}^{2} + \Delta \hat{K}_{II_{sr}}^{2}} \right)$ for each edge.
        We then define the graph representation for the node features $\xi_{s}$ and the edges features $\mathcal{E}_{sr}$ as
        \begin{flalign}
        && \xi_{s}^{t} = \left(  \hat{P}_{s}^{t}, \hat{N}_{s}^{t}, \hat{O}_{s}^{t}, \hat{K}_{I_{s}}^{t}, \hat{K}_{II_{s}}^{t}, \hat{\mathcal{F}_{s}}^{t}, \hat{\mathcal{B}}_{s}^{t} \right) && \{t \in \mathbf{\hat{T}}\} \ \ ; \{ s \in \mathbf{V} \}, \ \nonumber\\
        && \mathcal{E}_{sr}^{t}  = \left(\Delta \mathcal{\hat{X}}_{sr}^{t}, \Delta \mathcal{\hat{Y}}_{sr}^{t}, \Delta \mathcal{\hat{L}}_{sr}^{t}, \Delta {\hat{O}}_{sr}^{t}, \Delta {\hat{\Pi}_{sr}}^{t} \right) &&  \{t \in \mathbf{\hat{T}}\} \ \ ; \{(s,r,b_{sr}) \in \mathbf{E}\}.\label{eq:mod_vertex_edges}
        \end{flalign}

        Next, we implement the message-passing method for the ACCURATE framework (shown in Figure \ref{fig:MicroCrack-GNN_structure}a).
        For this, we used the Graph Isomorphism Network with Edges (GINE) model with aggregated weights. 
        The GINE model takes inputs of node features, edges connectivity array, and edges features to output a new one-hot encoded feature vector describing the latent space relations \cite{Hu2019GINE}.
        We define the GINE message-passing model for ACCURATE as
        \begin{flalign}
            &&\{q_{s}\}^{\hat{T}}  \longleftarrow GINE\left(\{ \xi_{s} \}^{\hat{T}}, \{ e_{sr} \}^{\hat{T}}, \{ \mathcal{E}_{sr} \}^{\hat{T}} \right) &&  \{s \in \mathbf{V}\}.
            \label{eq:GINE_one-hot_encoded}
        \end{flalign}
        To enable TL to train new problem-specific cases with a much smaller dataset, we use both one-hot encoded feature vectors from equation (\ref{eq:one-hot-encoded}) and equation (\ref{eq:GINE_one-hot_encoded}) in the remaining prediction steps.
        We take advantage of the pre-trained weights from $\mu_G$ shown in equation (\ref{eq:one-hot-encoded}), and implement TL for generating the one-hot encoded feature vector in MicroCrack-GNN, $\{ p_{s} \}$.
        This approach provides a generalized GNN architecture for XFEM fracture problems where new node and edge features can be included in future works involving other scenarios.    

    \subsection{K-GNN, C-GNN, and P-GNN}\label{subsubsect:K-GNN}
        
        As shown in Figure \ref{fig:MicroCrack-GNN_structure}b, the next step was an MLP with two prediction outputs initially, followed by another MLP with a single prediction output.
        The first MLP, K-intensity-factor-GNN ({$K$-GNN}), was a regression MLP for predicting the Mode-I and Mode-II stress intensity factors for all crack-tips, $(\hat{K}_{I})_{s}$ and $(\hat{K}_{II})_{s}$, at future time-steps.
        The second MLP, Classifier-GNN ({$C$-GNN}), was a classifier MLP for predicting the quasi-static behavior of all crack-tip (i.e., for predicting propagating and non-propagating crack-tips), $(\hat{Q}_{s}$) at future time-steps.
        The inputs to $K$-GNN and $C$-GNN involved both one-hot encoded feature vectors, generated following the same procedures described in Section \ref{subsect:GraphTheory} and Section \ref{subsubsec:XFEM_GraphTheory}. 
        Since we applied TL of $\mu_{G}$, both $K$-GNN and $C$-GNN required significantly smaller training datasets to achieve high accuracies.% compared to MicroCrack-GNN
        We describe the resulting initial input graphs in equation (\ref{eq:initial_MLPs}), where the first set of inputs are the one-hot encoded vertex-edge-vertex feature vectors from equation (\ref{eq:one-hot-encoded}) and the one-hot encoded feature vector from equation (\ref{eq:GINE_one-hot_encoded}) along with their groups of nearest-neighbors in time sequence {$\hat{\mathbf{T}}$}.
        We note that the predicted Mode-I and Mode-II stress intensity factors from {$K$-GNN} can be used to compute the linear elastic fracture mechanics (LEFM) stress distribution in the domain by the principle of superposition \cite{Bower1991Applied}.
        \begin{flalign}
            &&\left(\{\hat{K}_{I_{s}}\}, \{\hat{K}_{II_{s}}\}\right)^{T+1}  \longleftarrow \textit{K-GNN}\left[\{ p_{s}, q_{s} \}_{t=T-3}^{T}|\hat{N}_{s}\right], \nonumber\\ 
            &&\{\hat{Q}_{s}\}^{T+1}  \longleftarrow \textit{C-GNN}\left[\{ p_{s}, q_{s} \}_{t=T-3}^{T}|\hat{N}_{s}\right] &&  \{s \in \mathbf{V}\}.
            \label{eq:initial_MLPs}
        \end{flalign}
        
        Furthermore, we use the outputs from the $K$-GNN and $C$-GNN as input to an additional MLP network with a single prediction output (Figure \ref{fig:MicroCrack-GNN_structure}c). 
        The final MLP, Propagate-GNN ($P$-GNN), is a regression MLP for predicting the future $x$- and $y$-coordinate positions of all crack-tips.  
        We use the predicted stress intensity factors, $\hat{K}_{I}^{T+1}$ and $\hat{K}_{II}^{T+1}$, as well as the predicted quasi-static parameter, $\hat{Q}^{T+1}$, as part of the input to $P$-GNN.
        %This approach was also implemented in MicroCrack-GNN. 
        Because the propagating crack-tips in quasi-static fracture problems are driven by crack-tips where the effective stress intensity factor is greater than or equal to the critical stress intensity factor, the predicted $\hat{K}_{I}^{T+1}$, $\hat{K}_{II}^{T+1}$ and $\hat{Q}^{T+1}$ aid the final model in predicting future crack-tip positions. 
        The $P$-GNN model is described as 
        \begin{flalign}
            &&\{\hat{P}_{s}\}^{T+1}  \longleftarrow \textit{P-GNN}\left[\{ p_{s}, q_{s} \}_{t=T-3}^{T}|\hat{N}_{s}, \{ \hat{K}_{I_{s}}, \hat{K}_{II_{s}}, \hat{Q}_{s} \}^{T+1} \right] &&  \{s \in \mathbf{V}\}.
            \label{eq:Propagate_GNN}
        \end{flalign}
        
    \subsection{Order of transfer learning application}\label{subsubsect:Transfer_Learning_Order}
    
        {In Figure \ref{fig:MicroCrack-GNN_structure}c we show the order in which TL was applied.
        The pre-trained graph embeddings ($p_{s}$, and $q_{s}$) resulting from the standard case study involving vertical domains with fixed crack lengths and crack orientations were transferred as follows. 
        First, we used the graph embeddings for TL of arbitrary crack lengths.
        Next, we transferred the resulting new graph embeddings for the case of arbitrary crack orientations.
        Following this sequential approach, we also tuned the ACCURATE framework for new domain effects; square domains and horizontal domains, respectively.
        Lastly, we extended the framework's embeddings for new loading cases of shear as shown in Figure \ref{fig:MicroCrack-GNN_structure}c.  
        The resulting graph embeddings were able to emulate both tension and shear loadings for new domain configurations, with random crack lengths and crack orientations.
        We present a detailed error analysis for each case study in the following sections.
        }

\section{Results}\label{sec:Results}

    As described previously in Section \ref{subsect:Matlab}, the test datasets involved {10} simulations for each case study: (i) standard case with vertical domain, (ii) arbitrary length, (iii) arbitrary angle, (iv) square domain, (v) horizontal domain, and (vi) shear loading.
    This resulted in a total of {60} simulations for the following qualitative and quantitative error analyses.
    First, we present the framework's accuracy with respect to the number of training samples used for TL.  
    We then evaluate the framework's capability to emulate stress distribution and crack propagation for each case study used in the TL steps shown in Figure \ref{fig:MicroCrack-GNN_structure}c.
    For each of these case studies, we present detailed error analyses for predictions of effective stress intensity factors and crack paths.
    Then, we show ACCURATE's ability to simulate stress and crack propagation evolution for unseen cases involving new domain dimensions with arbitrary crack lengths and crack orientations subjected to tension and shear loads.  
    Lastly, we compare the required simulation times for XFEM versus ACCURATE. 
    
    \subsection{Framework's error versus number of TL training samples}\label{subsect:TL_training_samples_error} 
    
        We used 20 simulations as TL training samples in this work (i.e., approximately 50$\times$ smaller compared to the MicroCrack framework).
        The following analysis shows the effects of the number of TL training samples on the resulting model's accuracy.
        First, we generated a total of 50 simulations for the case study involving shear loading.
        We then chose 7 random groups involving 5, 10, 15, 20, 30, 40, and 50 TL training simulations.  
        For each group, we applied TL to ACCURATE for a total of 10 training epochs. 
        We chose the case of shear loading due to its higher transfer space complexity compared to other cases for tensile load, square domain, horizontal domain, arbitrary crack orientation, and arbitrary crack length.

        \begin{figure}
            \centering 
            \includegraphics[width=0.5\linewidth]{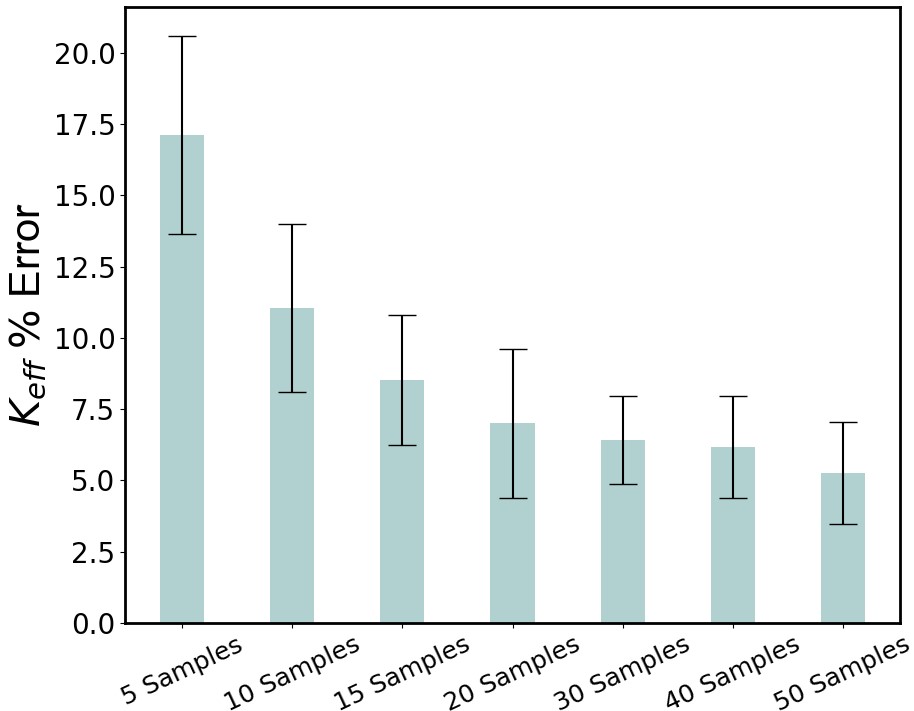}
            \centering
            \caption{Number of TL training samples versus error in stress intensity factor for 5, 10, 15, 20, 30, 40, and 50 shear loading training simulations.}
            \label{fig:TL_training_samples_error}
        \end{figure} 
        
        Figure \ref{fig:TL_training_samples_error} shows the obtained $\%$ errors in the stress intensity factor for each group (i.e., 5, 10, 15, 20, 30, 40, and 50 TL training simulations).
        We note that the highest error of $20.10 \pm 3.33 \%$ was obtained when using 5 TL simulations, as expected.
        The lowest error was obtained for 50 TL simulations at $5.27 \pm 1.78 \%$.
        We observe from Figure \ref{fig:TL_training_samples_error} that the error decreases with the increasing number of TL training samples.
        Thus, further increasing the number of TL training samples would result in a minor decrease in the error.    
        Additionally, using larger training datasets would result in longer training times.   
        Therefore, for each case study included in this work, we used a total of 20 samples during TL to achieve good prediction accuracy while decreasing training time.
        We emphasize that each simulation involves up to 101 time steps; using 20 training samples results in a TL dataset of up to 1,940 inputs (described in Section \ref{subsect:Matlab})

    \subsection{Prediction of Mode-I and Mode-II stress intensity factors}\label{subsect:stress_evolution}
    
\begin{figure}
            \begin{subfigure}[t]{1\textwidth}
                \begin{subfigure}[t]{0.32\textwidth}
                    \centering
                    \includegraphics[width=\linewidth]{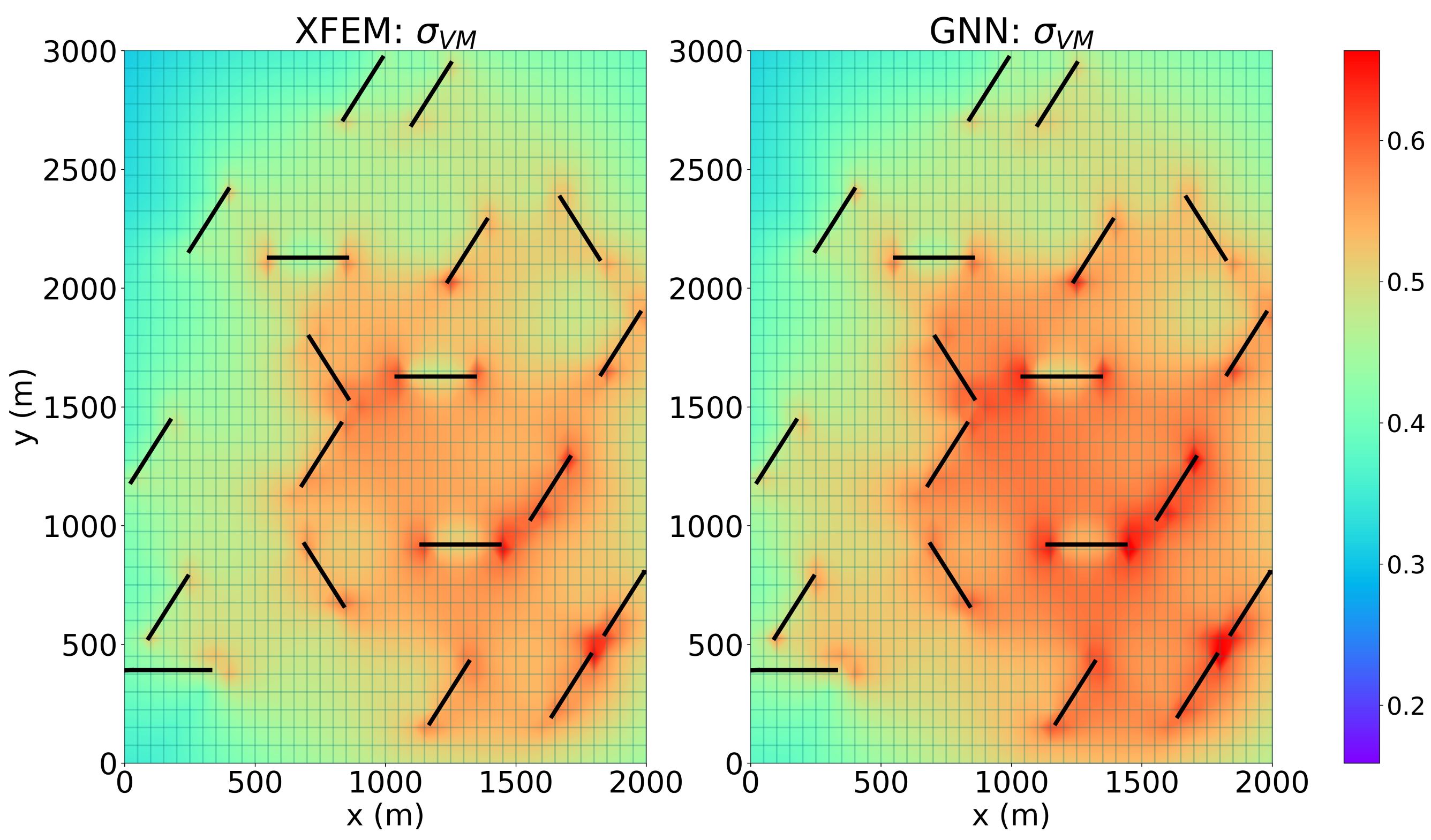}
                    \caption{Tensile load: t=$1\%$}
                \end{subfigure}
                \begin{subfigure}[t]{0.32\textwidth}
                    \centering
                    \includegraphics[width=\linewidth]{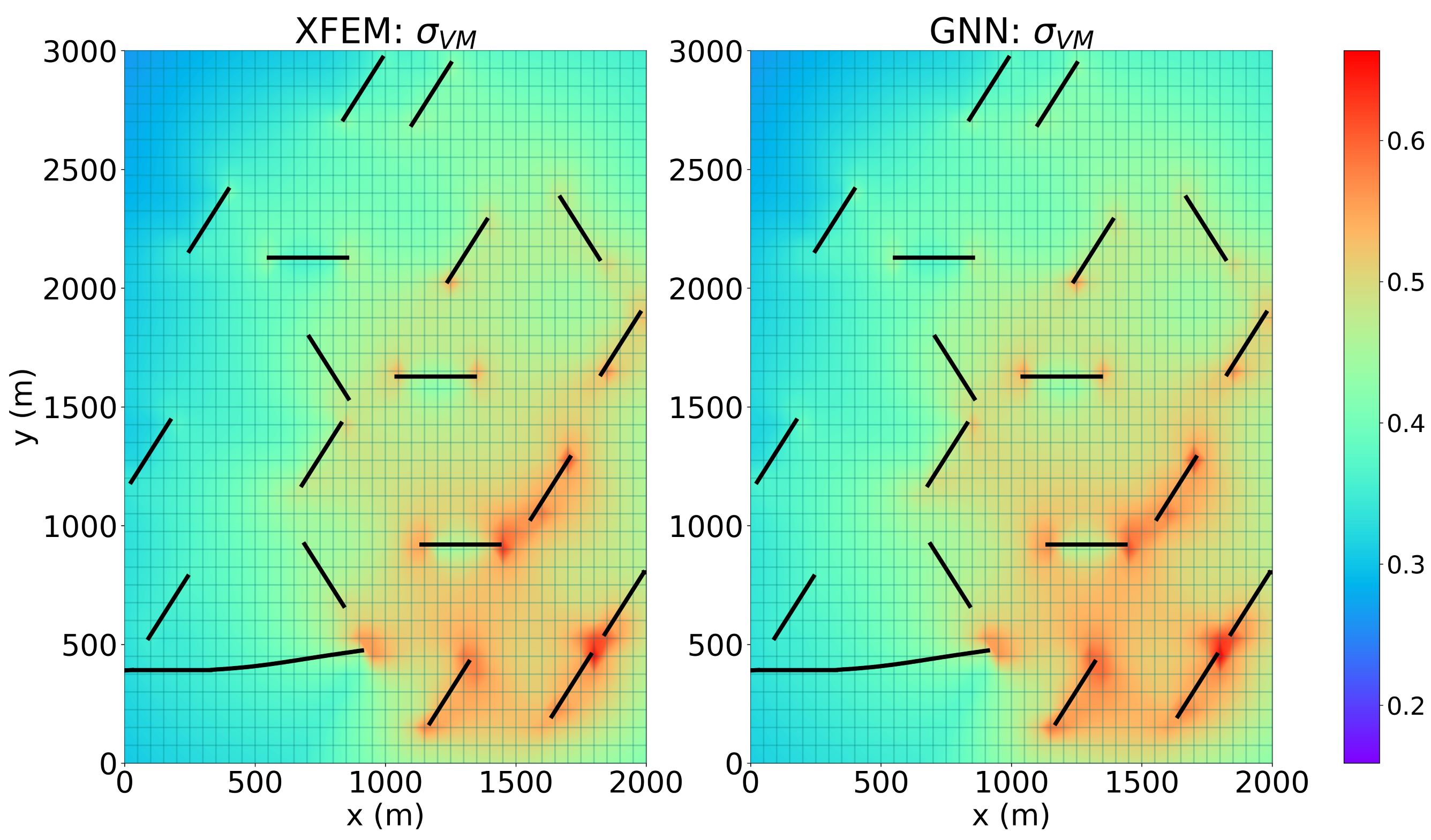}
                    \caption{Tensile load: t=$45\%$}
                \end{subfigure}
                \begin{subfigure}[t]{0.32\textwidth}
                    \centering
                    \includegraphics[width=\linewidth]{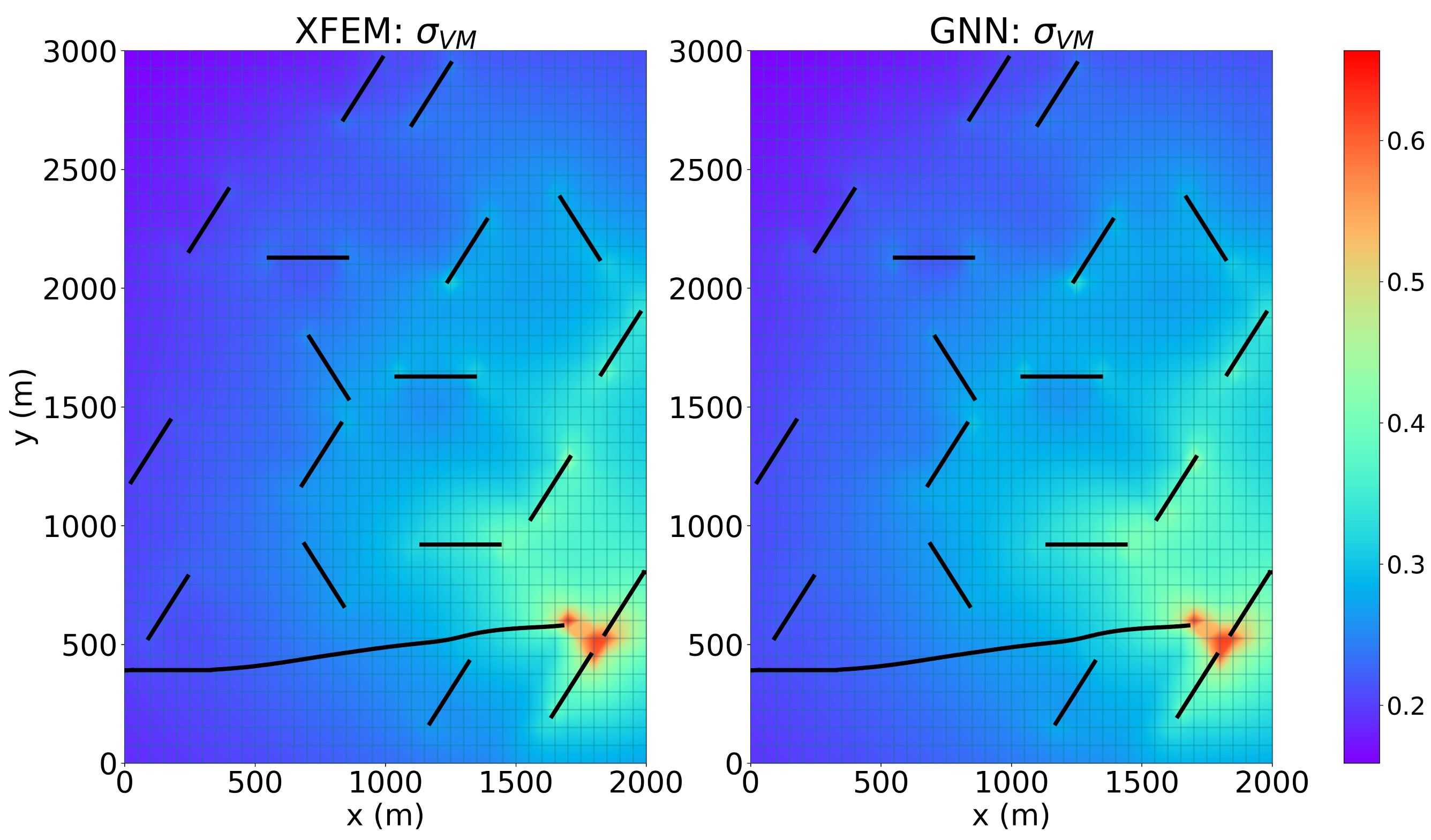}
                    \caption{Tensile load: t=$90\%$}
                \end{subfigure}
            \label{subfig:Vertical_StressContour}
            \end{subfigure}
            \begin{subfigure}[b]{1\textwidth}
                \begin{subfigure}[b]{0.32\textwidth}
                    \centering
                    \includegraphics[width=\linewidth]{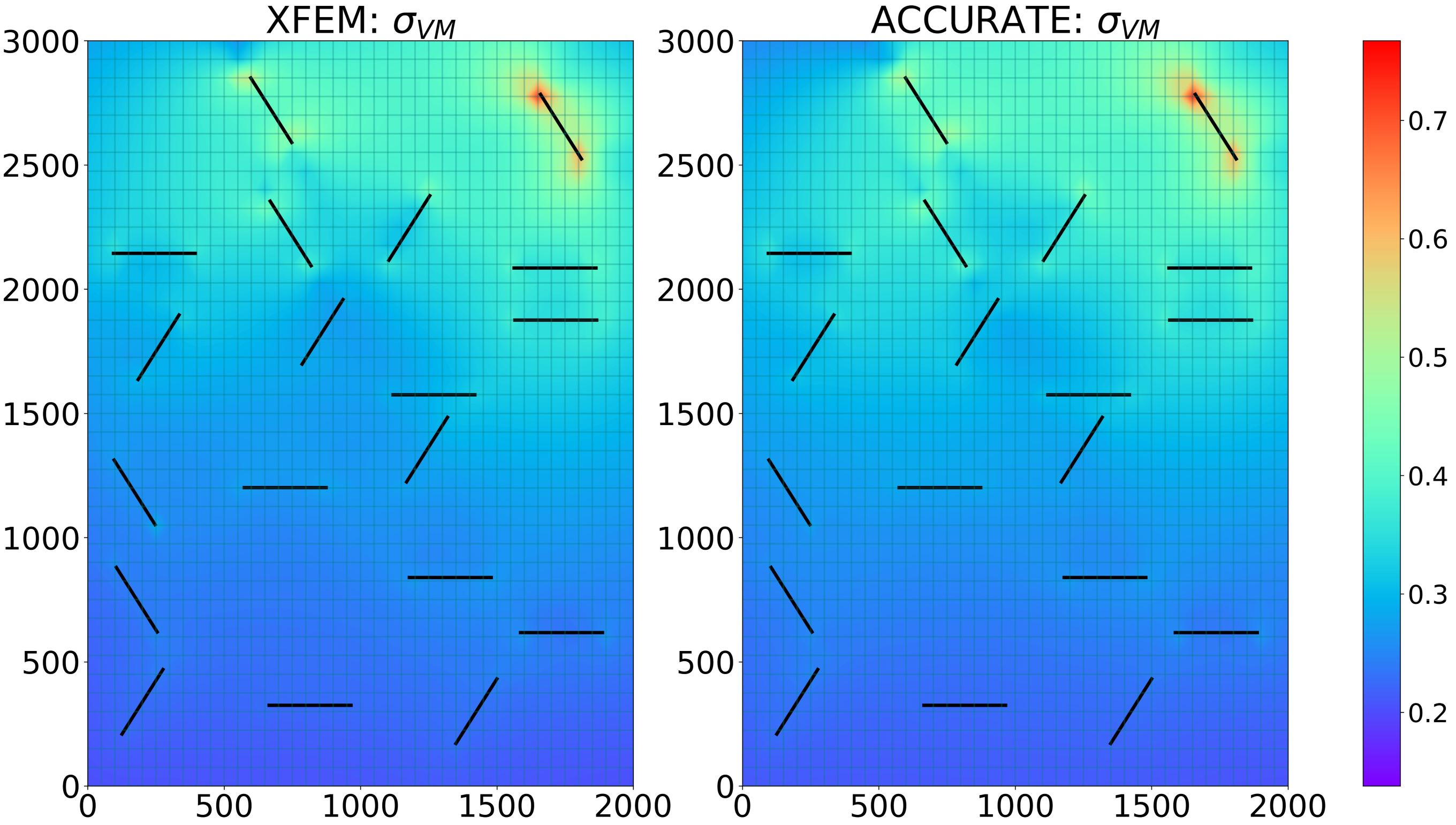}
                    \caption{Shear load: t=$1\%$}
                \end{subfigure}
                \begin{subfigure}[b]{0.32\textwidth}
                    \centering
                    \includegraphics[width=\linewidth]{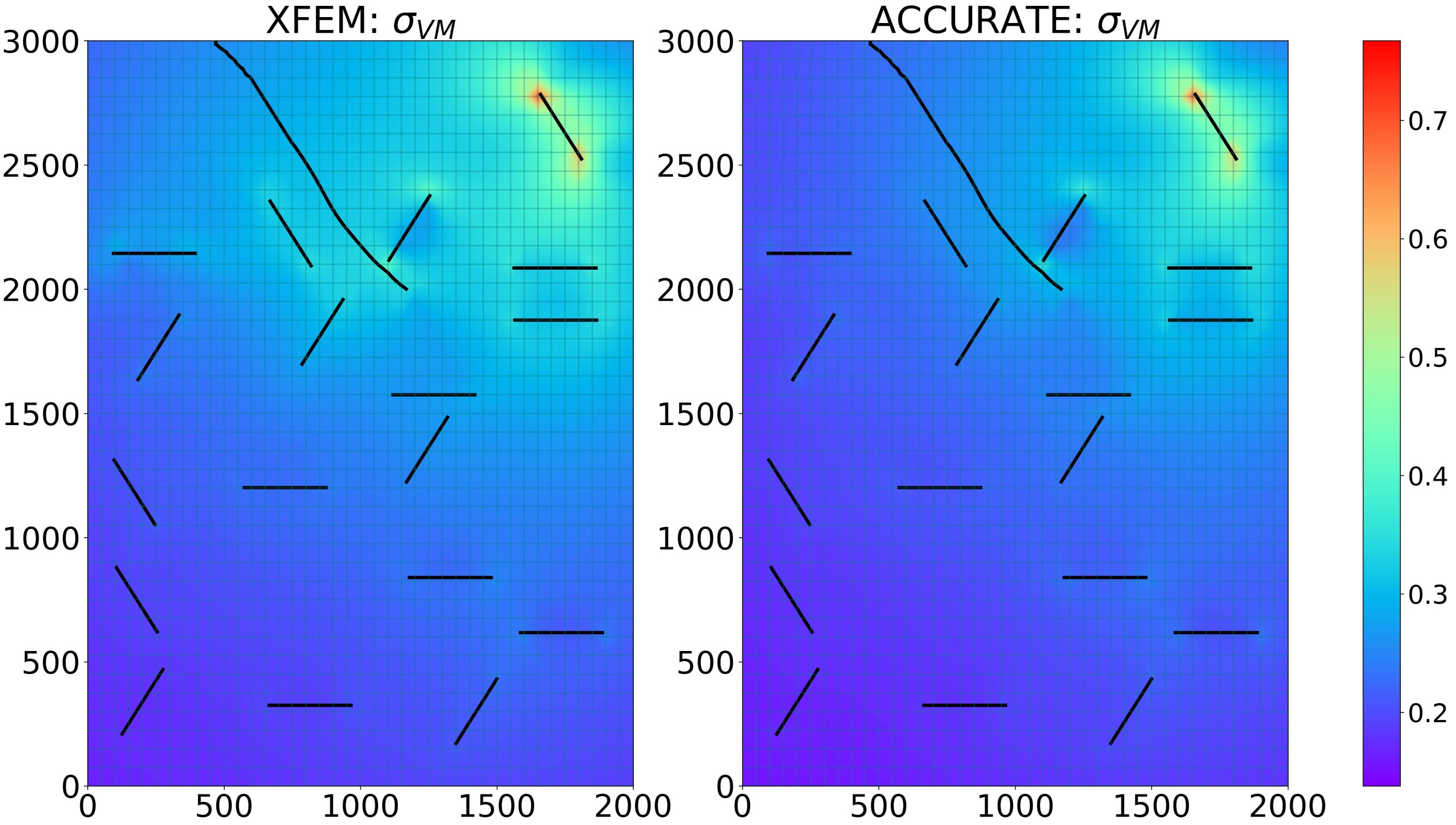}
                    \caption{Shear load: t=$45\%$}
                \end{subfigure}
                \begin{subfigure}[b]{0.32\textwidth}
                    \centering
                    \includegraphics[width=\linewidth]{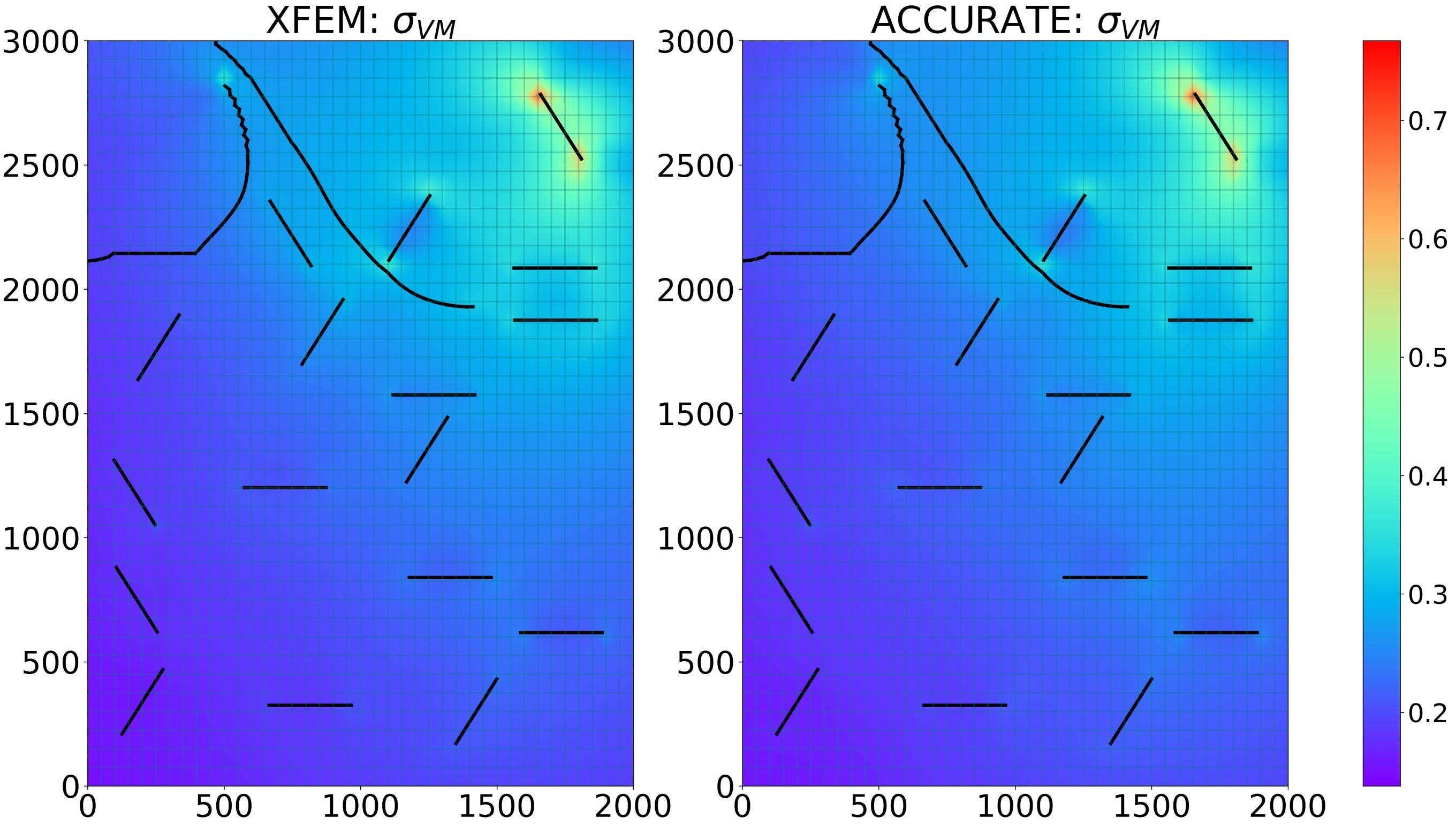}
                    \caption{Shear load: t=$90\%$}
                \end{subfigure}
            \label{subfig:Shear_StressContour}
            \end{subfigure}
            \caption{von Mises stress evolution (MPa) from $t=1\%$ to $t=90\%$ for  (a-c) test case under tensile loading, and (d-f) test case under shear loading.}
            \label{fig:Stress_evolution_LoadSamples}
        \end{figure}

        We perform a qualitative analysis on the framework's ability to emulate the stress evolution using TL.
        To compute the stress, we make use of the Linear Elastic Fracture Mechanics (LEFM) equations along with the predicted Mode-I and Mode-II stress intensity factors.
        In Figure \ref{fig:Stress_evolution_LoadSamples}, we show the time evolution of the von Mises stress for two loading types; a vertical domain subjected to tension (Figures \ref{fig:Stress_evolution_LoadSamples}a-\ref{fig:Stress_evolution_LoadSamples}c), and shear load (Figures \ref{fig:Stress_evolution_LoadSamples}d-\ref{fig:Stress_evolution_LoadSamples}f), from $t=1\%$ to $t=90\%$.
        For tensile loading, the Mode-I stress intensity factors play a critical role in determining the crack tips with the highest stress interactions, as well as the direction in which these crack tips tend to propagate.
        For shear loading, the Mode-II stress intensity factors play a higher role in determining propagating crack tips, and resulting crack path evolution.
        This can be seen in Figure \ref{fig:Stress_evolution_LoadSamples}, where the cracks subjected to tension tend to propagate horizontally, while the crack tips subjected to shear loading tend to propagate diagonally. 
        Additionally, for tensile loading, the cracks with an initial orientation of $0^{o}$ are more likely to propagate, while for shear loading the crack tips with diagonal orientations cause higher stress interactions between their neighboring cracks.
        Comparing the XFEM surrogate model versus ACCURATE qualitatively for these cases, both time evolution are nearly identical.
        Therefore, the ACCURATE framework is able to generate good stress evolution prediction for cases involving tension and shear loads in vertical domains ($2000mm \times 3000mm$) with fixed crack length and crack angles. 

\begin{figure}
            \begin{subfigure}[t]{1\textwidth}
                \begin{subfigure}[t]{0.32\textwidth}
                    \centering
                    \includegraphics[width=\linewidth]{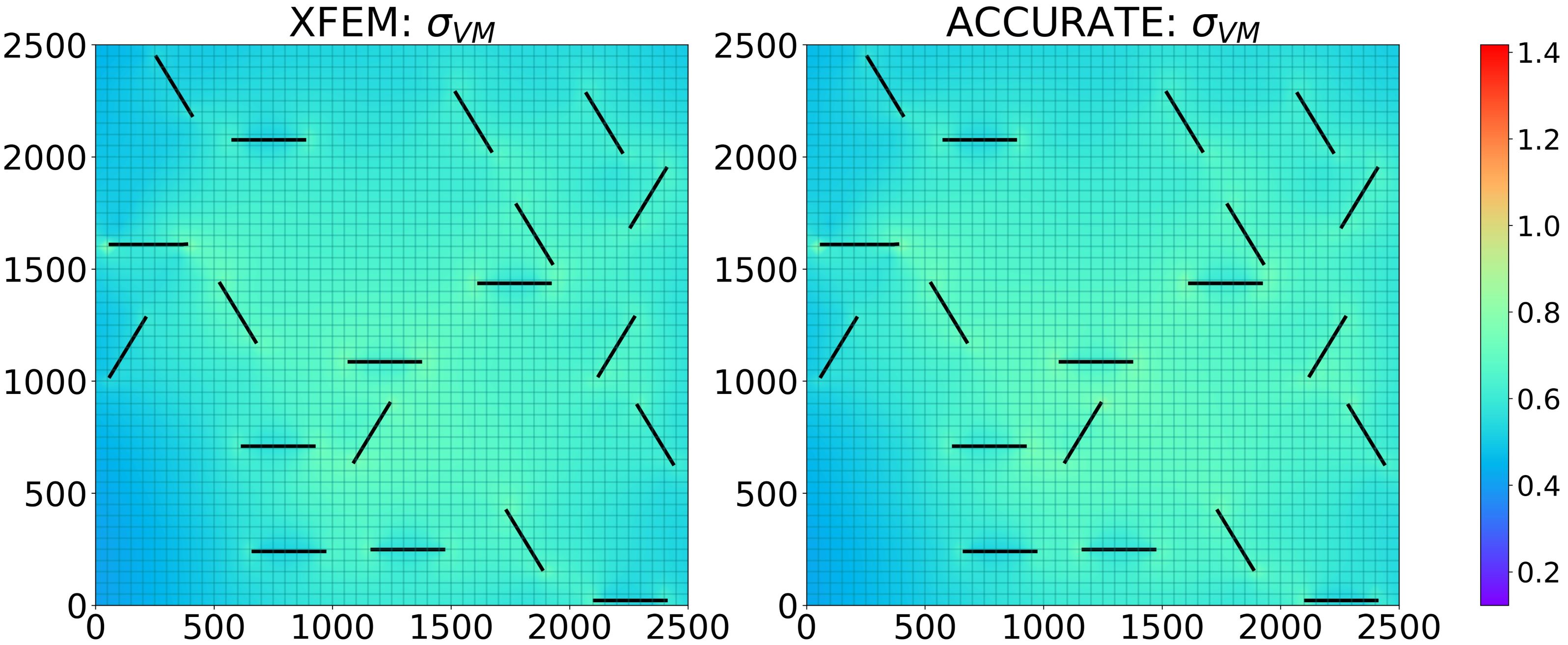}
                    \caption{Square domain: t=$1\%$}
                \end{subfigure}
                \begin{subfigure}[t]{0.32\textwidth}
                    \centering
                    \includegraphics[width=\linewidth]{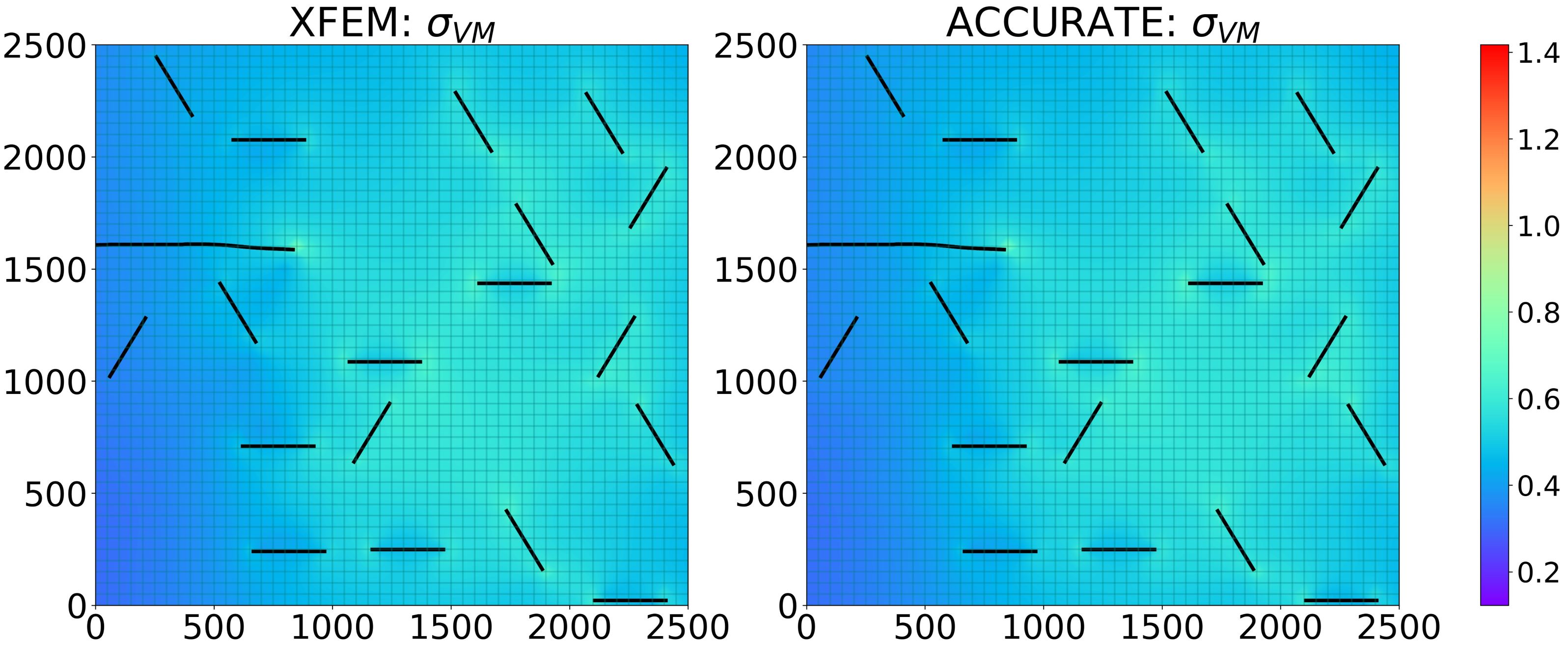}
                    \caption{Square domain: t=$45\%$}
                \end{subfigure}
                \begin{subfigure}[t]{0.32\textwidth}
                    \centering
                    \includegraphics[width=\linewidth]{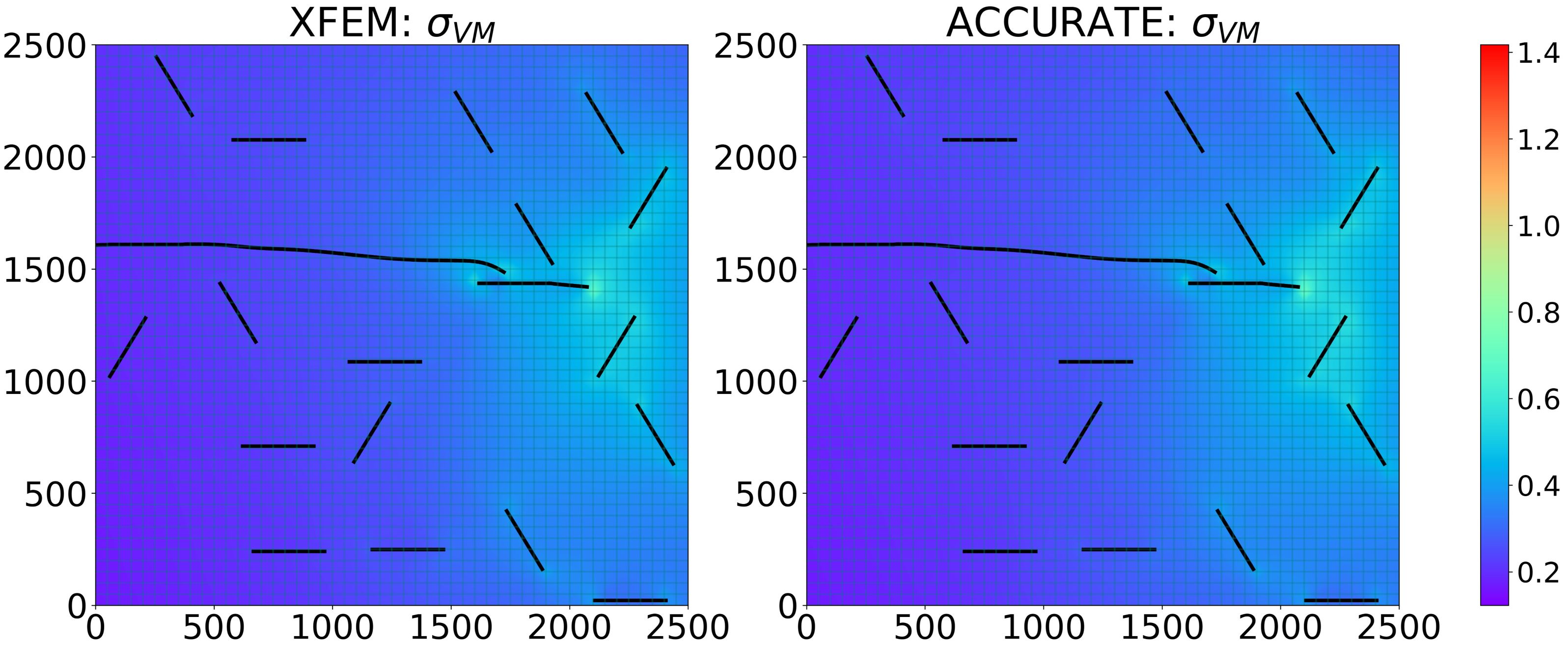}
                    \caption{Square domain: t=$90\%$}
                \end{subfigure}
            \label{subfig:Square_StressContour}
            \end{subfigure}
            \begin{subfigure}[b]{1\textwidth}
                \begin{subfigure}[b]{0.32\textwidth}
                    \centering
                    \includegraphics[width=\linewidth]{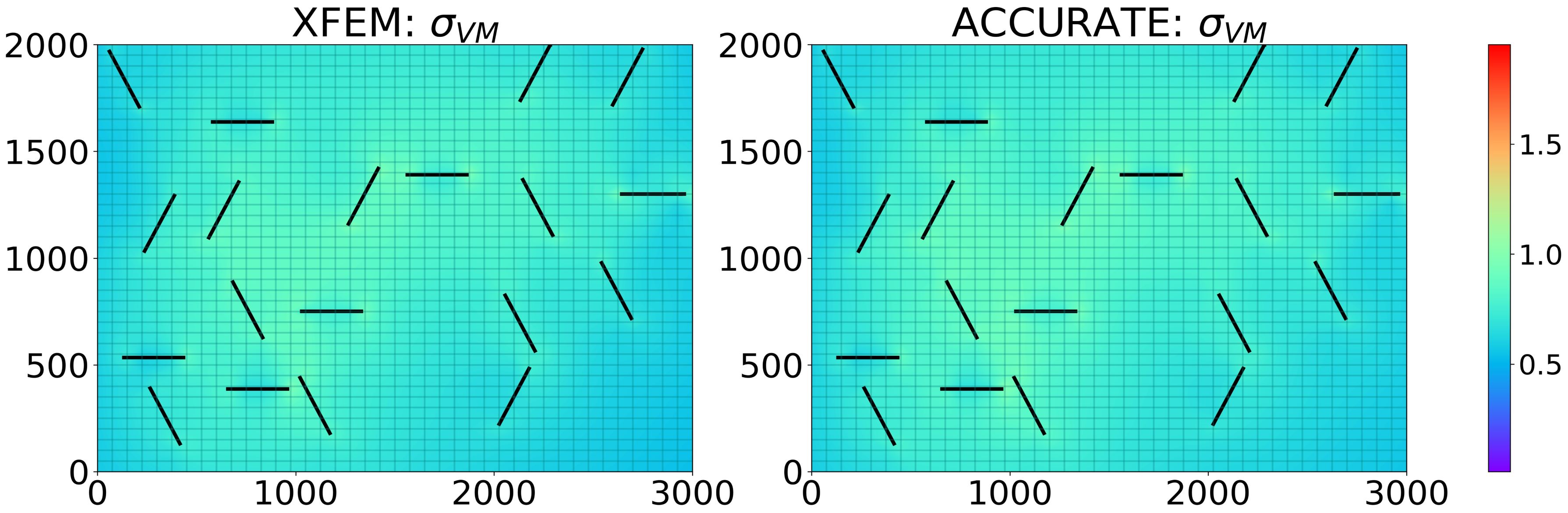}
                    \caption{Horizontal domain: t=$1\%$}
                \end{subfigure}
                \begin{subfigure}[b]{0.32\textwidth}
                    \centering
                    \includegraphics[width=\linewidth]{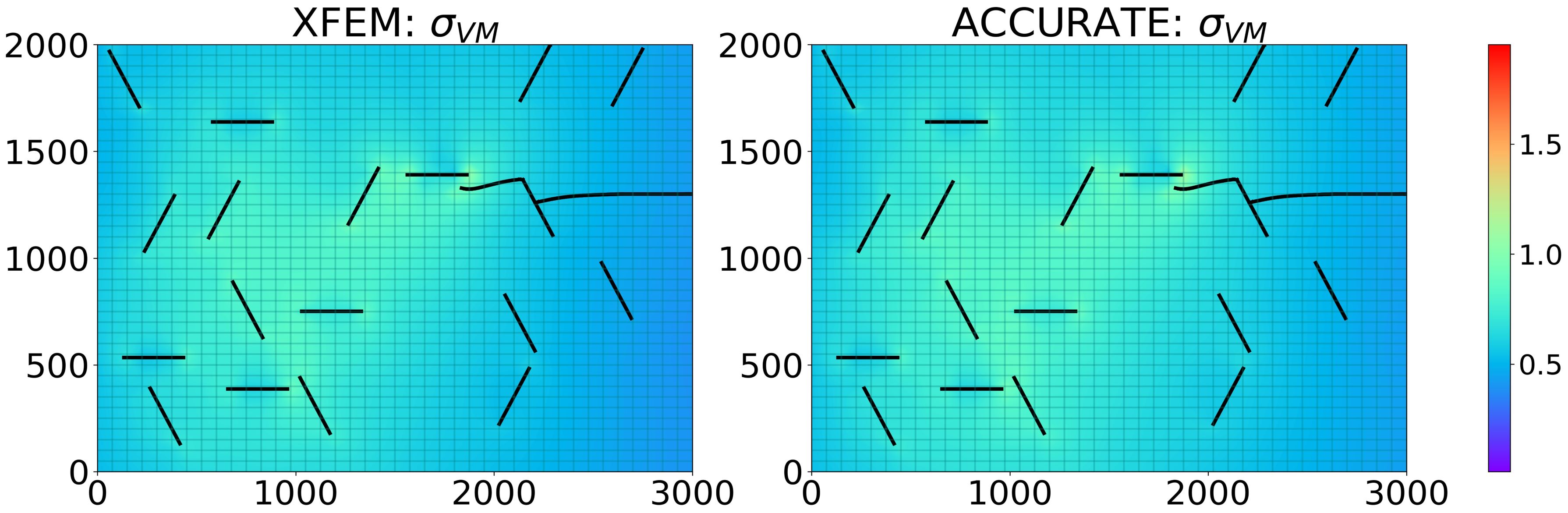}
                    \caption{Horizontal domain: t=$45\%$}
                \end{subfigure}
                \begin{subfigure}[b]{0.32\textwidth}
                    \centering
                    \includegraphics[width=\linewidth]{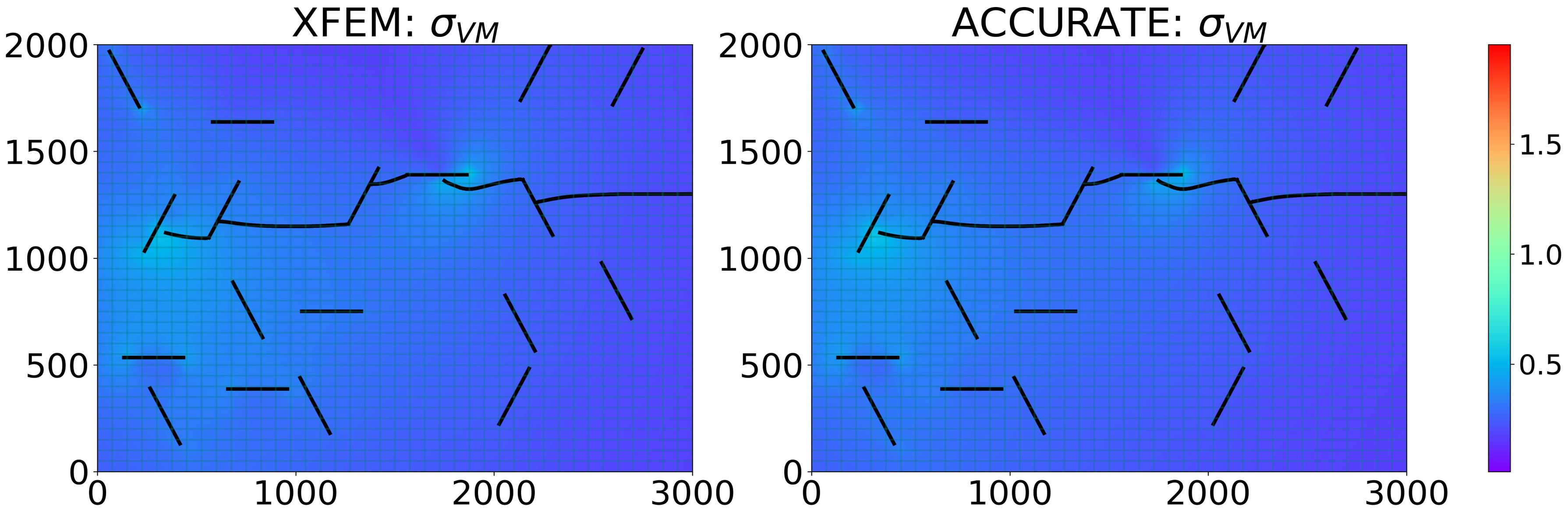}
                    \caption{Horizontal domain: t=$90\%$}
                \end{subfigure}
            \label{subfig:Horizontal_StressContour}
            \end{subfigure}
            \caption{von Mises stress evolution (MPa) from $t=1\%$ to $t=90\%$ for  (a-c) test case with square domain, and (d-f) test case with horizontal domain.}
            \label{fig:Stress_evolution_DomainSamples}
        \end{figure}

        Following a similar approach to the varying load cases, we analyzed the performance of ACCURATE with varying domain sizes.
        As mentioned in Section \ref{subsect:Matlab}, we introduced two domains: (i) square domains ($2500mm \times 2500mm$), and (ii) horizontal domains  ($3000mm \times 2000mm$).
        We used a fixed crack length of $300mm$ and a fixed set of crack orientations of $0^{o}$, $60^{o}$, and $120^{o}$ for these case studies. 
        As a result, implementing TL for varying domain geometries, parameterized by vertical, square, or horizontal, would provide a universal framework capable of emulating fracture in new domains. 
        Figure \ref{fig:Stress_evolution_DomainSamples} shows the time evolution of von Mises stress from $t=1\%$ to $t=90\%$ for the square domain case (Figures \ref{fig:Stress_evolution_DomainSamples}a-\ref{fig:Stress_evolution_DomainSamples}c), and for the horizontal domain case (Figures \ref{fig:Stress_evolution_DomainSamples}d-\ref{fig:Stress_evolution_DomainSamples}f). 
        For the square domain, we see that for $t=1\%$ to $t=90\%$, the stress distributions generated by ACCURATE are qualitatively identical to the physics-based XFEM simulator.
        Similarly, for the horizontal domain, the stress evolution generated by ACCURATE shows good agreement with the XFEM model.
        We perform a detailed quantitative error analysis for these cases in the following section.

\begin{figure}
            \begin{subfigure}[t]{1\textwidth}
                \begin{subfigure}[t]{0.32\textwidth}
                    \centering
                    \includegraphics[width=\linewidth]{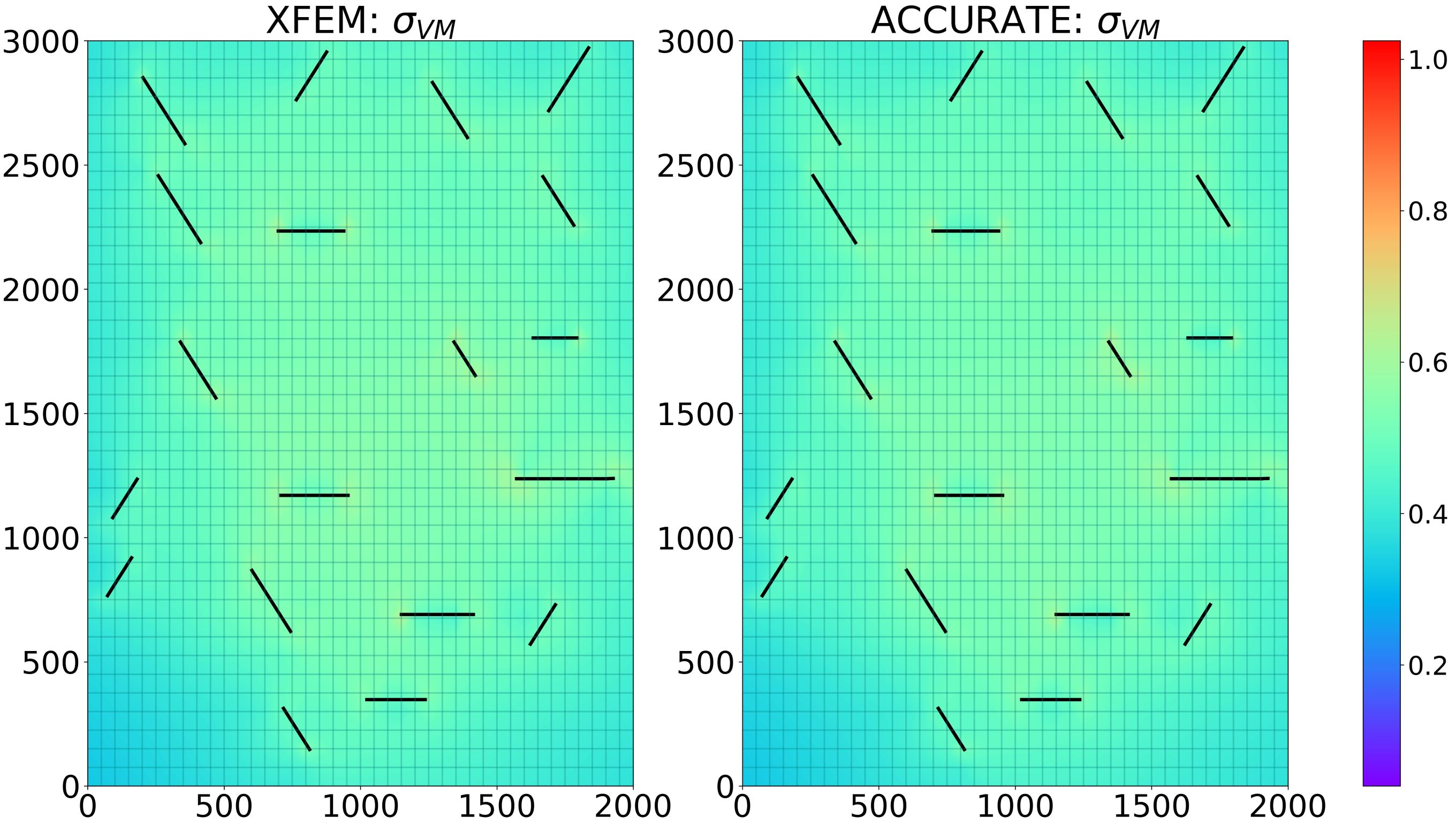}
                    \caption{Arbitrary length: t=$1\%$}
                \end{subfigure}
                \begin{subfigure}[t]{0.32\textwidth}
                    \centering
                    \includegraphics[width=\linewidth]{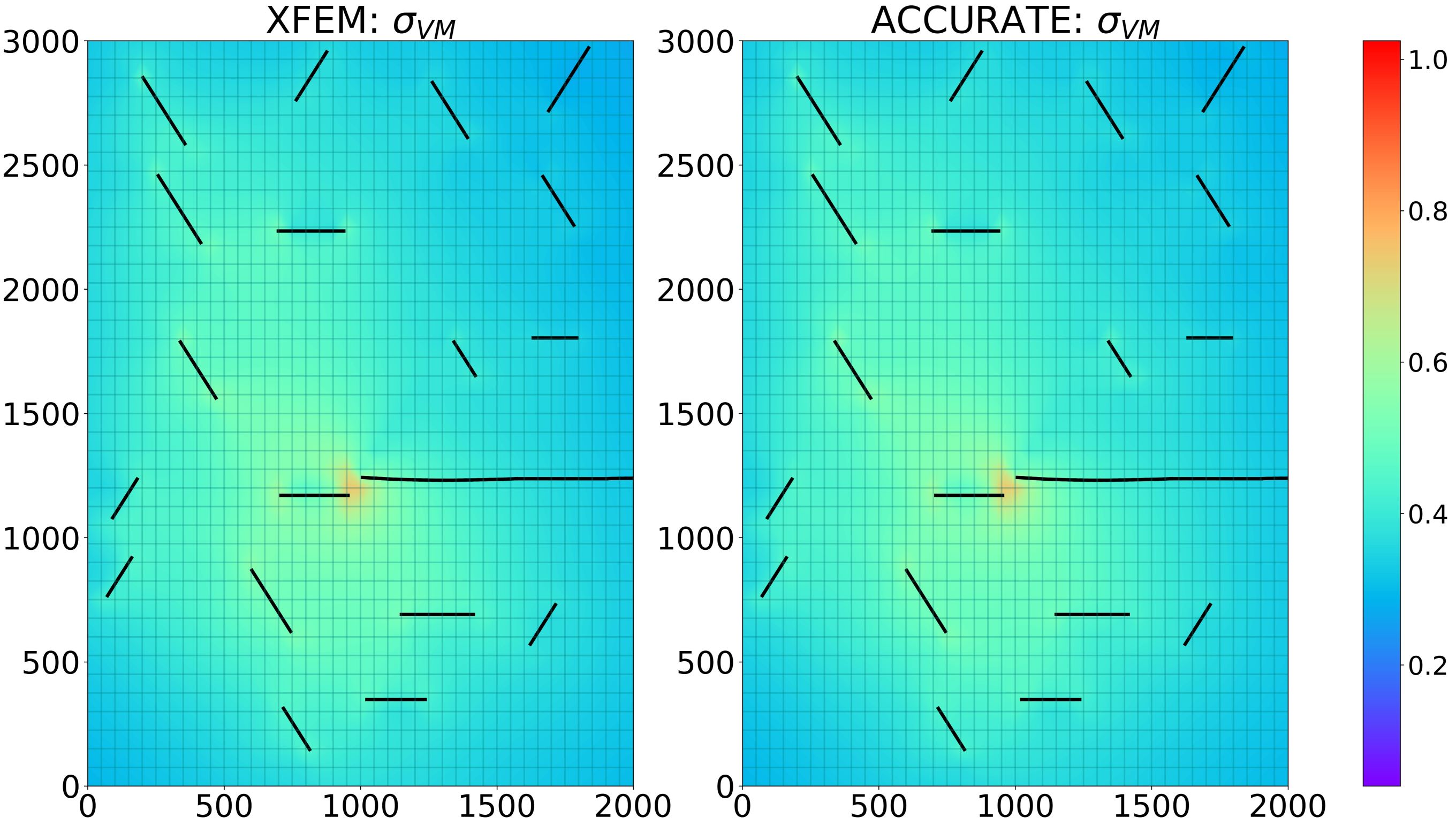}
                    \caption{Arbitrary length: t=$45\%$}
                \end{subfigure}
                \begin{subfigure}[t]{0.32\textwidth}
                    \centering
                    \includegraphics[width=\linewidth]{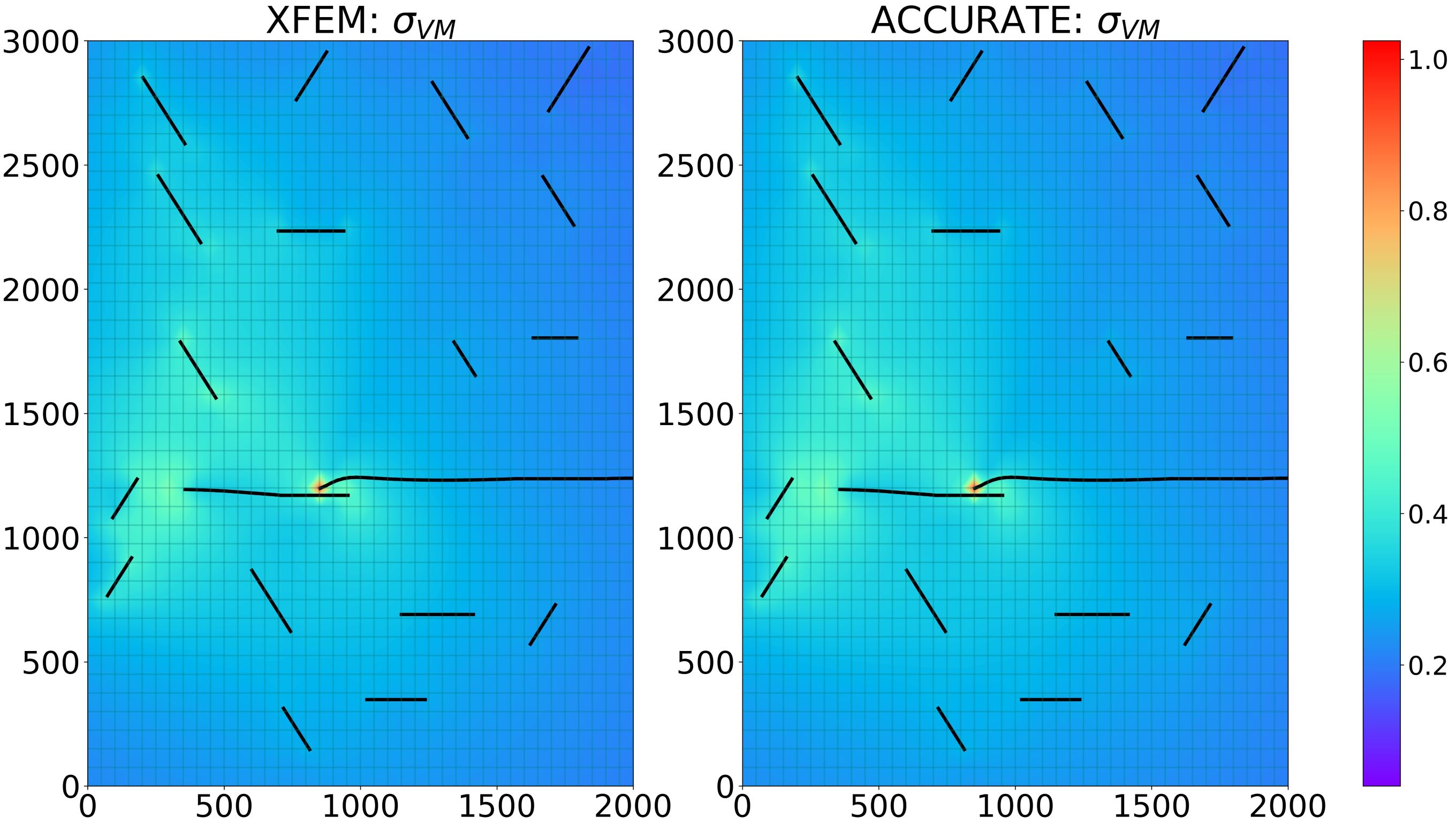}
                    \caption{Arbitrary length: t=$90\%$}
                \end{subfigure}
            \label{subfig:Length_StressContour}
            \end{subfigure}
            \begin{subfigure}[b]{1\textwidth}
                \begin{subfigure}[b]{0.32\textwidth}
                    \centering
                    \includegraphics[width=\linewidth]{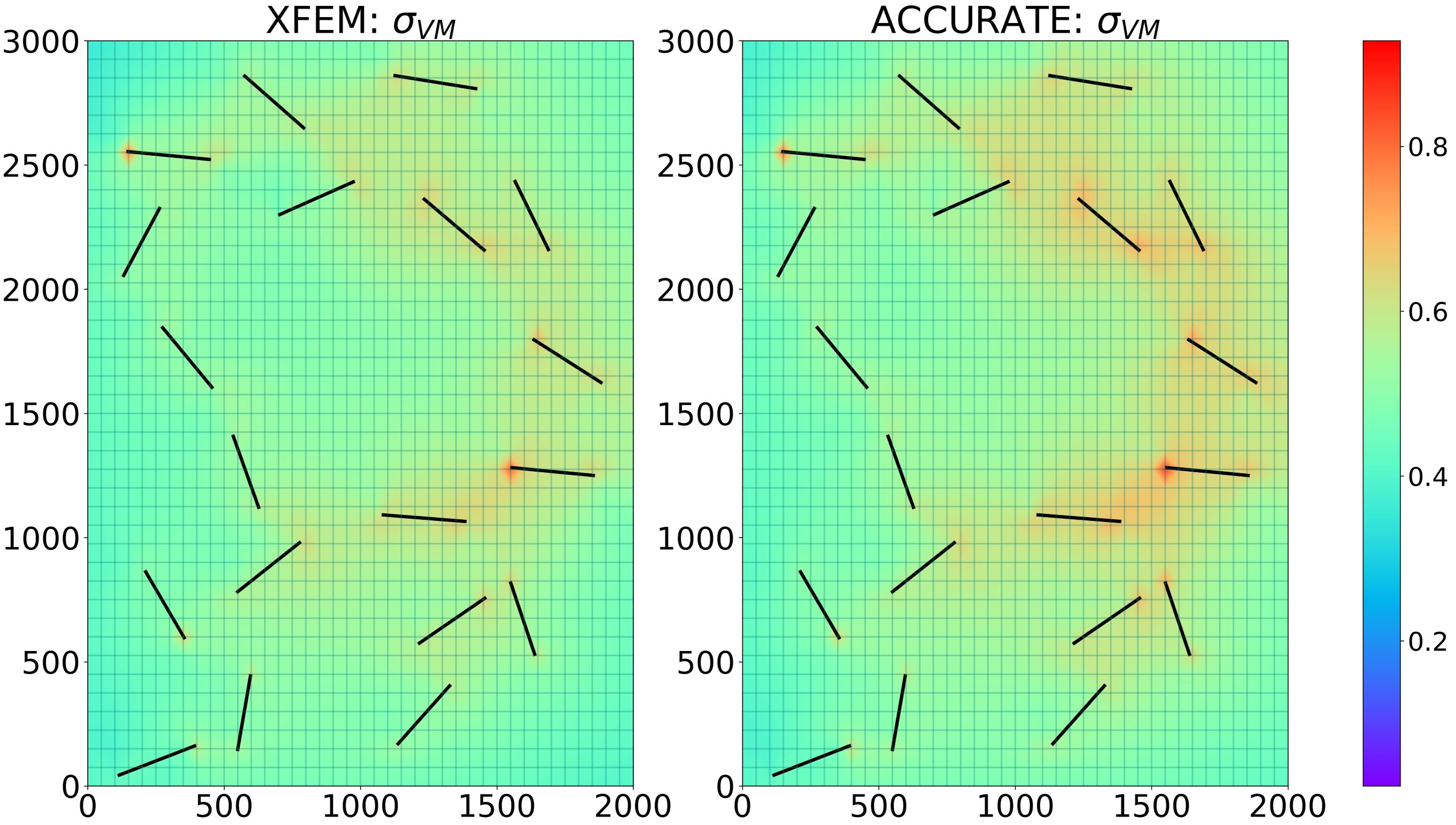}
                    \caption{Arbitrary angle: t=$1\%$}
                \end{subfigure}
                \begin{subfigure}[b]{0.32\textwidth}
                    \centering
                    \includegraphics[width=\linewidth]{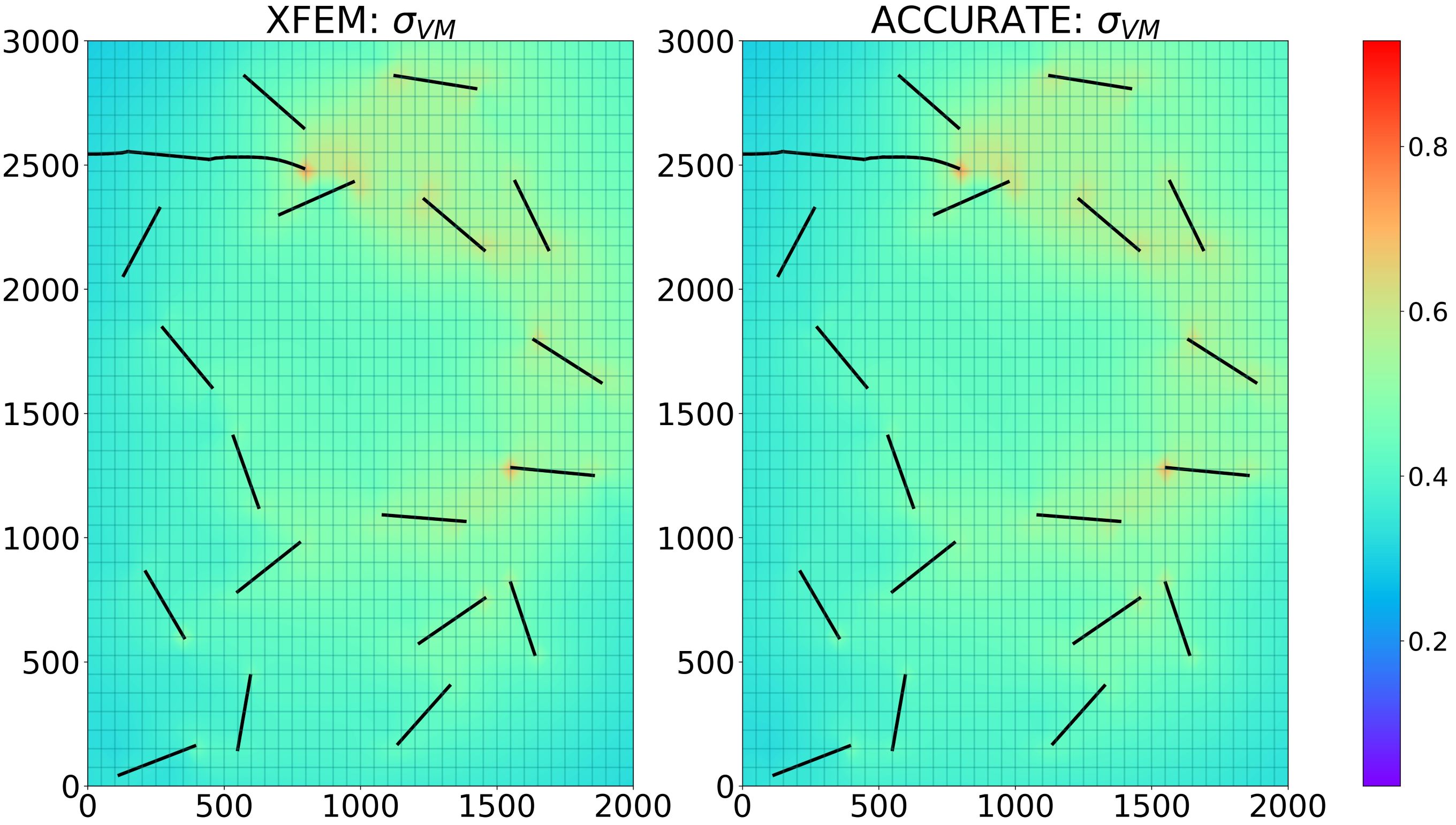}
                    \caption{Arbitrary angle: t=$45\%$}
                \end{subfigure}
                \begin{subfigure}[b]{0.32\textwidth}
                    \centering
                    \includegraphics[width=\linewidth]{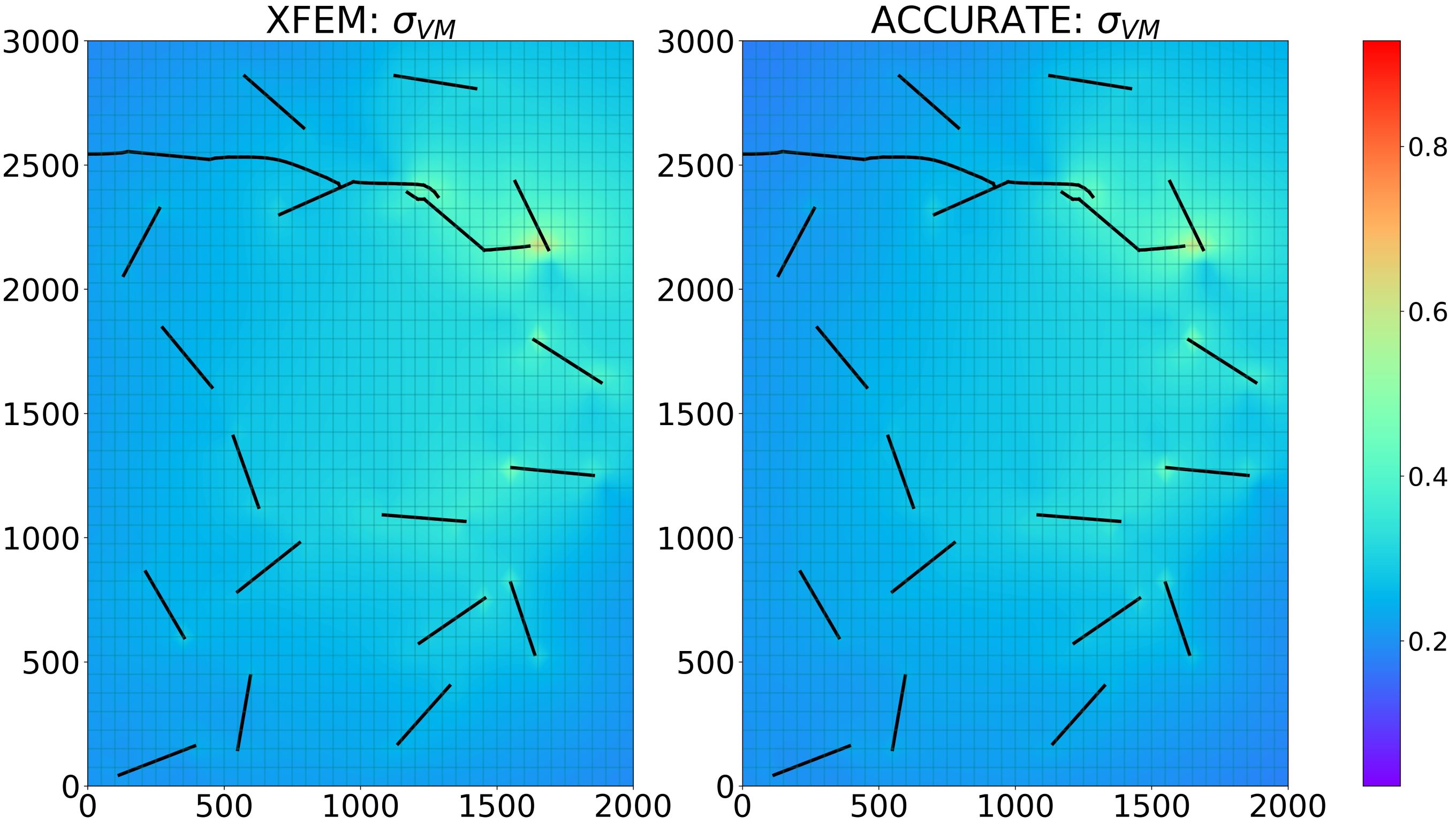}
                    \caption{Arbitrary angle: t=$90\%$}
                \end{subfigure}
            \label{subfig:Angle_StressContour}
            \end{subfigure}
            \caption{von Mises stress evolution (MPa) from $t=1\%$ to $t=90\%$ for  (a-c) test case with arbitrary crack length, and (d-f) test case with arbitrary crack orientation.}
            \label{fig:Stress_evolution_LengthAngle}
        \end{figure}

        Lastly, we evaluated the ACCURATE framework for predicting the stress evolution in cases involving cracks with arbitrary lengths, and arbitrary angles.
        We note that the domain size was fixed as vertical ($2000mm \times 3000mm$) for these case studies.
        We show the resulting stress evolution in Figures \ref{fig:Stress_evolution_LengthAngle}a-\ref{fig:Stress_evolution_LengthAngle}c, and Figures \ref{fig:Stress_evolution_LengthAngle}d-\ref{fig:Stress_evolution_LengthAngle}f for arbitrary crack lengths and arbitrary crack orientations, respectively.
        For arbitrary crack lengths, the ACCURATE framework shows good prediction agreement to the XFEM stresses during all time steps shown. 
        These results suggest that the ACCURATE framework originally trained for vertical domains subjected to tension, is able to transfer knowledge for cases with new crack lengths. 
        Additionally, for the case involving arbitrary angles, although Figures \ref{fig:Stress_evolution_LengthAngle}d-\ref{fig:Stress_evolution_LengthAngle}f show ACCURATE to predict overall good stress distributions, the GNN framework shows slight errors at locations of high stress interactions. 
        For instance, at $t=1\%$ in Figure \ref{fig:Stress_evolution_LengthAngle}d, the highest stress interaction is observed for the right-most crack at approximately $\{x=1600mm,y=1200mm\}$.
        The predicted stress distribution at this location can be seen as slightly higher than that computed by the XFEM-based simulator, thus, resulting in a higher stress distribution by the ACCURATE framework. 
        This may be due to the simulation involving propagation, interaction, and coalescence of a crack with orientation close to $90^{o}$ (right-most crack at approximately $\{x=1700mm,y=2300mm\}$).
        Although an error is observed for this case, ACCURATE is able to predict the crack tips with the highest stress distributions throughout all time steps shown.

    \subsection{Prediction of microcrack propagation and coalescence}\label{subsect:crack_growth} 
    
\begin{figure}
                \begin{subfigure}[c]{0.49\textwidth}
                    \centering 
                    \includegraphics[width=\linewidth]{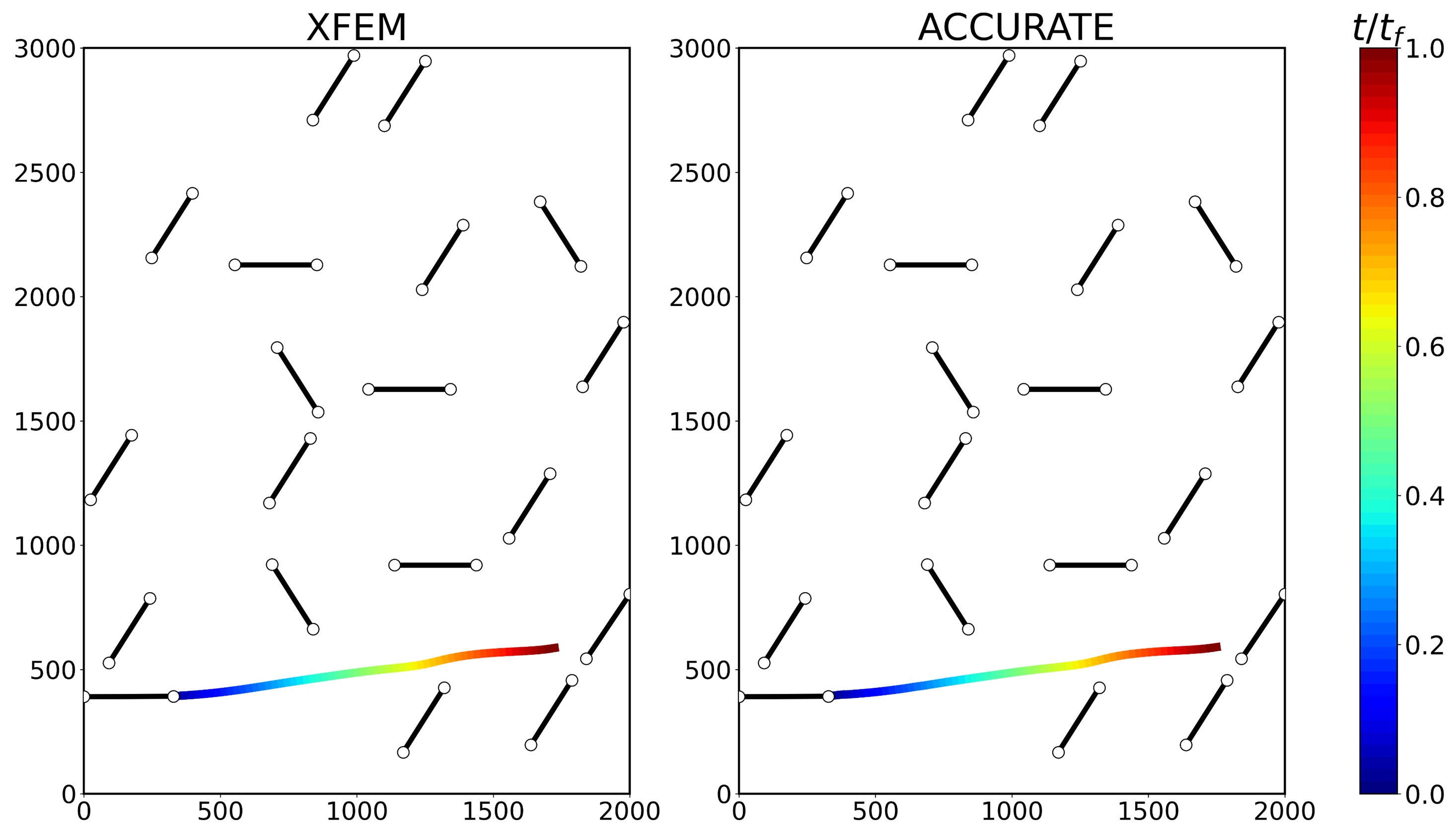}
                    \caption{Tensile load - Crack path}
                    \label{subfig:TensileLoad_CrackPath}
                \end{subfigure}
                \begin{subfigure}[c]{0.49\textwidth}
                    \centering 
                    \includegraphics[width=\linewidth]{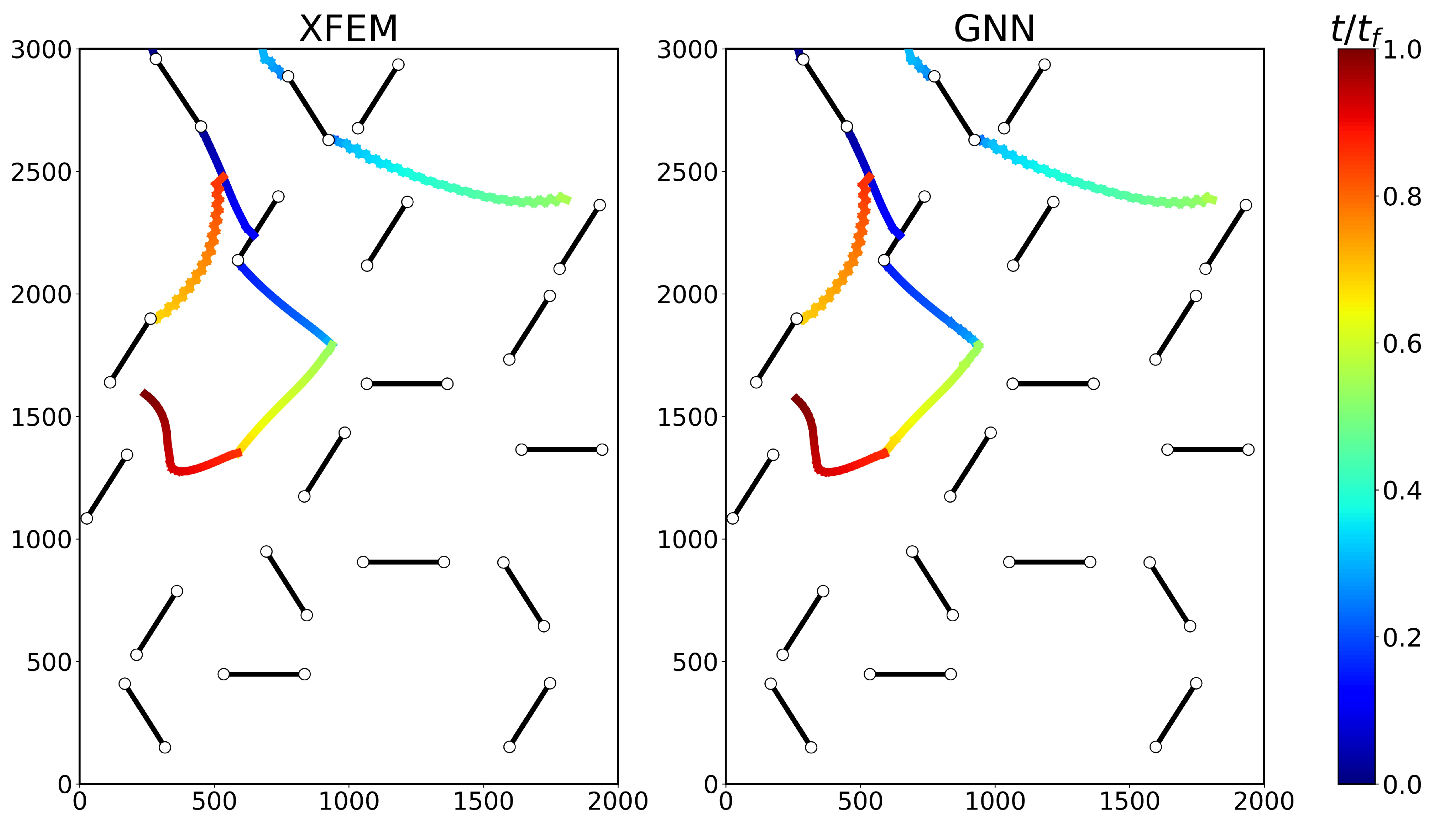}
                    \caption{Shear load - Crack path}
                    \label{subfig:ShearLoad_CrackPath}
                \end{subfigure}
                \caption{Crack path evolution for (a) test case under Tensile load, and (b) test case under Shear loading.}    
                \label{fig:CrackPath_evolution_LoadingSamples}
            \end{figure}

        Next, we perform a similar qualitative analysis %similar to Section \ref{subsect:stress_evolution} 
        for the predicted crack paths (using \textit{P}-GNN) of each simulation shown in Section \ref{subsect:stress_evolution}.
        Figures \ref{subfig:TensileLoad_CrackPath} and \ref{subfig:ShearLoad_CrackPath} show the evolution of crack growth for the case studies involving a vertical domain subjected to tension and shear loads, respectively.
        From Figure \ref{subfig:TensileLoad_CrackPath}, the predicted evolution of crack growth for the vertical domain subjected to tension is qualitatively indistinguishable from that generated by the XFEM model.
        This is consistent with the Microcrack-GNN which performed predictions of crack growth for a vertical domain with high accuracy. 
        Similarly, for the vertical domain subjected to shear load shown in Figure \ref{subfig:ShearLoad_CrackPath} the predicted crack path evolution by ACCURATE is also qualitatively identical to the XFEM-based simulator.
        In Section \ref{subsect:stress_evolution}, we showed the GNN framework's ability to predict stress evolution for cases involving shear loads. 
        Because ACCURATE is able to predict the stress intensity factors at future time steps prior to predicting the future crack-tip positions, as shown in equation (\ref{eq:Propagate_GNN}), we take advantage of these prior predictions in the \textit{P}-GNN model by including the predicted $\hat{K}_{I}^{T+1}$ and $\hat{K}_{II}^{T+1}$ as part of \textit{P}-GNN's input.
        Therefore, the high accuracy achieved when predicting the stress evolution aids \textit{P}-GNN to achieve high accuracy for the case study involving shear loading.

\begin{figure}
                \begin{subfigure}[c]{0.49\textwidth}
                    \centering 
                    \includegraphics[width=\linewidth]{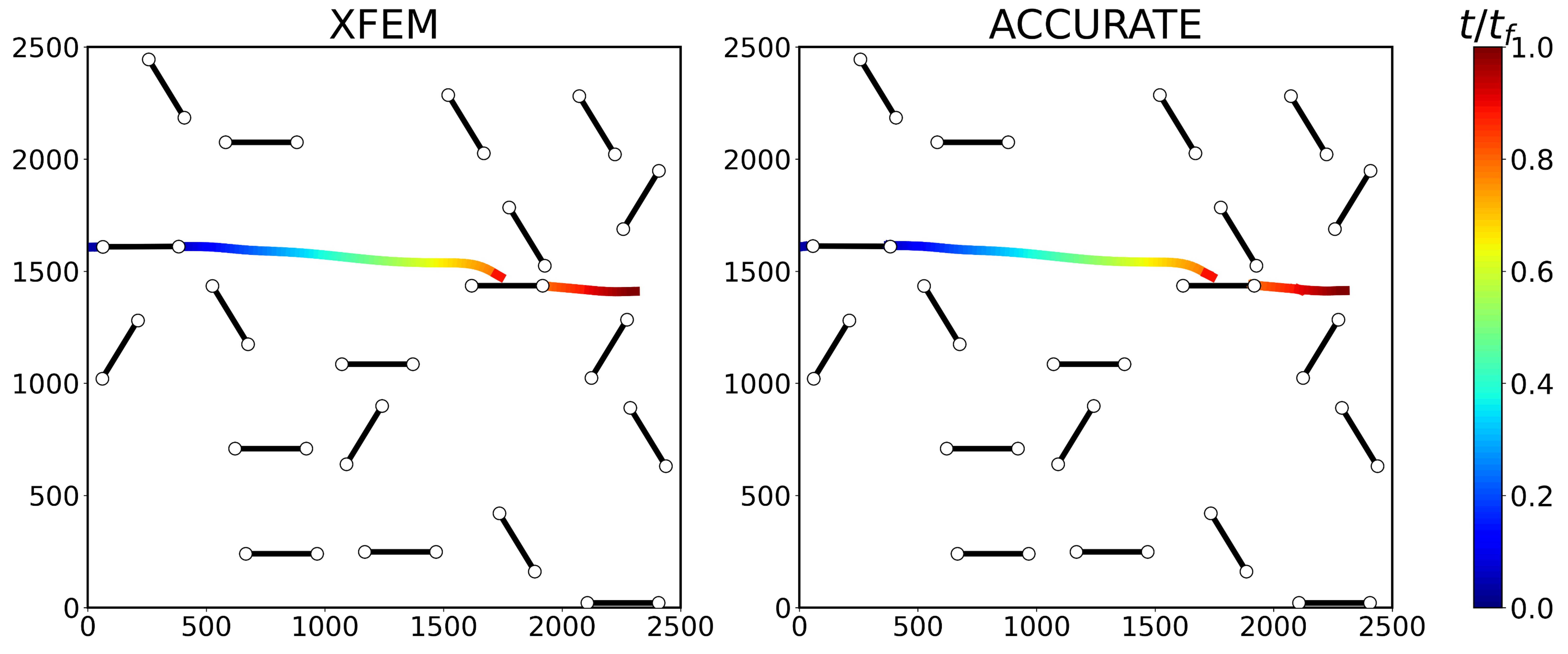}
                    \caption{Square domain - Crack path}
                    \label{subfig:SquareDomain_CrackPath}
                \end{subfigure}
                \begin{subfigure}[c]{0.49\textwidth}
                    \centering 
                    \includegraphics[width=\linewidth]{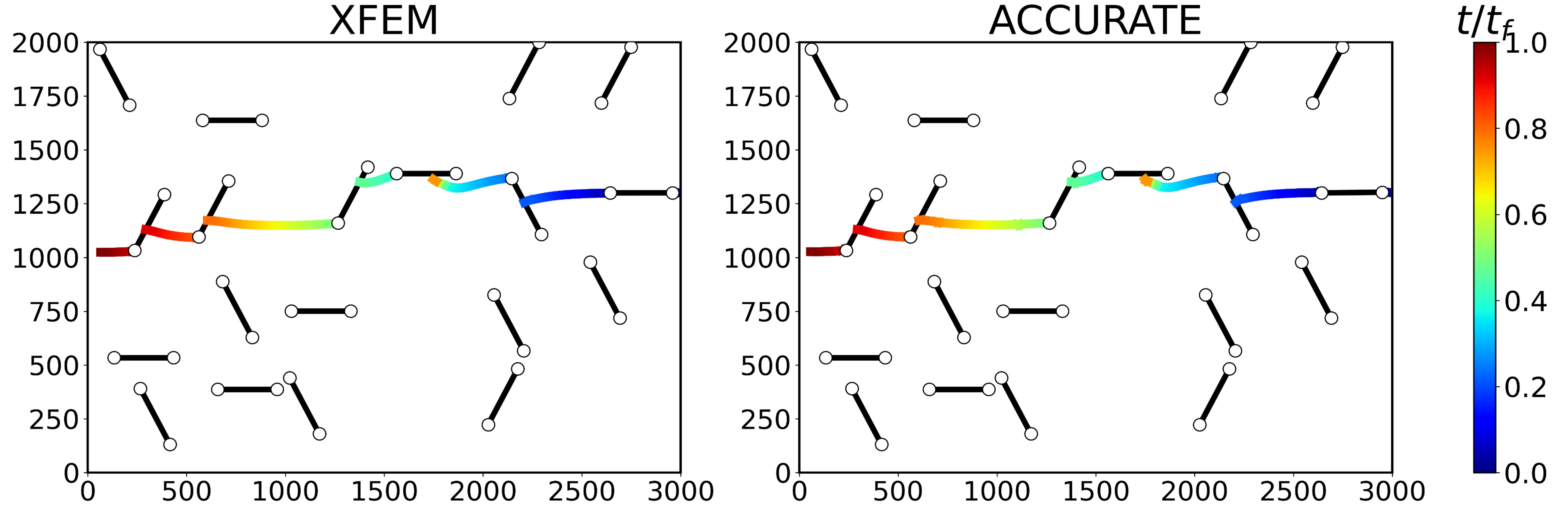}
                    \caption{Horizontal domain - Crack path}
                    \label{subfig:HorizontalDomain_CrackPath}
                \end{subfigure}
                \caption{Crack path evolution for (a) test case with Square domain, and (b) test case with Horizontal domain.}
                \label{fig:CrackPath_evolution_DomainSamples}
            \end{figure}

            Figures \ref{subfig:SquareDomain_CrackPath} - \ref{subfig:HorizontalDomain_CrackPath} show the evolution of crack growth of the square domain and horizontal domain cases, respectively, for the XFEM simulator versus ACCURATE.
            Similar to the predicted stress evolution of the square domain case, the predicted crack growth evolution is qualitatively identical to the XFEM crack path evolution.
            For the horizontal domain case study, we obtained good crack path predictions compared to XFEM.
            A detailed error analysis is described in Section \ref{subsec:results_Keff_CProp}.
    
\begin{figure}
                \begin{subfigure}[c]{0.49\textwidth}
                    \centering 
                    \includegraphics[width=\linewidth]{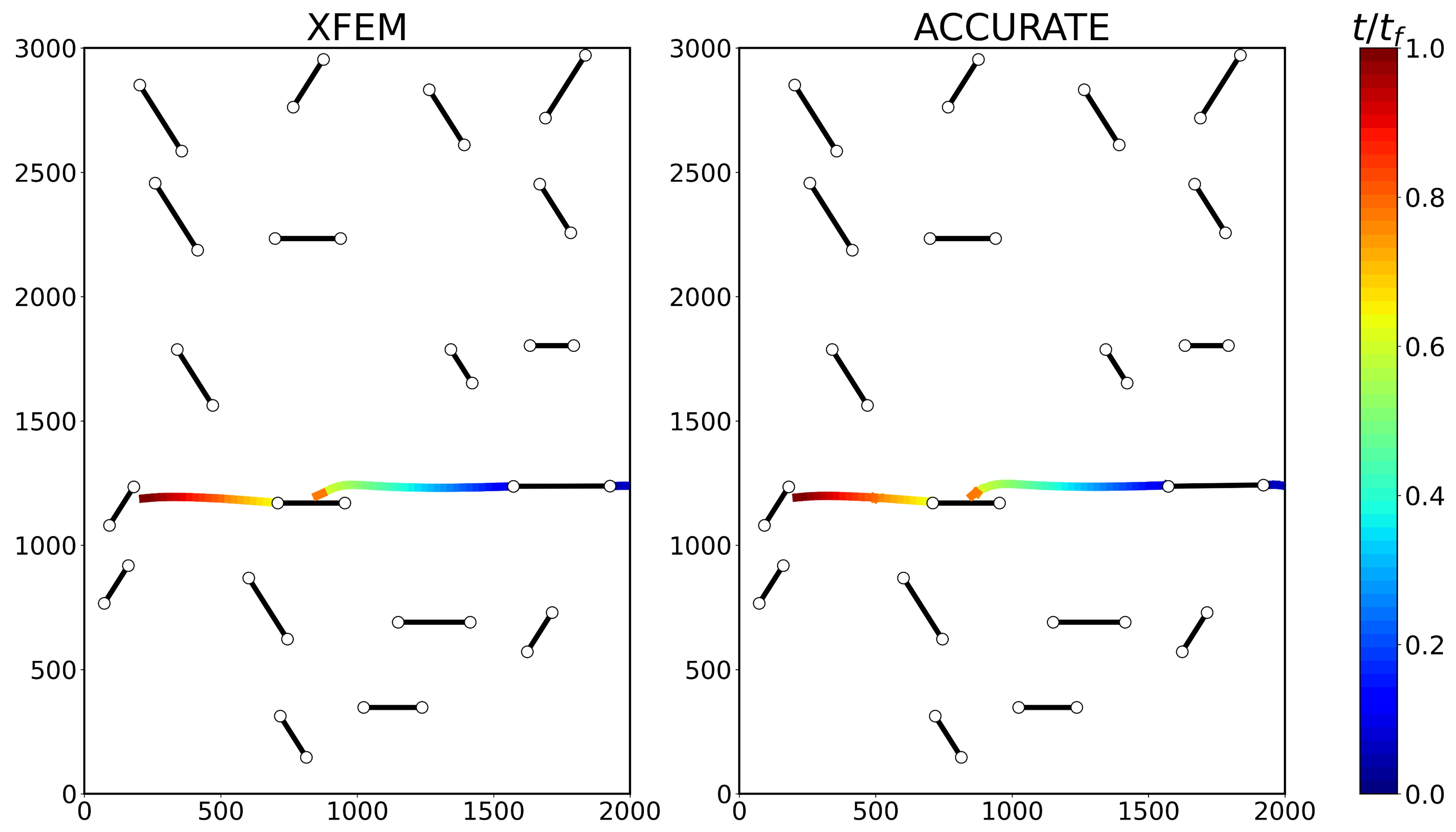}
                    \caption{Arbitrary crack length - Crack path}
                    \label{subfig:Length_CrackPath}
                \end{subfigure}
                \begin{subfigure}[c]{0.49\textwidth}
                    \centering 
                    \includegraphics[width=\linewidth]{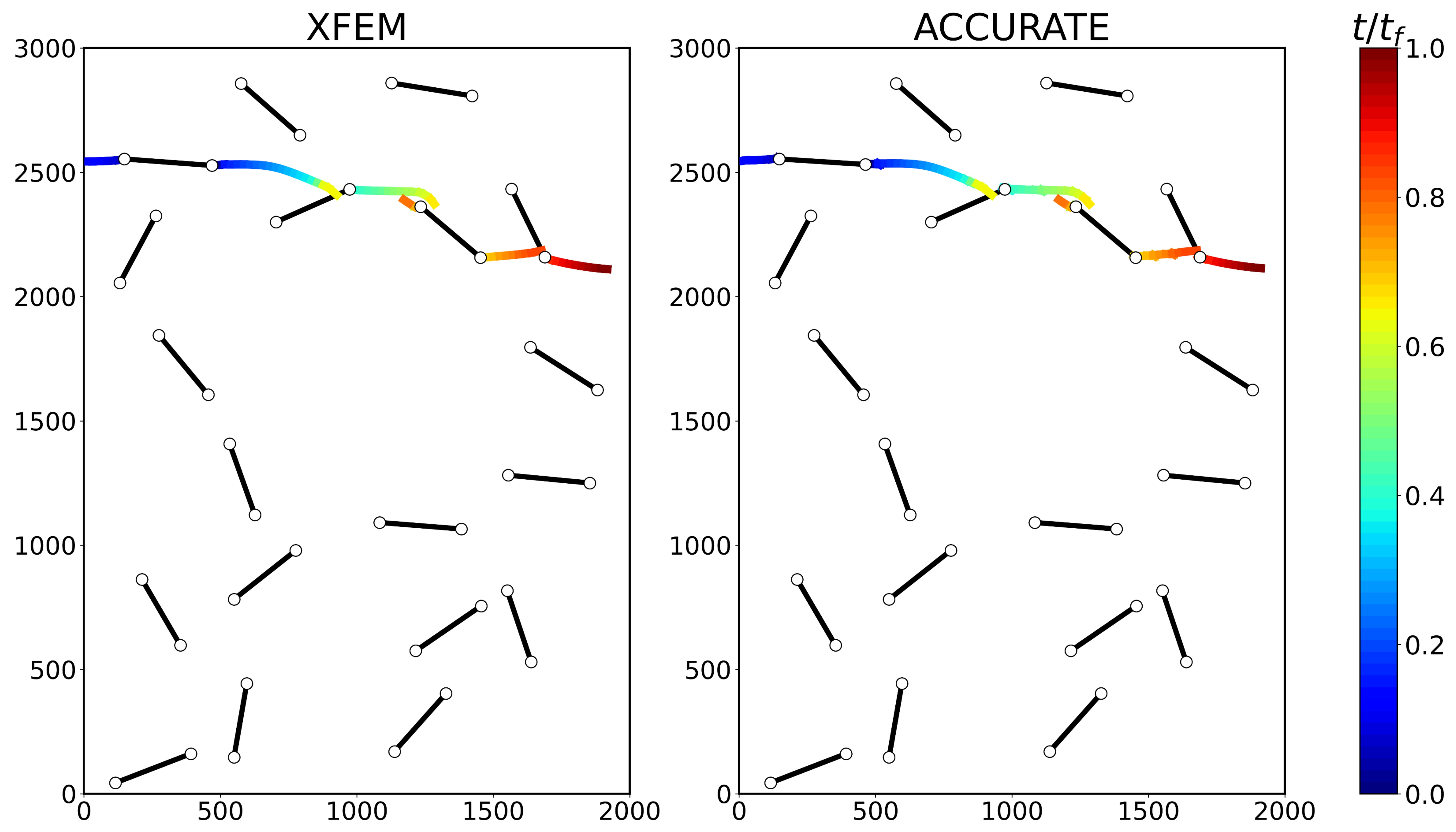}
                    \caption{Arbitrary crack angle - Crack path}
                    \label{subfig:Angle_CrackPath}
                \end{subfigure}
                \caption{Crack path evolution for (a) test case with Arbitrary crack length, and (b) test case with Arbitrary crack orientation.}
                \label{fig:CrackPath_evolution_LengthAngle}
            \end{figure}

            Lastly, Figures \ref{subfig:Length_CrackPath} - \ref{subfig:Angle_CrackPath} show the XFEM versus predicted crack growth evolution for the case studies involving arbitrary length and arbitrary angles, respectively.
            Similarly to the accuracy of the predicted stresses in the arbitrary length case, the predicted crack growth for this case shows good agreement with the XFEM simulator. 
            This result shows the ACCURATE framework is capable of emulating both the stress and crack growth of cases with arbitrary crack lengths.
            Additionally, the simulation with arbitrary crack angles qualitatively shows good prediction accuracy.
            We emphasize that while the simulation showing the most observable difference in predicted stress versus XFEM was for the case of arbitrary angle during the initial time-step ($t=1\%$), the \textit{K}-GNN model still captured the regions with the highest stress between interacting crack-tips, as well as the crack-tips with the highest stress intensity factors across the remaining time-steps. 
            These overall good predictions by ACCURATE may facilitate \textit{P}-GNN to predict future crack-tip positions with good accuracy compared to the XFEM model as shown in Figure \ref{subfig:Angle_CrackPath}.
            Therefore, Figures \ref{fig:CrackPath_evolution_LoadingSamples} - \ref{fig:CrackPath_evolution_LengthAngle} show a qualitative result for the capability of the ACCURATE framework to predict crack growth evolution of cases with variable configurations using TL on very small datasets.
            As supplementary material, we have included 7 animations of stress evolution and crack growth for each problem-specific configuration used in the TL sequence.
            
    \subsection{Errors on effective stress intensity factor and crack path}\label{subsec:results_Keff_CProp}
    
\begin{figure} %[htpb]
            %\centering
            \begin{subfigure}[c]{0.49\textwidth}
                \centering
                \begin{subfigure}[t]{1\textwidth}    
                  \centering \includegraphics[width=0.98\linewidth]{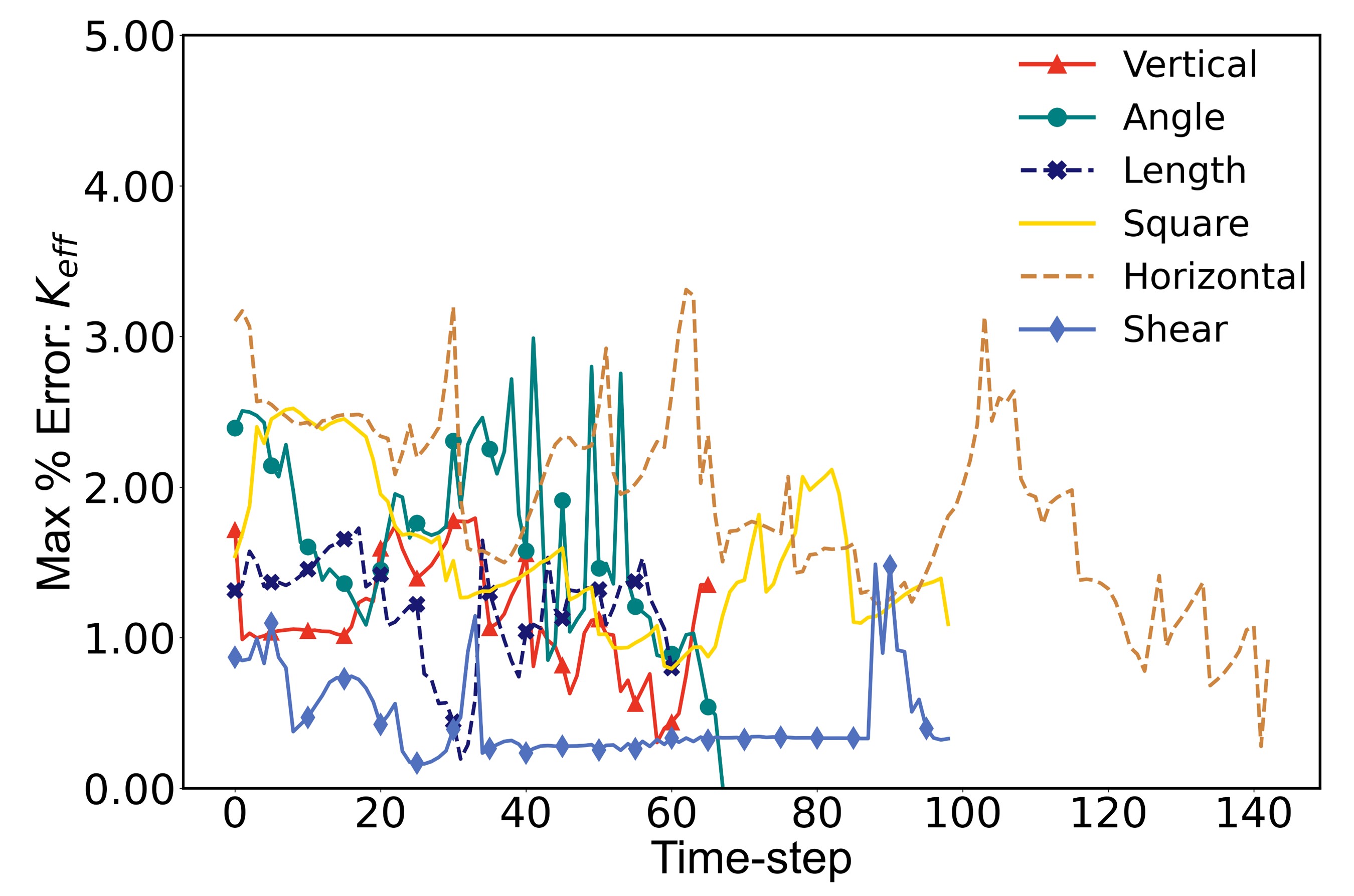}
                    \caption{Maximum error in $K_{eff}$ vs. time for each case study}
                    \label{fig:Keff_timewise_errors}
                \end{subfigure}
            \end{subfigure}
            %\hfill
            \begin{subfigure}[c]{0.49\textwidth}
                \centering
                \begin{subfigure}[t]{1\textwidth}    
                  \centering
                  \includegraphics[width=0.98\linewidth]{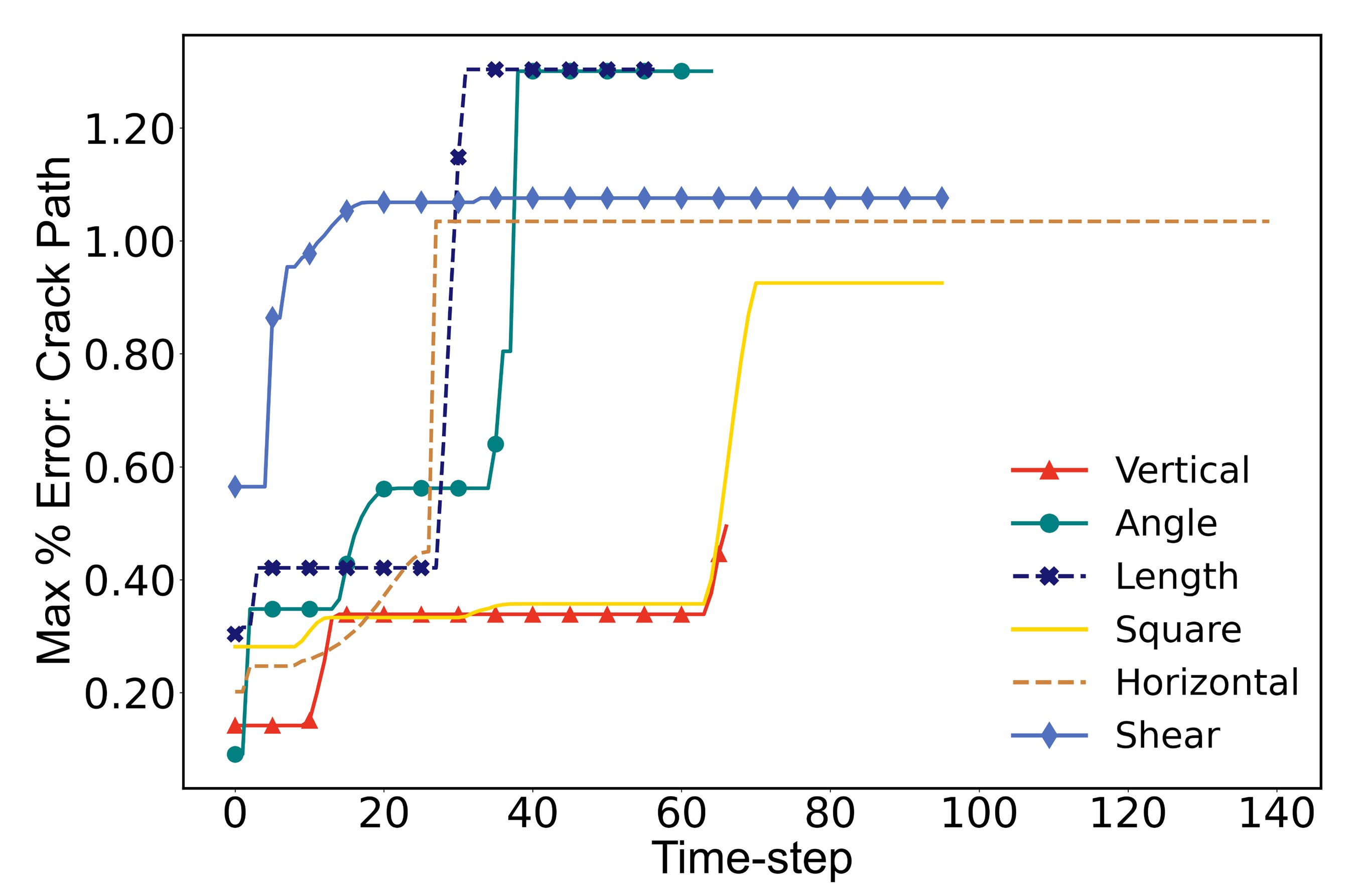}
                    \caption{Maximum error in crack path vs. time for each case study}
                    \label{fig:CProp_timewise_error}
                \end{subfigure}
            \end{subfigure}
            \caption{Maximum timewise percent errors in effective stress intensity factor and crack path for simulations shown in Sections \ref{subsect:stress_evolution} - \ref{subsect:crack_growth}.}
            \label{fig:Timewise_errors}
        \end{figure}

        For each simulation shown in Sections \ref{subsect:stress_evolution} and \ref{subsect:crack_growth}, we implement a detailed analysis of the errors in effective stress intensity factors and crack path as a function of time.
        Similar to MicroCrack-GNN, the error in the predicted Mode-I and Mode-II stress intensity factors, $K_{I}$ and $K_{II}$, is performed by computing the error in the resulting effective stress intensity factors $K_{eff}$.
        In Figures \ref{fig:Stress_evolution_LoadSamples} - \ref{fig:Stress_evolution_LengthAngle}, the highest errors in stress were observed at the locations of propagating crack-tips, and their nearest neighboring crack-tips. 
        Additionally, in quasi-static fracture problems the propagating crack-tips are determined by the crack tips where the effective stress intensity factor is greater than or equal to the critical stress intensity factor, i.e., $K_{eff} \geq K_{c}$.
        As a result, Figure \ref{fig:Keff_timewise_errors} shows the evolution of maximum percent error in $K_{eff}$ from the predicted $K_{I}$ and $K_{II}$, for the simulations shown in Section \ref{subsect:stress_evolution}.
        The error in $K_{eff}$ is computed as shown in equation (\ref{eq:Keff_error}). 
        \begin{flalign}
            && {K_{eff}}_{\% error} = \max_{s\in N_{crt}^t} \left( \frac{\left| {K_{eff}}_{\ Pred}^{t} - {K_{eff}}_{\ True}^{t} \right|}{{K_{eff}}_{\ True}^{t}} \right)_{s} \times 100 &&  \{t = 1, 2, \dots ,  T_{f}\},
        \label{eq:Keff_error}
        \end{flalign}
        where $N_{crt}^t$ is the number of cracks tips with $K_{eff}\geq K_{crt}$ at any given time $t$, ${K_{eff}}^t_{Pred}$ is the predicted $K_{eff}$ at time $t$ by ACCURATE, and ${K_{eff}}^t_{True}$ is the true $K_{eff}$ at time $t$ by the XFEM fracture model.
        While Figure \ref{fig:Keff_timewise_errors} depicts a fluctuating trend for the $K_{eff}$ errors, the average errors across all time-steps correspond to $1.13 \pm 0.36\%$, $0.46 \pm 0.27\%$, $1.55 \pm 0.49\%$, $1.89 \pm 0.61\%$, $1.20 \pm 0.34\%$, and $1.64 \pm 0.62\%$ for the vertical domain, shear load, square domain, horizontal domain, arbitrary length, and arbitrary angle, respectively.  
        The highest error spikes are seen for the horizontal domain and arbitrary crack angle at approximately $3.2\%$ and $3.0\%$, respectively, while the lowest $K_{eff}$ error is obtained for the simulation of Shear loading.  
        Moreover, the maximum $\%$ errors for the predicted crack paths are shown in Figure \ref{fig:CProp_timewise_error}.
        Unlike the fluctuating trend in error obtained for $K_{eff}$, the timewise maximum error in the crack path remains constant in time until a new maximum error is reached.
        The maximum crack path error is computed using the maximum difference between the predicted crack growth at each time step and the XFEM model's crack growth.
        The highest crack growth errors in time correspond to the cases of arbitrary crack length and arbitrary crack orientation at approximately 1.35$\%$, while the lowest error is seen for the standard case of vertical domain with fixed crack lengths and crack orientations at approximately 0.5 $\%$.
        The average errors across all time steps in Figure \ref{fig:CProp_timewise_error} are $0.31 \pm 0.08\%$, $1.04 \pm 0.12\%$, $0.51 \pm 0.26\%$, $0.89 \pm 0.29\%$, $0.85 \pm 0.44\%$, and $0.82 \pm 0.42\%$ for the vertical domain, shear load, square domain, horizontal domain, arbitrary length, and arbitrary angle, respectively.  
        %Thus, the Shear load case showed the highest maximum crack path error both at a given time-step, and across all time-steps of the simulation.

\begin{figure} %[htpb]
            %\centering
            \begin{subfigure}[c]{0.49\textwidth}
                \centering
                \begin{subfigure}[t]{1\textwidth}    
                  \centering \includegraphics[width=0.98\linewidth]{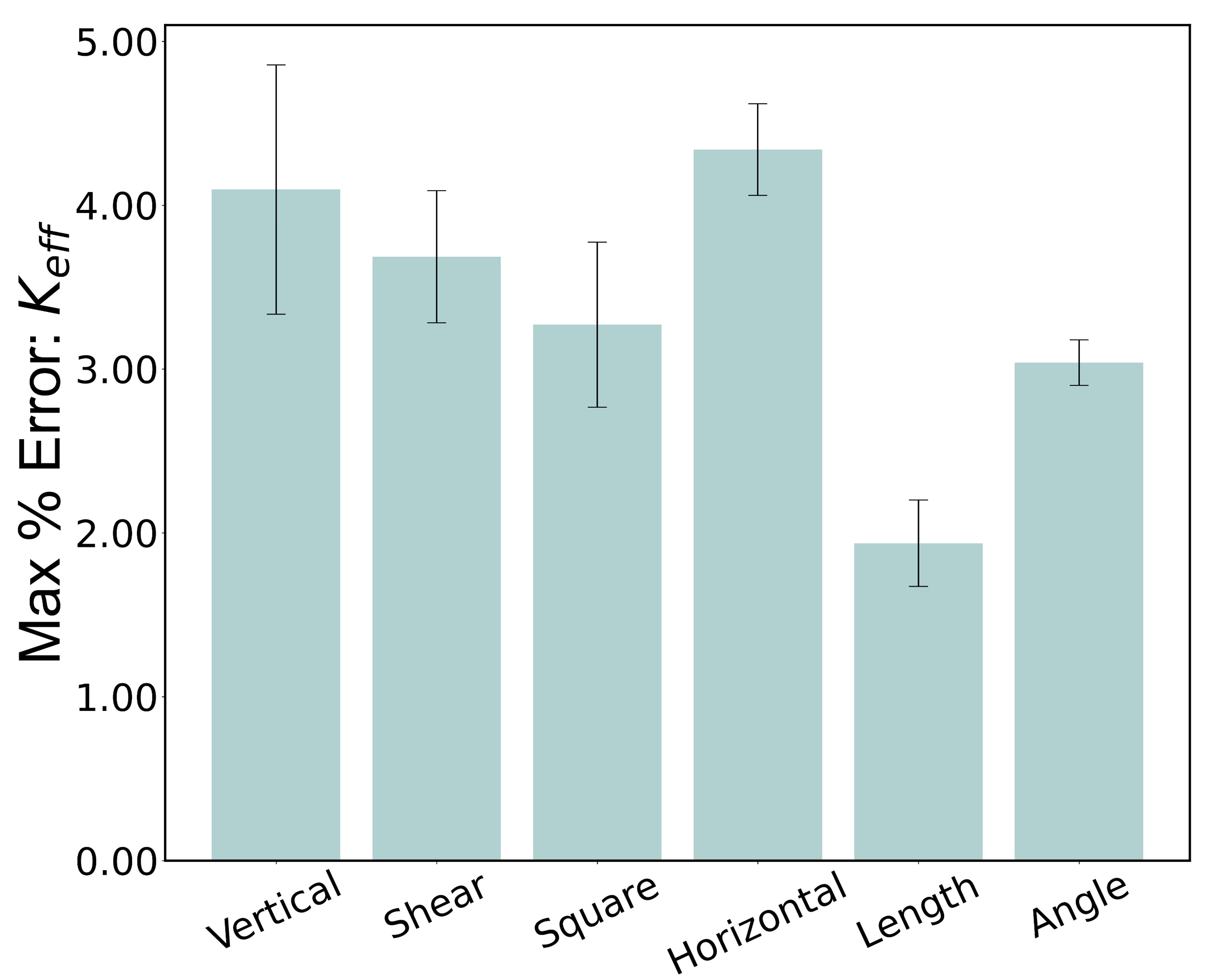}
                    \caption{Maximum error in predicted $K_{eff}$ for each case study}
                    \label{fig:Keff_Max_errors}
                \end{subfigure}
            \end{subfigure}
            %\hfill
            \begin{subfigure}[c]{0.49\textwidth}
                \centering
                \begin{subfigure}[t]{1\textwidth}    
                  \centering
                  \includegraphics[width=0.98\linewidth]{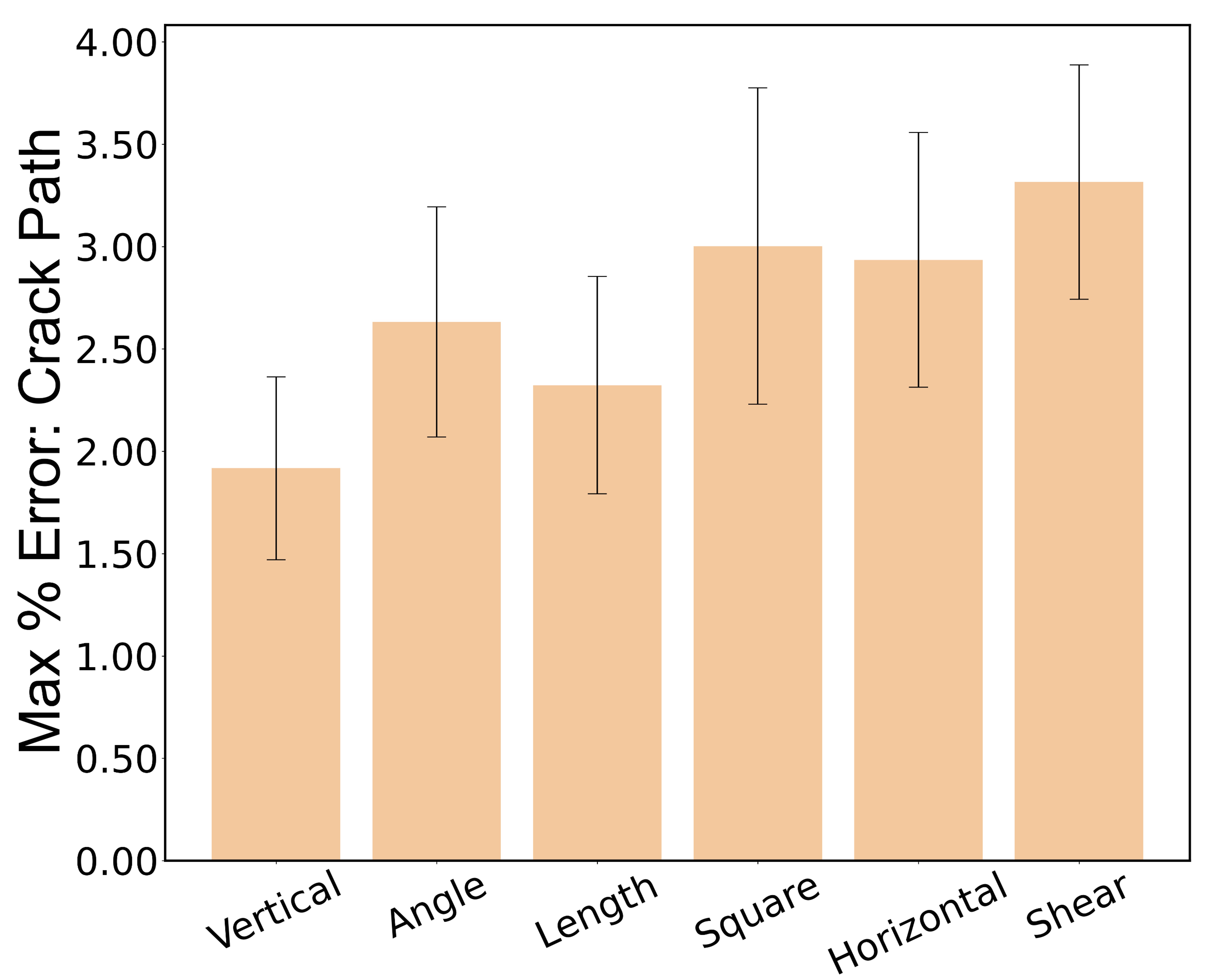}
                    \caption{Maximum error in predicted crack path for each case study}
                    \label{fig:Cprop_Max_errors}
                \end{subfigure}
            \end{subfigure}
            \caption{Maximum percent error in predicted stress intensity factor and crack path across all test simulations for each case studies}
            \label{fig:Max_errors}
        \end{figure}

        Next, we show the maximum errors in $K_{eff}$ and crack path for each case study.
        First, the maximum errors in all the test simulations of each case study are computed at each time step following a similar approach as the timewise errors shown in Figure \ref{fig:Timewise_errors}.
        We then take the maximum error across time for all simulations in each case study and pick the simulation resulting in the highest error.
        The resulting highest maximum $K_{eff}$ errors and crack path errors for each case study are shown in Figures \ref{fig:Keff_Max_errors} and \ref{fig:Cprop_Max_errors}, respectively. 
        For $K_{eff}$, the case study with the highest error is the horizontal domain at $4.34 \pm 0.28 \%$, while the lowest error was obtained for the case of arbitrary length at $1.94 \pm 0.26 \%$.
        
        Furthermore, the maximum crack path errors shown in Figure \ref{fig:Cprop_Max_errors} depict high prediction accuracy by \textit{P}-GNN across each case study.
        All cases resulted in maximum errors lower than $4\%$.
        The highest crack path error was obtained for shear load at $3.32 \pm 0.57 \%$, while the lowest error was obtained for the standard vertical domain case at $1.92 \pm 0.45 \%$. 
        From Figures \ref{fig:Timewise_errors} and \ref{fig:Max_errors}, we see that the ACCURATE framework achieved good prediction accuracy in time across each case study for predictions of stress intensity factors and crack growth.
        The implementation of TL with only 20 simulations to new arbitrary initial crack configurations, boundary dimensions, and shear loading scenarios showed to provide ACCURATE with accuracies of approximately $95.5\%$ for stress intensity factors, and $96.5\%$ for crack growth predictions.

    \subsection{Unseen Cases}\label{subsec:Unseen_Cases}
    
        In this section, we present a key contribution of the proposed ACCURATE framework in its ability to emulate stresses and crack growth for new unseen scenarios not introduced during TL.
        From Figure \ref{fig:MicroCrack-GNN_structure}c, we see that during the TL updates the ACCURATE framework was introduced to arbitrary crack lengths and arbitrary crack orientations for vertical domains subjected to tension only. 
        However, arbitrary crack length and orientation effects were not introduced during the TL update for square and horizontal domains, nor for shear loading.
        Similarly, the TL update for the square and horizontal domains did not include shear loading effects or vice versa.
        To show ACCURATE's ability for predicting new unseen cases without the need for additional TL implementations, we generate the following four new unseen case studies: (i)-(ii) $\{2500mm \times 2000mm\}$ domain with arbitrary crack lengths and crack orientations subjected to tension and shear loads, respectively, and (iii)-(iv) $\{2500mm \times 3000mm\}$ domain with arbitrary crack lengths and crack orientations subjected to tension and shear, respectively.
        The purpose of this approach was to present the framework's ability to handle systems involving arbitrary crack orientations and lengths without the need for retraining or generating new computationally expensive and time-consuming simulations.
        
\begin{figure}
            \begin{subfigure}[t]{1\textwidth}
                \begin{subfigure}[t]{0.32\textwidth}
                    \centering
                    \includegraphics[width=\linewidth]{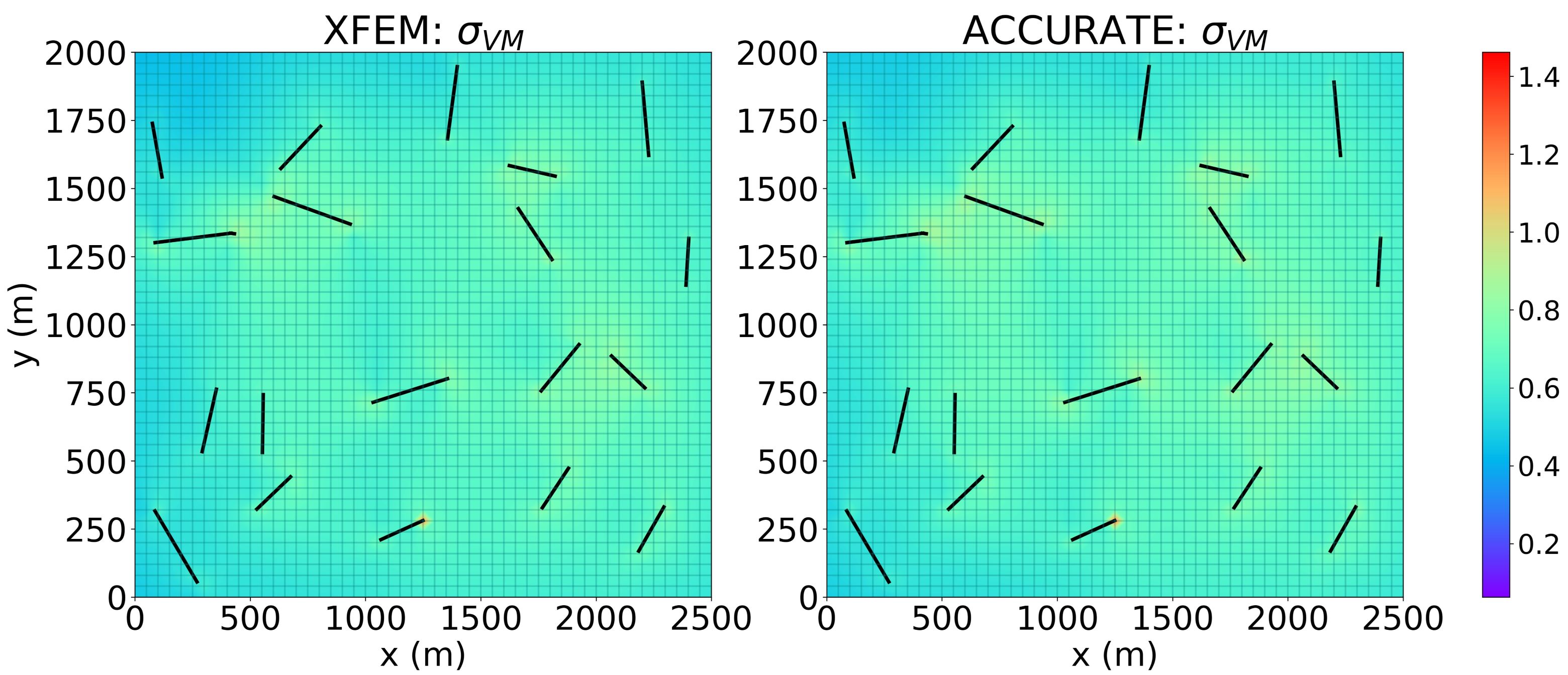}
                    \caption{Tension - 2500$\times$2000: t=$1\%$}
                \end{subfigure}
                \begin{subfigure}[t]{0.32\textwidth}
                    \centering
                    \includegraphics[width=\linewidth]{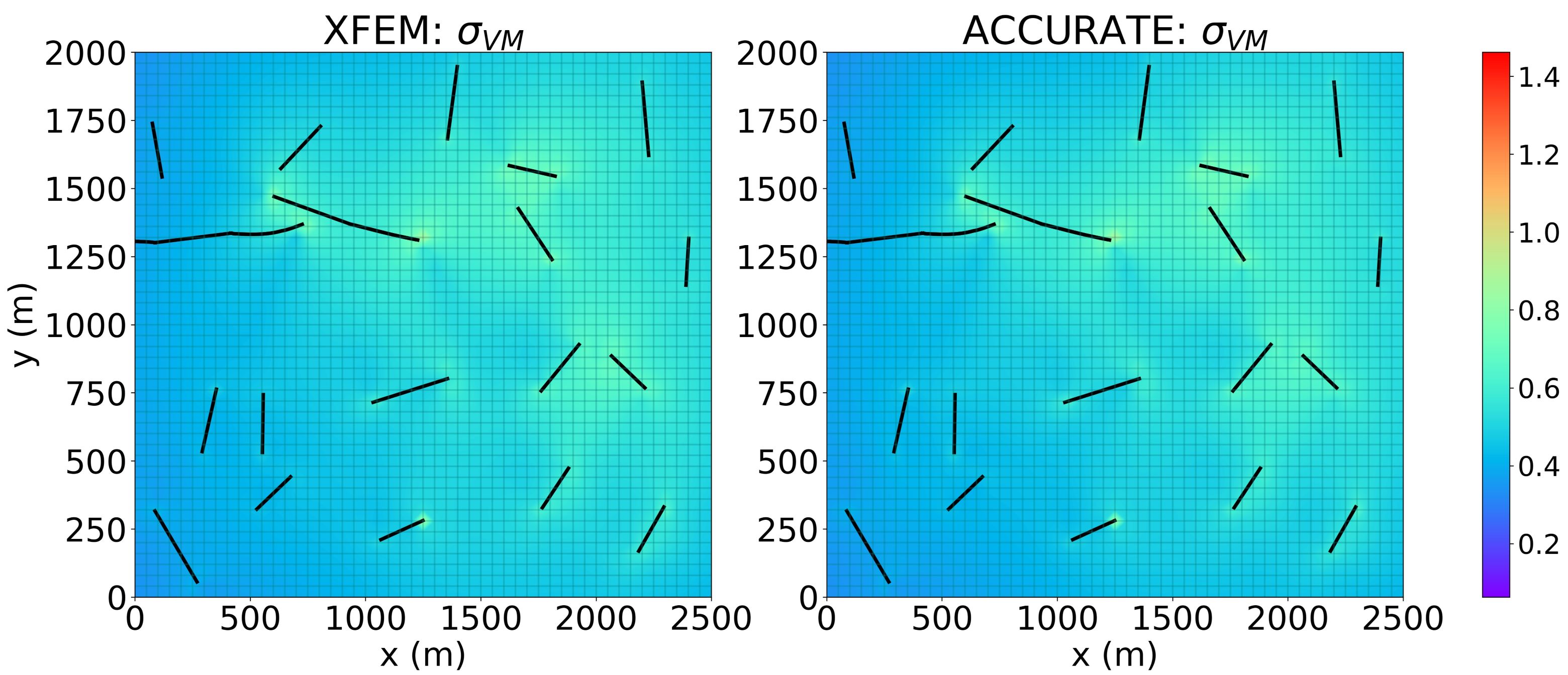}
                    \caption{Tension - 2500$\times$2000: t=$45\%$}
                \end{subfigure}
                \begin{subfigure}[t]{0.32\textwidth}
                    \centering
                    \includegraphics[width=\linewidth]{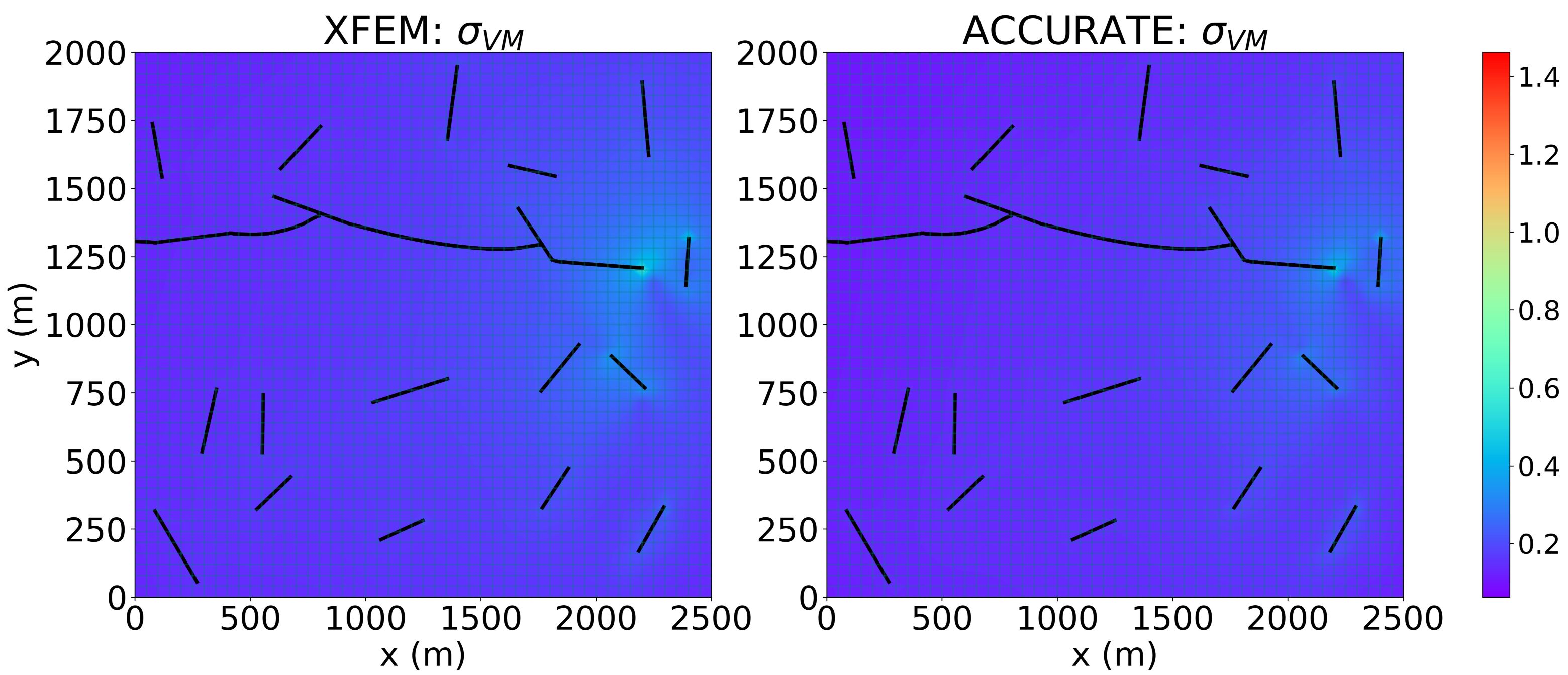}
                    \caption{Tension - 2500$\times$2000: t=$90\%$}
                \end{subfigure}
            \label{subfig:TensionUnseen2500x2000_StressContour}
            \end{subfigure}
            \begin{subfigure}[b]{1\textwidth}
                \begin{subfigure}[b]{0.32\textwidth}
                    \centering
                    \includegraphics[width=\linewidth]{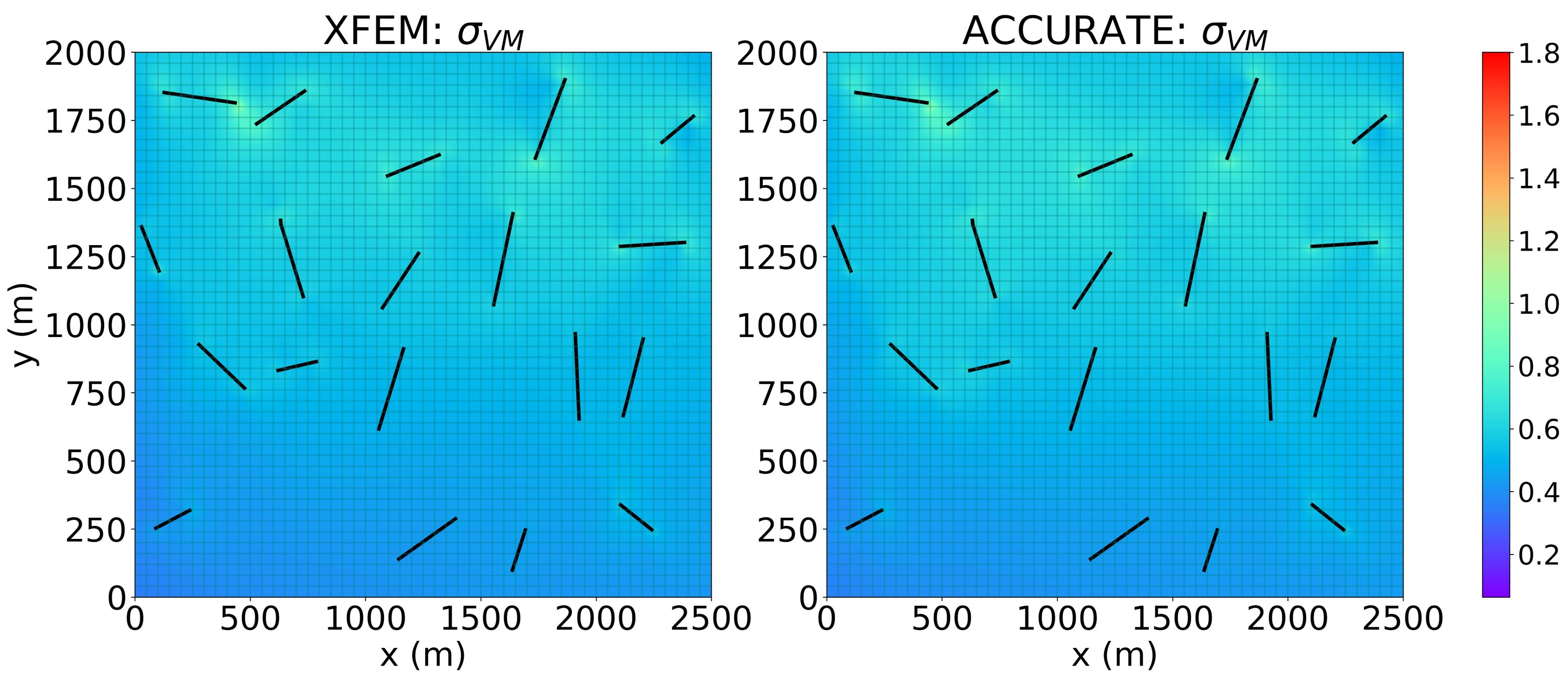}
                    \caption{Shear - 2500$\times$2000: t=$1\%$}
                \end{subfigure}
                \begin{subfigure}[b]{0.32\textwidth}
                    \centering
                    \includegraphics[width=\linewidth]{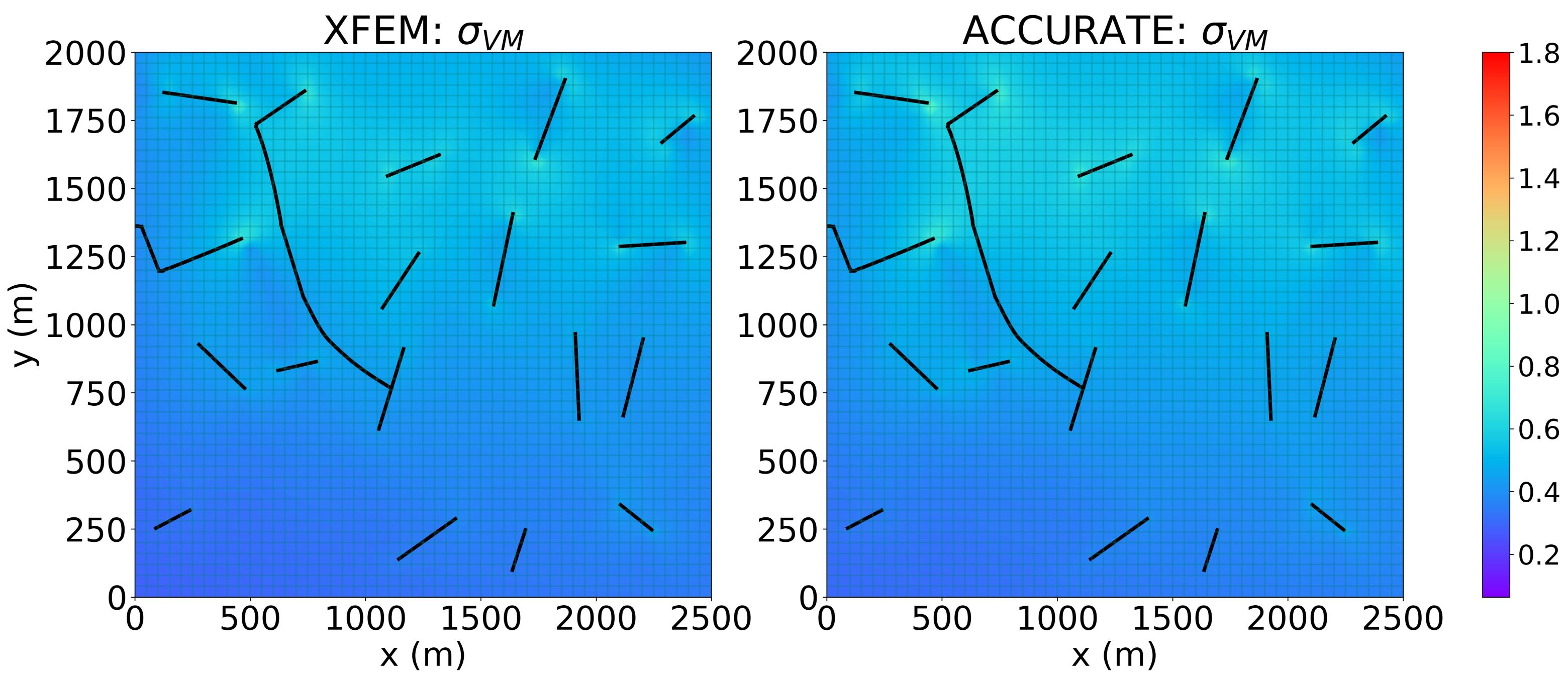}
                    \caption{Shear - 2500$\times$2000: t=$45\%$}
                \end{subfigure}
                \begin{subfigure}[b]{0.32\textwidth}
                    \centering
                    \includegraphics[width=\linewidth]{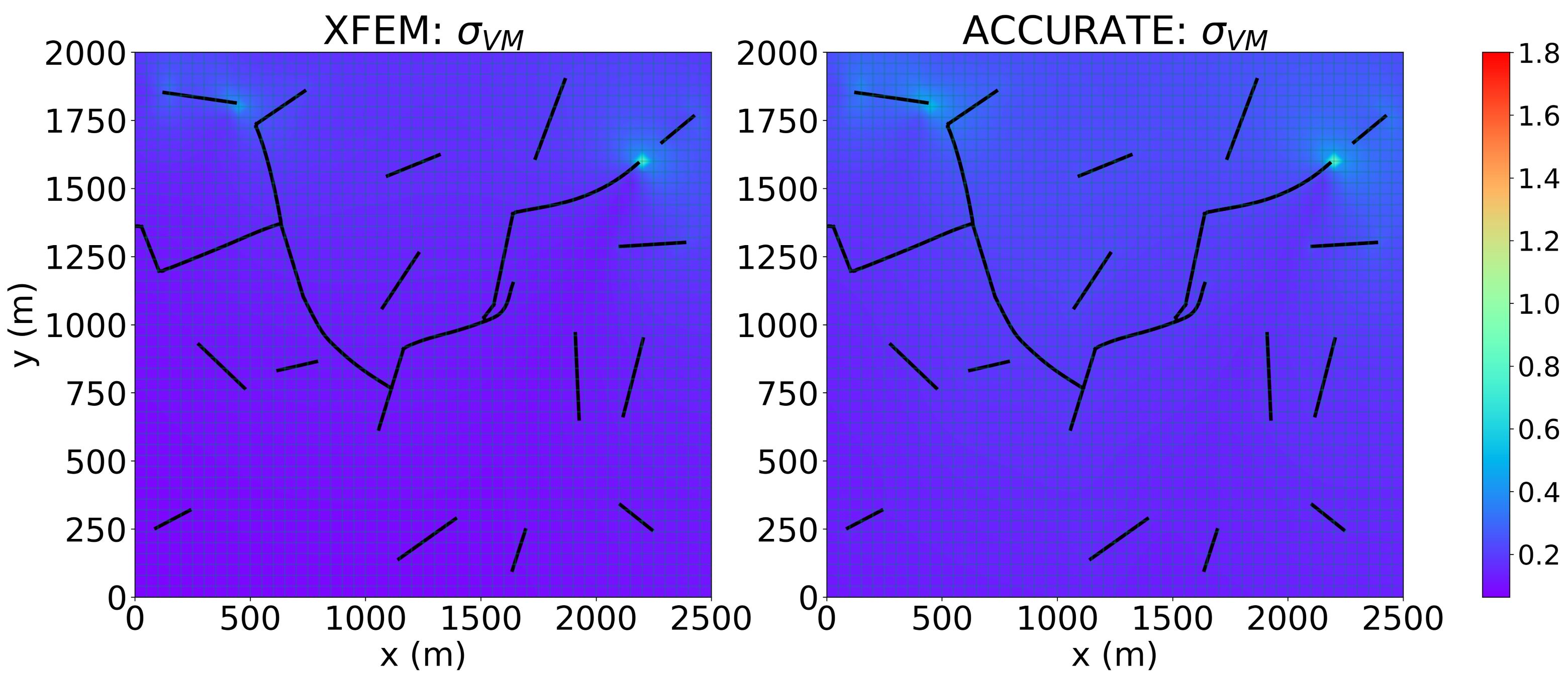}
                    \caption{Shear - 2500$\times$2000: t=$90\%$}
                \end{subfigure}
            \label{subfig:ShearUnseen2500x2000_StressContour}
            \end{subfigure}
            \caption{von Mises stress evolution (MPa) from $t=1\%$ to $t=90\%$ for  (a-c) unseen case of 2500mm $\times$ 2000mm domain with arbitrary crack lengths and crack orientations subjected to tensile load, and (d-f) unseen case of 2500mm $\times$ 2000mm domain with arbitrary crack lengths and crack orientations subjected to shear load.}
            \label{fig:Stress_evolution_Unseen2500x2000}
        \end{figure}

\begin{figure}
            \begin{subfigure}[t]{1\textwidth}
                \begin{subfigure}[t]{0.32\textwidth}
                    \centering
                    \includegraphics[width=\linewidth]{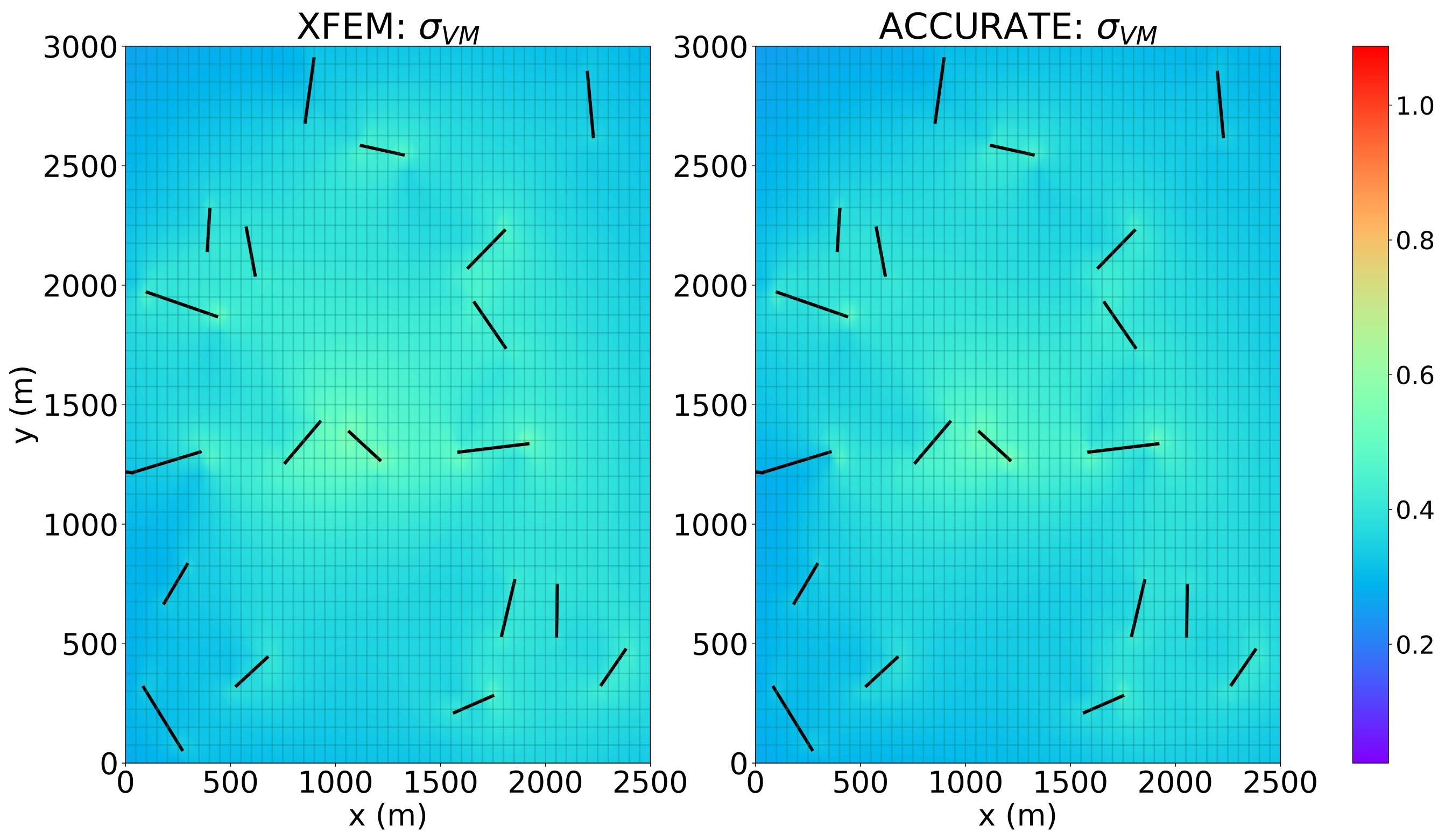}
                    \caption{Tension - 2500$\times$3000: t=$1\%$}
                \end{subfigure}
                \begin{subfigure}[t]{0.32\textwidth}
                    \centering
                    \includegraphics[width=\linewidth]{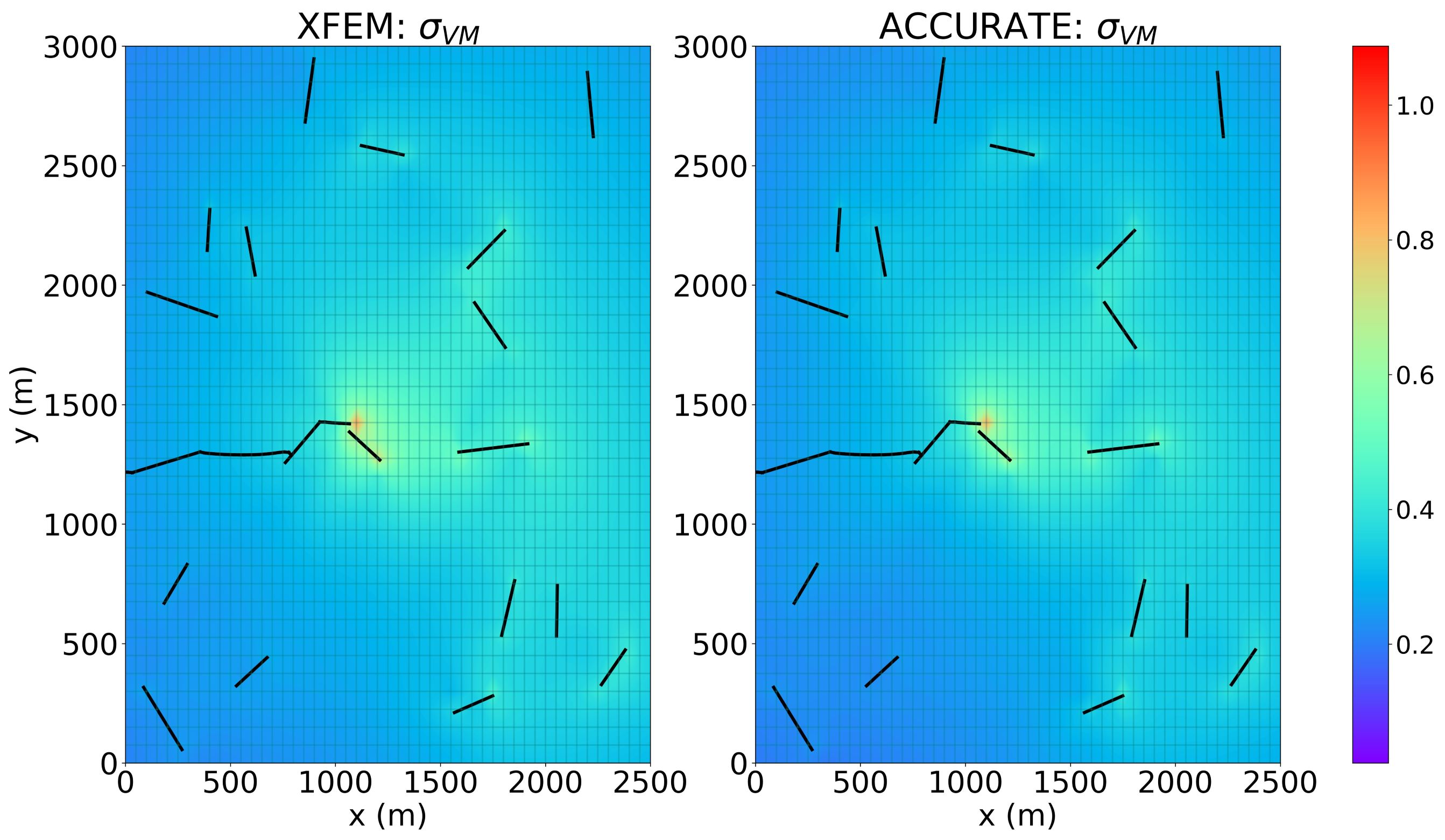}
                    \caption{Tension - 2500$\times$3000: t=$45\%$}
                \end{subfigure}
                \begin{subfigure}[t]{0.32\textwidth}
                    \centering
                    \includegraphics[width=\linewidth]{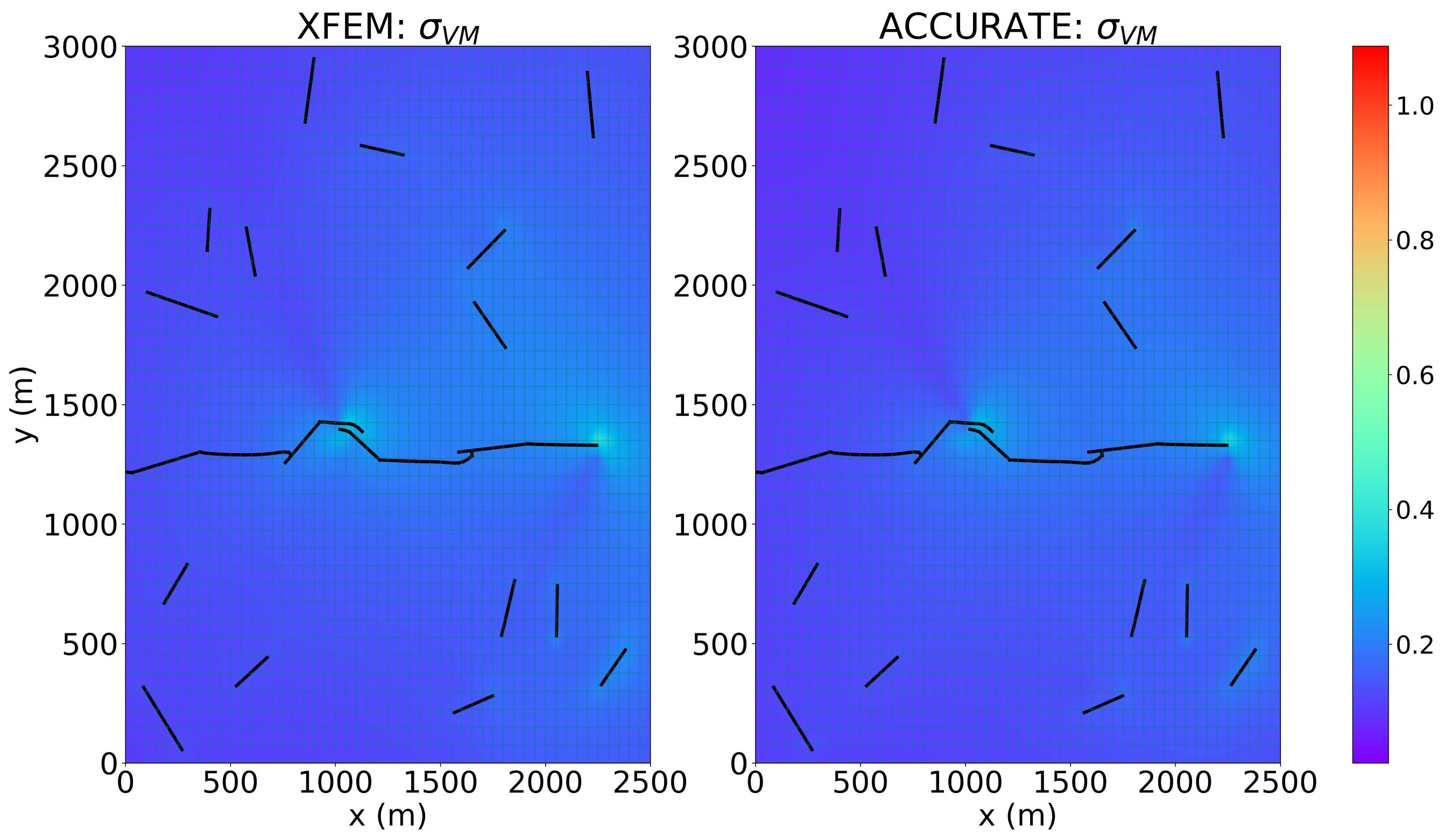}
                    \caption{Tension - 2500$\times$3000: t=$90\%$}
                \end{subfigure}
            \label{subfig:TensionUnseen2500x3000_StressContour}
            \end{subfigure}
            \begin{subfigure}[b]{1\textwidth}
                \begin{subfigure}[b]{0.32\textwidth}
                    \centering
                    \includegraphics[width=\linewidth]{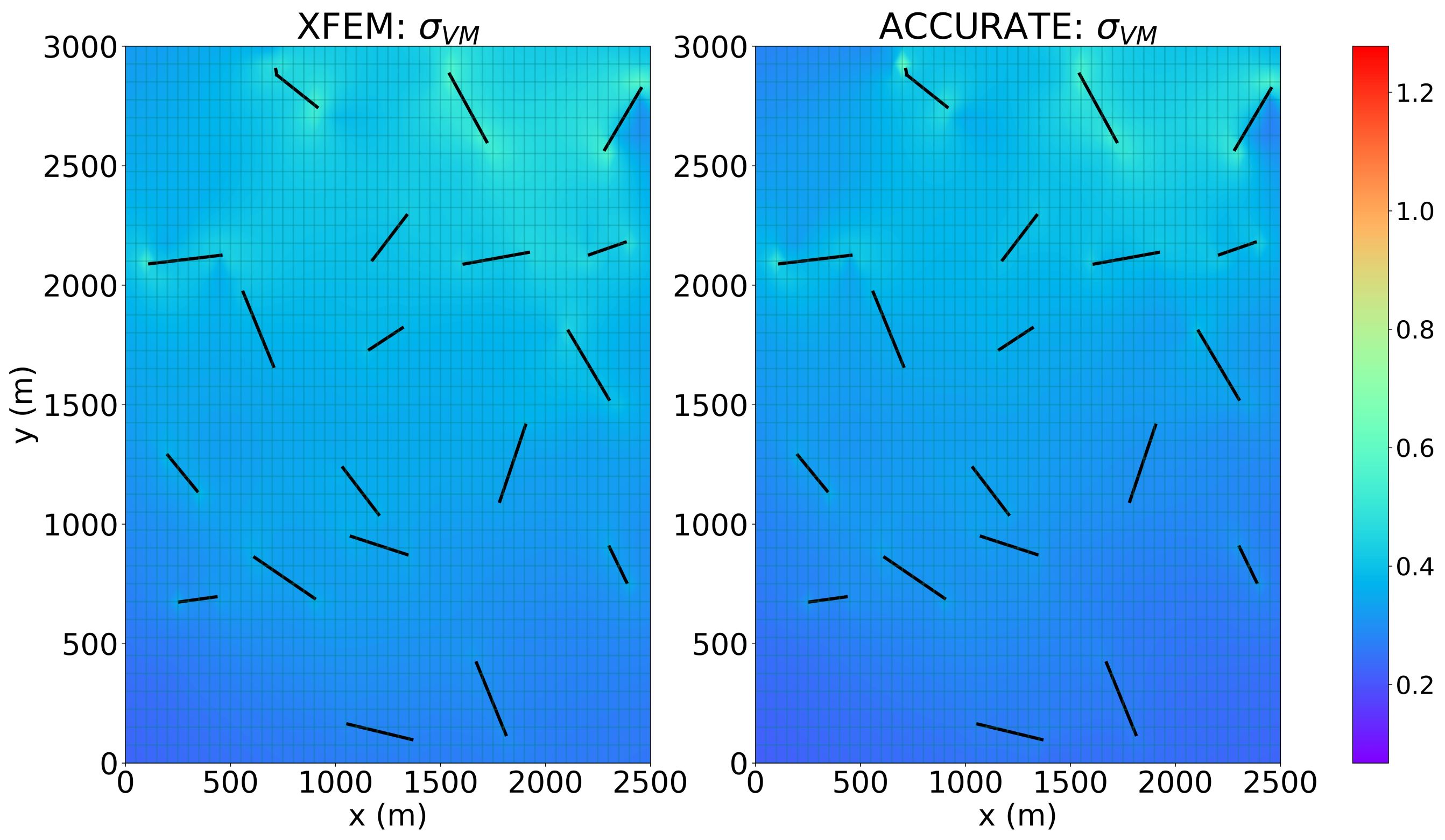}
                    \caption{Shear - 2500$\times$3000: t=$1\%$}
                \end{subfigure}
                \begin{subfigure}[b]{0.32\textwidth}
                    \centering
                    \includegraphics[width=\linewidth]{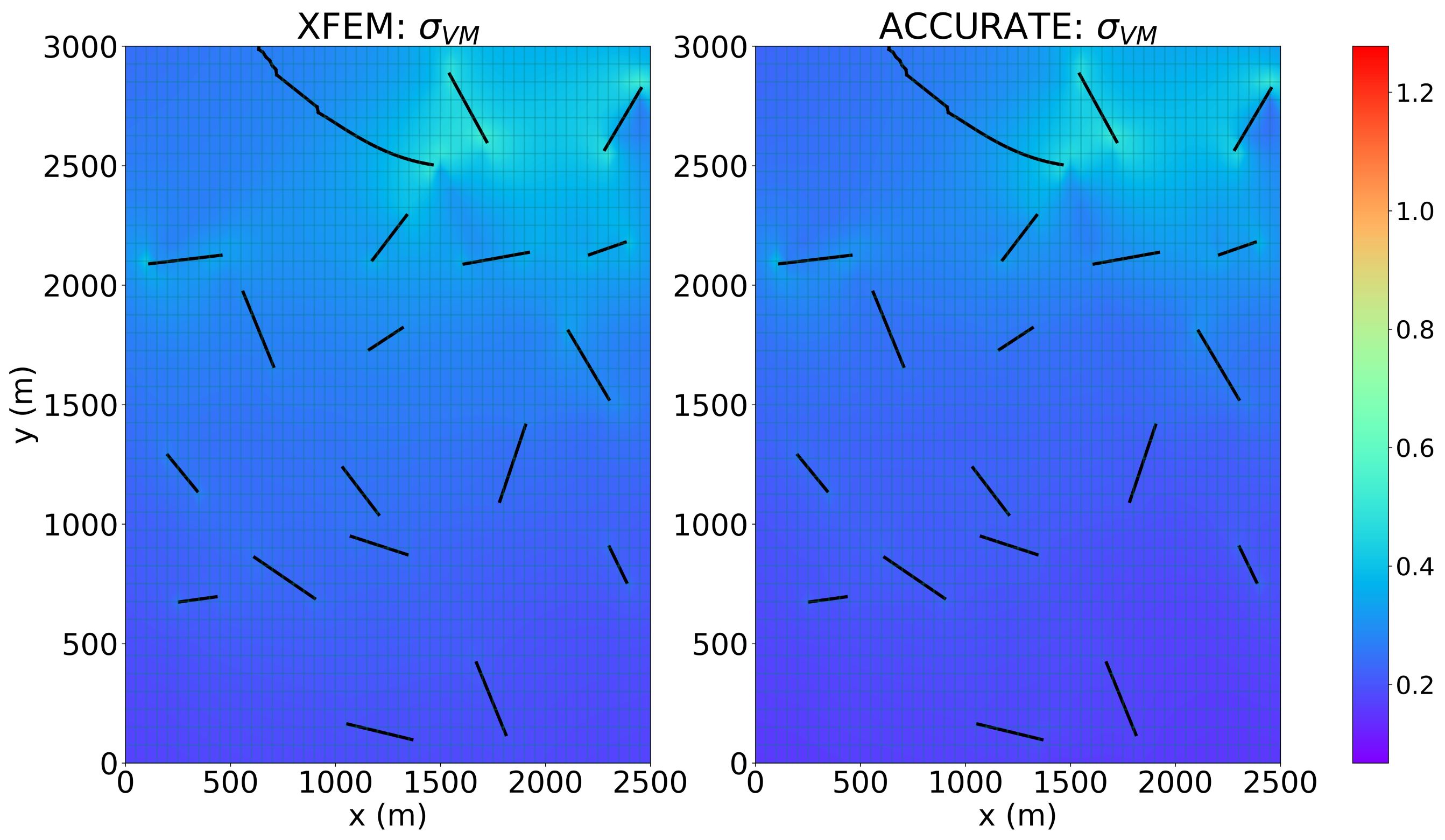}
                    \caption{Shear - 2500$\times$3000: t=$45\%$}
                \end{subfigure}
                \begin{subfigure}[b]{0.32\textwidth}
                    \centering
                    \includegraphics[width=\linewidth]{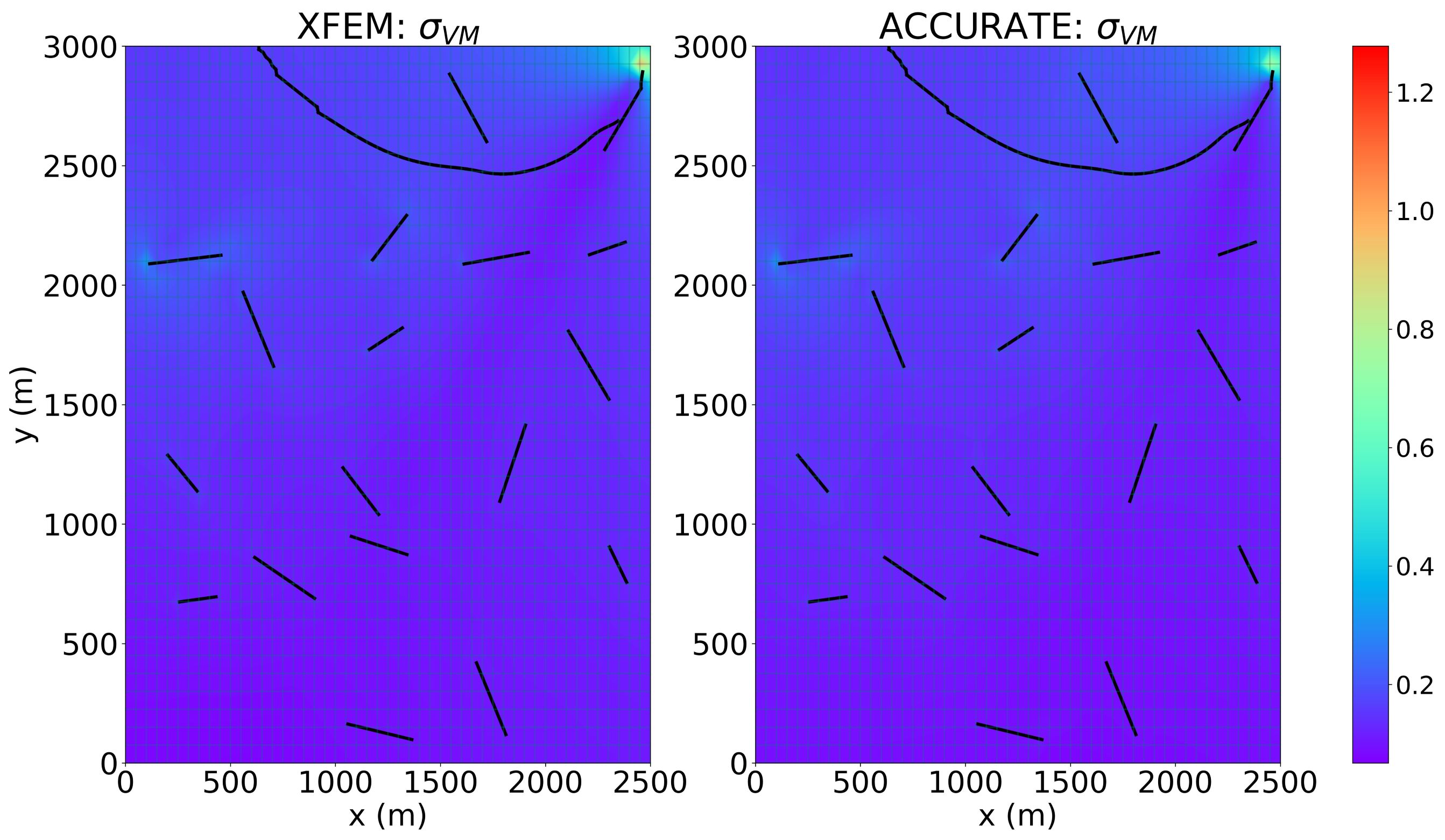}
                    \caption{Shear - 2500$\times$3000: t=$90\%$}
                \end{subfigure}
            \label{subfig:ShearUnseen2500x3000_StressContour}
            \end{subfigure}
            \caption{von Mises stress evolution (MPa) from $t=1\%$ to $t=90\%$ for  (a-c) unseen case of 2500mm $\times$ 3000mm domain with arbitrary crack lengths and crack orientations subjected to tensile load, and (d-f) unseen case of 2500mm $\times$ 3000mm domain with arbitrary crack lengths and crack orientations subjected to shear load.}
            \label{fig:Stress_evolution_Unseen2500x3000}
        \end{figure}

        We show the von Mises stress evolution for XFEM versus ACCURATE from $t=1\%$ to $t=90\%$ for the new unseen cases (i)-(ii) in Figures \ref{fig:Stress_evolution_Unseen2500x2000}a - \ref{fig:Stress_evolution_Unseen2500x2000}f, and Figures \ref{fig:Stress_evolution_Unseen2500x3000}a - \ref{fig:Stress_evolution_Unseen2500x2000}f, respectively.
        For both tension and shear loading cases shown, the predicted stress distribution is qualitatively indistinguishable from the XFEM stresses.
        We emphasize that these new boundary dimensions ($\{2500mm \times 2000mm\}$ and $\{2500mm \times 3000mm\}$) coupled with arbitrary crack lengths and orientations were never seen by the ACCURATE framework during training. 
        However, the framework was able to predict the stress intensity factors with good accuracy compared to the XFEM model, thus, generating qualitatively identical von Mises stress evolution. 

\begin{figure}
            \begin{subfigure}[c]{0.49\textwidth}
                \centering 
                \includegraphics[width=\linewidth]{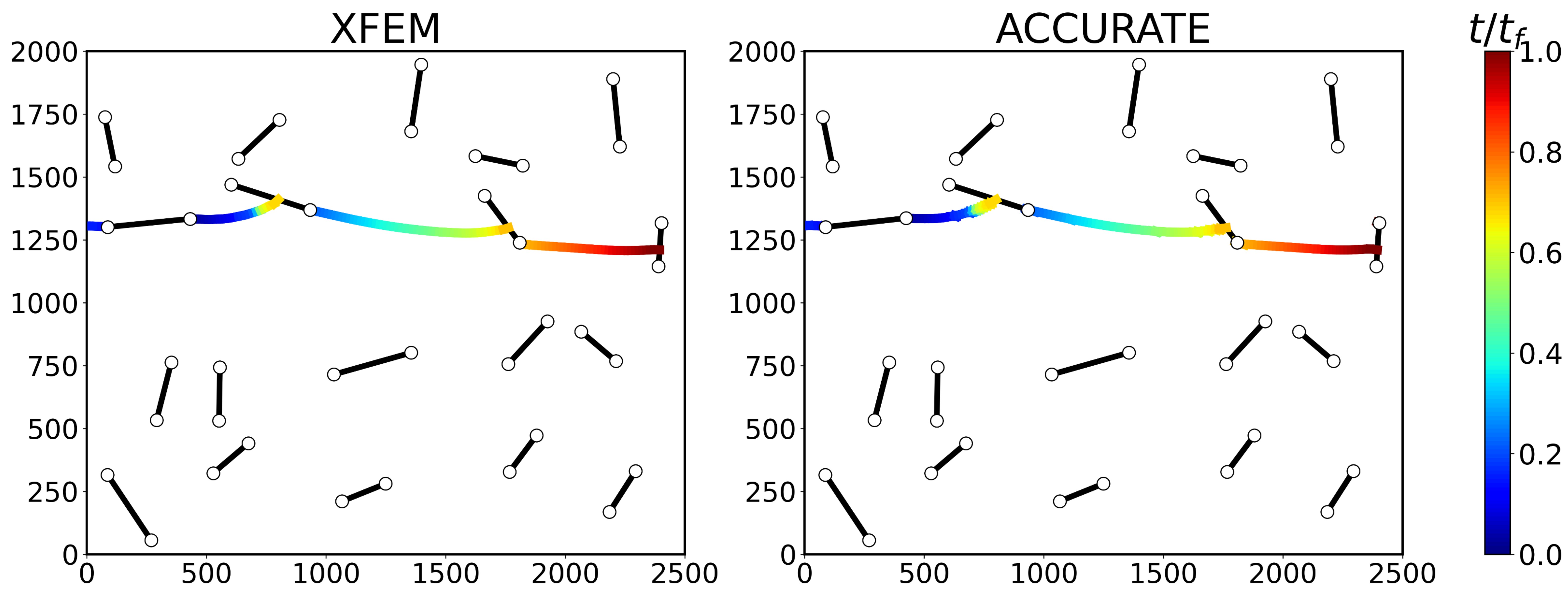}
                \caption{Tension: 2500$\times$2000}
                \label{subfig:TensionUnseen2500x2000_CrackPath}
            \end{subfigure}
            \begin{subfigure}[c]{0.49\textwidth}
                \centering 
                \includegraphics[width=\linewidth]{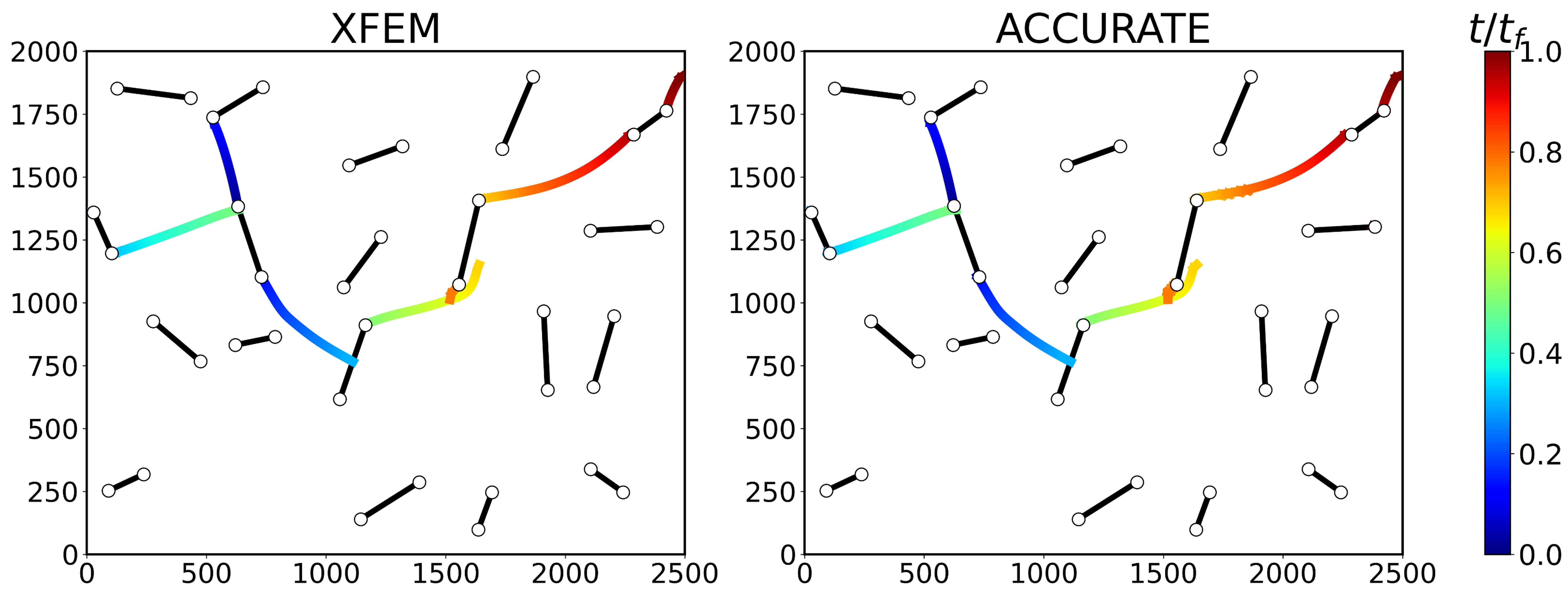}
                \caption{Shear: 2500$\times$2000}
                \label{subfig:ShearUnseen2500x2000_CrackPath}
            \end{subfigure}
            \caption{Crack path evolution for (a) unseen case of 2500mm $\times$ 2000mm domain with arbitrary crack lengths and crack orientations subjected to tensile load, and (b) unseen case of 2500mm $\times$ 2000mm domain with arbitrary crack lengths and crack orientations subjected to shear load.}    
            \label{fig:CrackPath_evolution_Unseen2500x2000}
        \end{figure}

\begin{figure}
            \begin{subfigure}[c]{0.49\textwidth}
                \centering 
                \includegraphics[width=\linewidth]{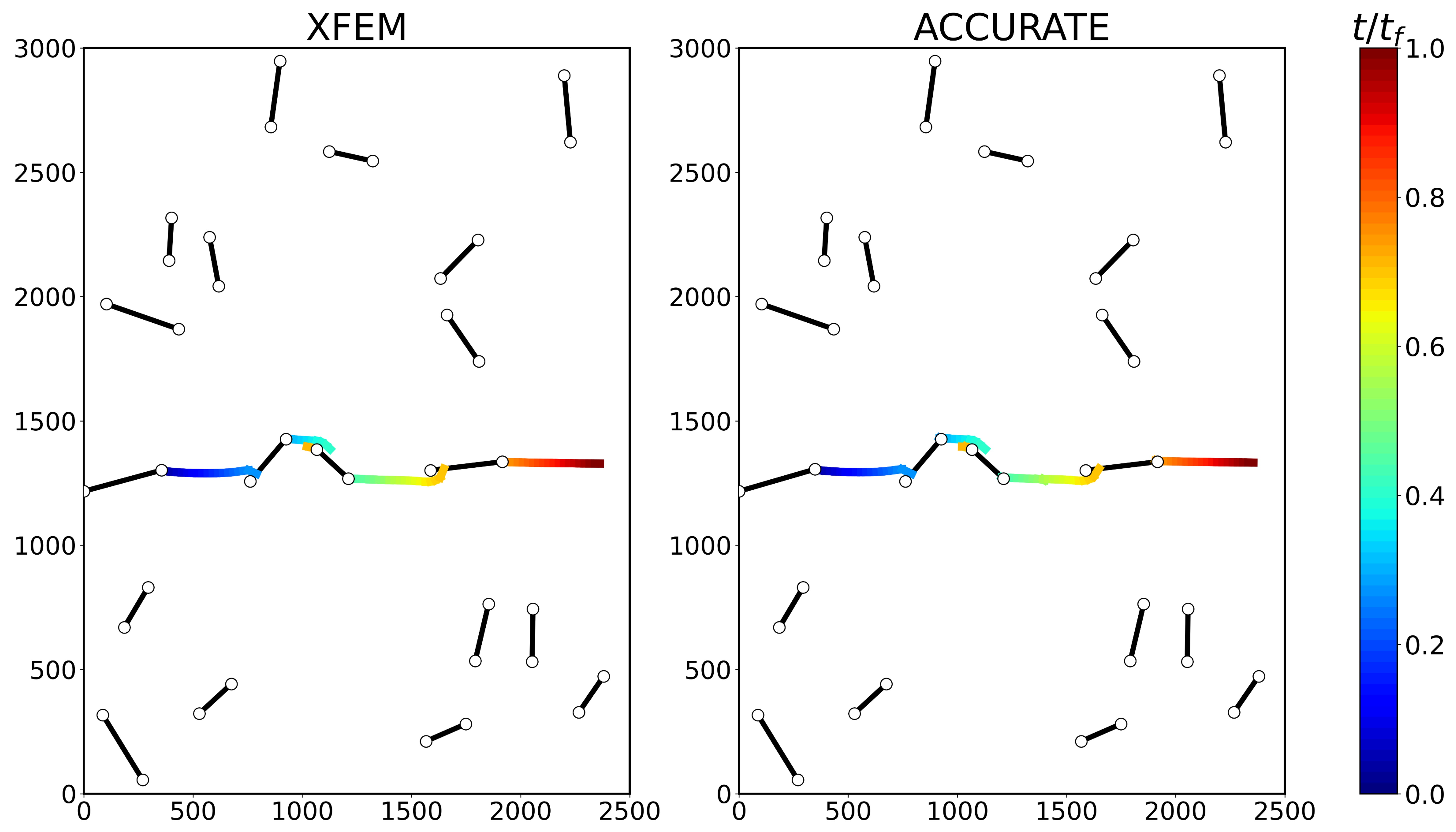}
                \caption{Tension: 2500$\times$3000}
                \label{subfig:TensionUnseen2500x3000_CrackPath}
            \end{subfigure}
            \begin{subfigure}[c]{0.49\textwidth}
                \centering 
                \includegraphics[width=\linewidth]{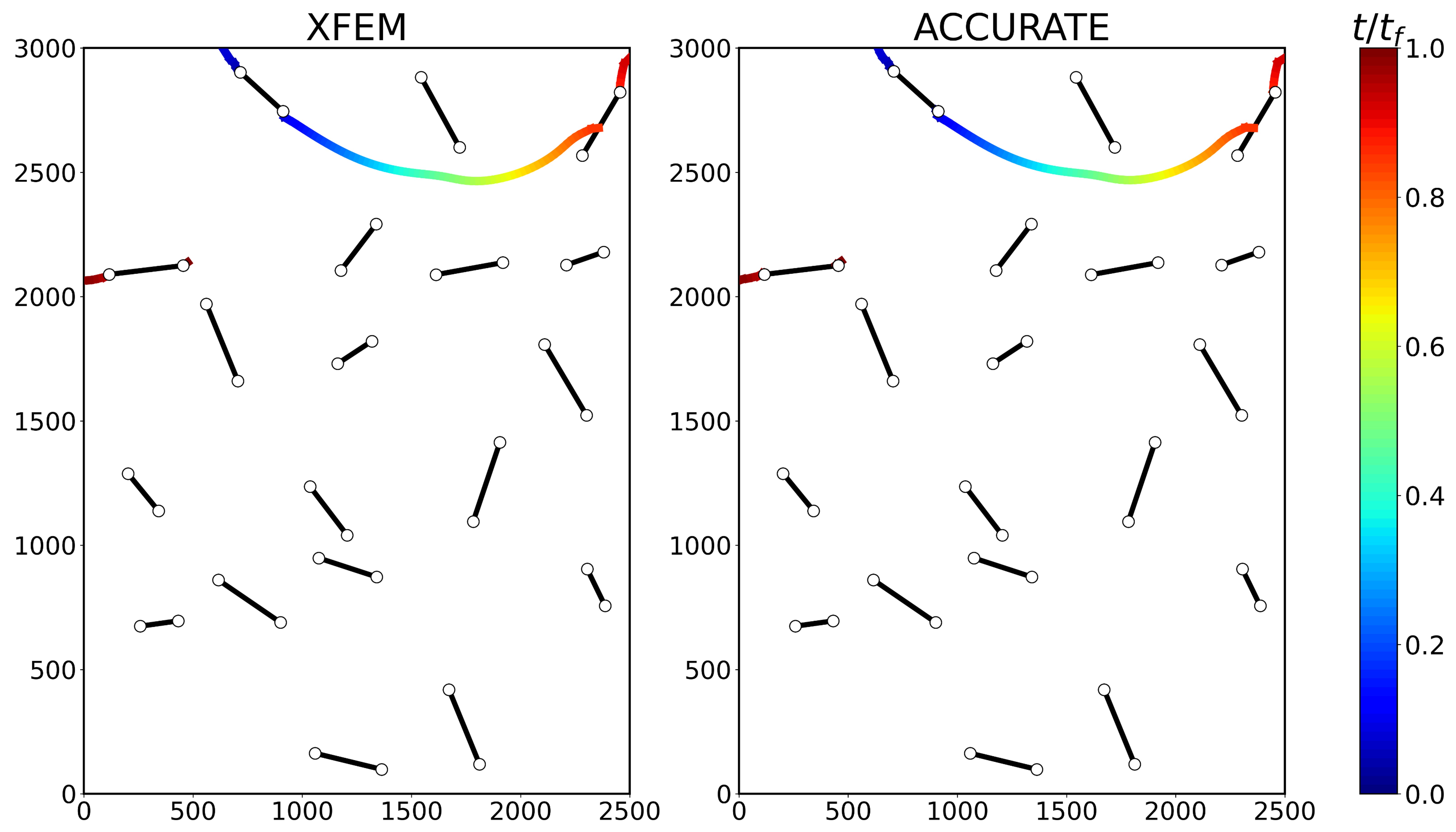}
                \caption{Shear: 2500$\times$3000}
                \label{subfig:ShearUnseen2500x3000_CrackPath}
            \end{subfigure}
            \caption{Crack path evolution for (a) unseen case of 2500mm $\times$ 3000mm domain with arbitrary crack lengths and crack orientations subjected to tensile load, and (b) unseen case of 2500mm $\times$ 3000mm domain with arbitrary crack lengths and crack orientations subjected to shear load.}    
            \label{fig:CrackPath_evolution_Unseen2500x3000}
        \end{figure}

        Next, we implement a similar qualitative analysis for \textit{P}-GNN predictions of crack growth for the new unseen case studies.
        We show the evolution of crack growth for unseen cases (i)-(ii) in Figures \ref{subfig:TensionUnseen2500x2000_CrackPath} and \ref{subfig:ShearUnseen2500x2000_CrackPath}, respectively.
        Similarly, we present the resulting crack growth evolution for unseen cases (iii)-(iv) in Figures \ref{subfig:TensionUnseen2500x3000_CrackPath} and \ref{subfig:ShearUnseen2500x3000_CrackPath}, respectively.
        The resulting crack paths for all unseen case studies agree with the predicted stress distributions shown in Figures \ref{fig:Stress_evolution_Unseen2500x2000} - \ref{fig:Stress_evolution_Unseen2500x3000}, by depicting visually identical crack paths to the XFEM model.
        we note that 6 animations involving stress evolution and crack growth for unseen cases can be seen in the supplementary material.
        
\begin{figure} %[htpb]
            %\centering
            \begin{subfigure}[c]{0.49\textwidth}
                \centering
                \begin{subfigure}[t]{1\textwidth}    
                  \centering \includegraphics[width=0.98\linewidth]{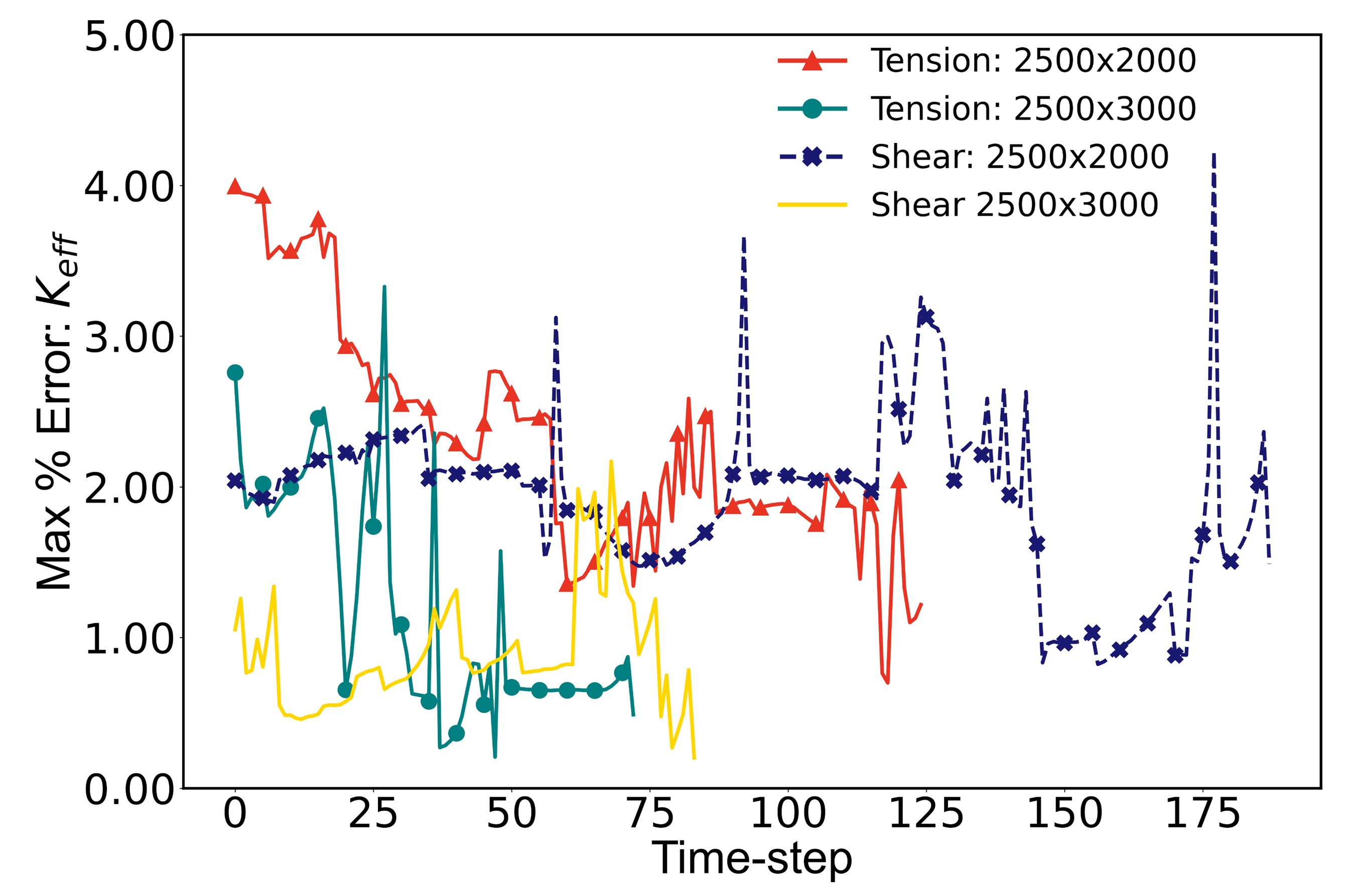}
                    \caption{Maximum error in $K_{eff}$ vs. time for unseen cases}
                    \label{fig:Unseen_Keff_timewise_errors}
                \end{subfigure}
            \end{subfigure}
            %\hfill
            \begin{subfigure}[c]{0.49\textwidth}
                \centering
                \begin{subfigure}[t]{1\textwidth}    
                  \centering
                  \includegraphics[width=0.98\linewidth]{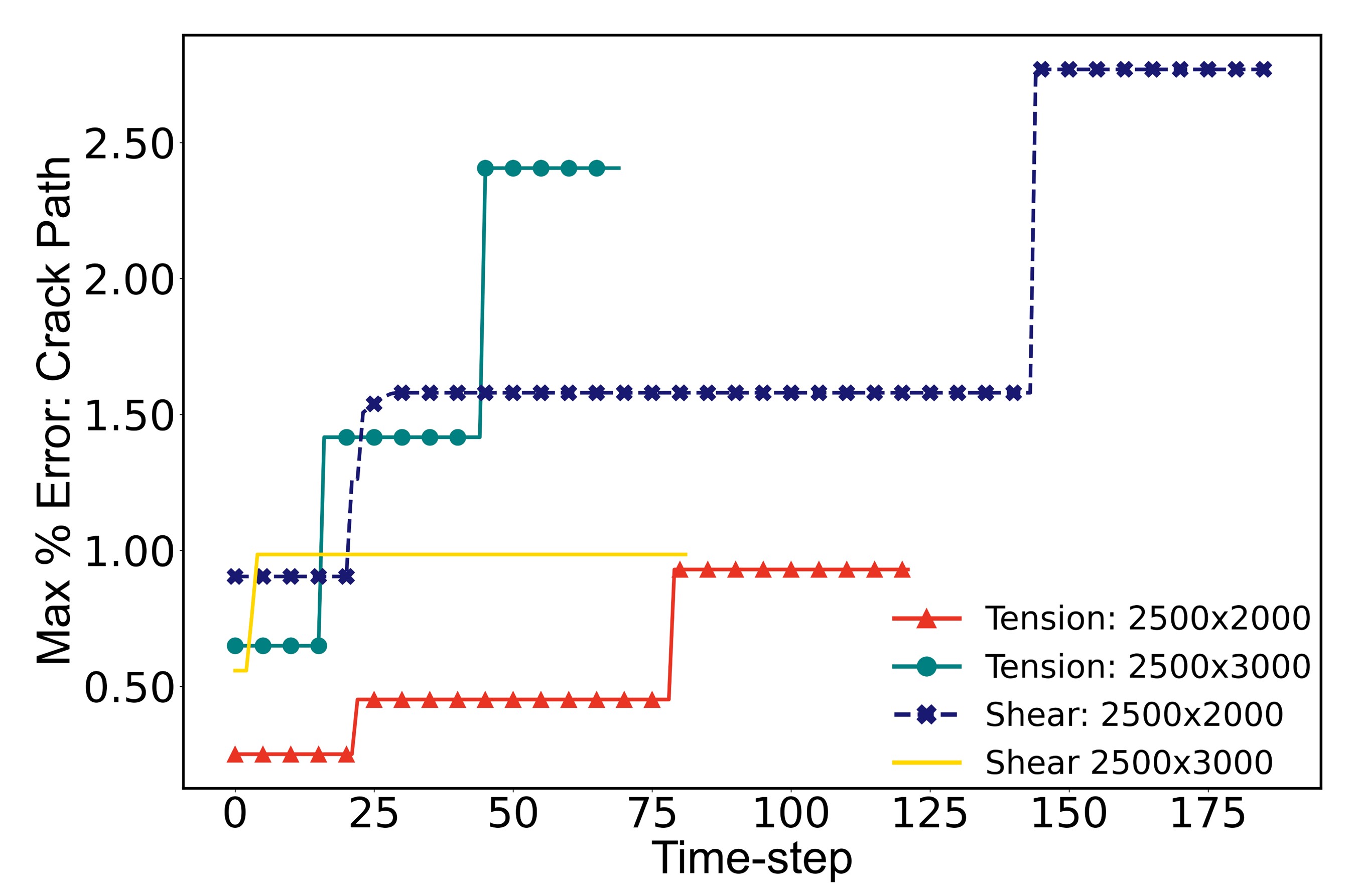}
                    \caption{Maximum error in crack path vs. time for unseen cases}
                    \label{fig:Unseen_CProp_timewise_error}
                \end{subfigure}
            \end{subfigure}
            \caption{Maximum timewise percent errors in effective stress intensity factor and crack path for each unseen case study}
            \label{fig:Unseen_Timewise_errors}
        \end{figure}

        We show the quantitative error analysis for the resulting maximum percent error in both $K_{eff}$ and crack path versus time for all unseen cases in Figure \ref{fig:Unseen_Keff_timewise_errors} - \ref{fig:Unseen_CProp_timewise_error}. 
        For errors in $K_{eff}$, the highest peak in error was obtained for the unseen case involving a $2500mm \times 2000mm$ domain subjected to shear at approximately $4.15\%$.
        The resulting average percent errors of $K_{eff}$ across all time steps for unseen cases (i) - (iV) were $2.31 \pm 0.75$,  $1.2 \pm 0.74$,  $1.92 \pm 0.55$, and $0.89 \pm 0.38$, respectively.
        These results show the ACCURATE framework's ability to emulate the stress evolution in new domain dimensions subjected to either tension or shear involving cracks of arbitrary lengths and orientations with high accuracy ($\approx 96\%$) compared to the XFEM model.
        Following a similar convention, for the errors in the crack path, the highest peak was seen for the $2500mm \times 2000mm$ domain subjected to tension at approximately $2.9\%$.
        The average crack path percent errors across all time steps for unseen cases (i) - (iV) were $0.58 \pm 0.27$,  $1.59 \pm 0.67$,  $1.77 \pm 0.58$, and $0.97 \pm 0.08$, respectively.
        The errors in the crack path along with Figure \ref{fig:Unseen_CProp_timewise_error} show that ACCURATE predicts crack growth and coalescence with very high accuracy ($\approx 97\%$) for the unseen cases. 
        A possible explanation for ACCURATE's ability to handle new arbitrary crack lengths and crack angles without requiring TL to achieve good accuracy may be directly related to both the normalization and randomization of the datasets.
        During TL of a case study involving vertical domains subjected to tension, at any instance in time two or more cracks may have already coalesced, thus, embodying a single larger crack, with an arbitrary crack length and arbitrary crack orientation.  
        Additionally, we note that ACCURATE involved normalization \cite{Singh2020Investigating,Ferreira2019Exploring} criteria for the positions of the crack tips.
        The x-coordinate crack-tip positions were normalized by the width of the domain (i.e., $2000mm$ for vertical domains, and $3000mm$ for horizontal domains), and the y-coordinates by the height of the domain (i.e., $3000mm$ for vertical domains, and $2000mm$ for horizontal domains).
        Therefore, these results show that although TL was applied independently for new boundary effects, arbitrary crack configurations, and shear loadings, by implementing a sequential order the framework was able to generalize across the combination of these variations without additional TL.

    \subsection{Analysis time VS. number of microcracks}\label{subsec:simulation_time} 
    
\begin{figure} %[htpb]
            %\centering
            \begin{subfigure}[c]{0.49\textwidth}
                \centering
                \begin{subfigure}[t]{1\textwidth}    
                  \centering \includegraphics[width=0.98\linewidth]{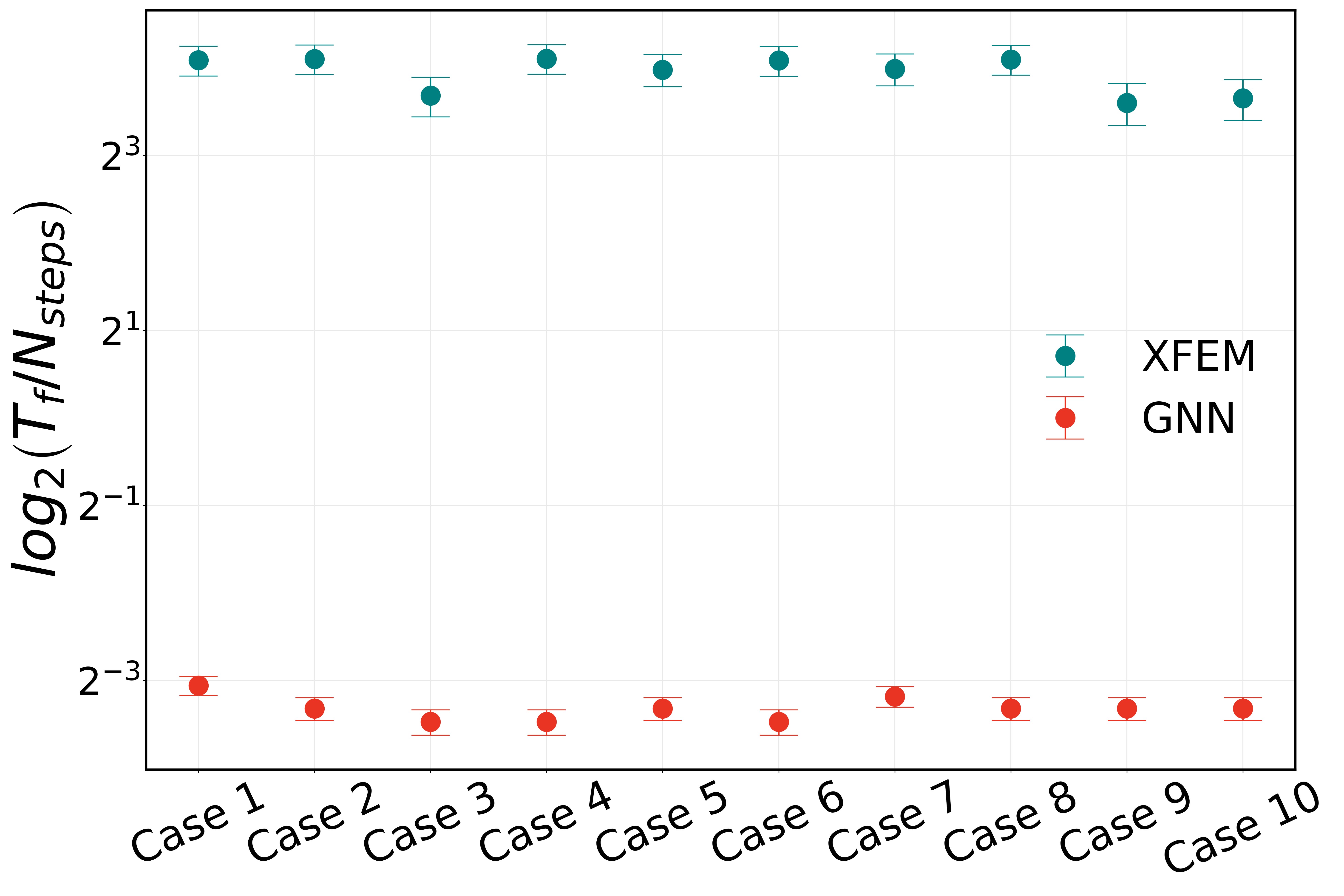}
                    \caption{Simulation time of all test cases}
                    \label{fig:SimulationTime_scatter}
                \end{subfigure}
            \end{subfigure}
            %\hfill
            \begin{subfigure}[c]{0.49\textwidth}
                \centering
                \begin{subfigure}[t]{1\textwidth}    
                  \centering
                  \includegraphics[width=0.98\linewidth]{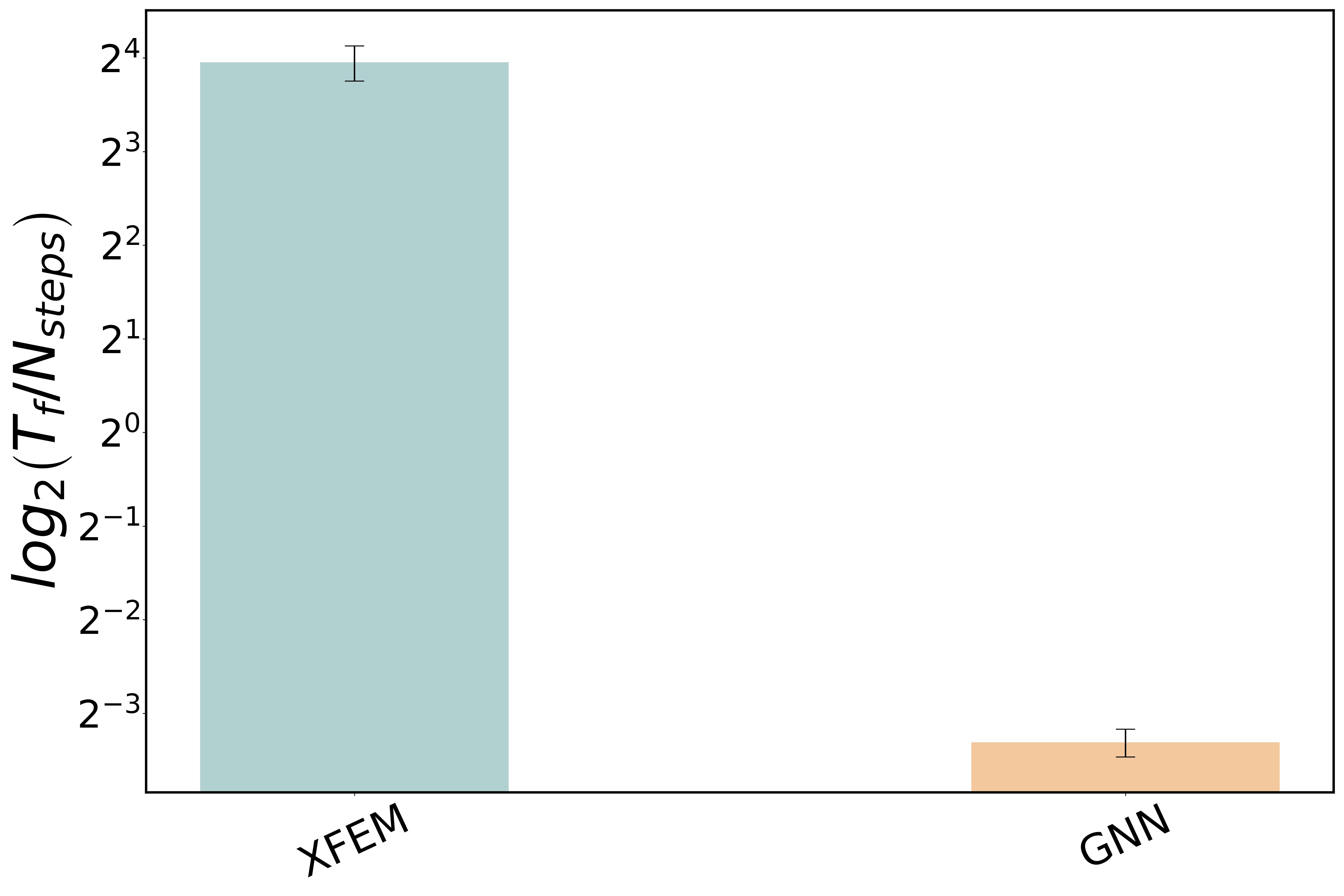}
                    \caption{Average simulation time across test cases}
                    \label{fig:Simulation_time_bar}
                \end{subfigure}
            \end{subfigure}
            \caption{Simulation time (seconds per time-step) comparison for XFEM surrogate model versus ACCURATE for (a) all simulations in the test dataset, and (b) average time across all test simulations.}
            \label{fig:Simulation_Time}
        \end{figure}

        To evaluate the computational speed of the developed GNN framework, we obtained the total simulation time of the XFEM surrogate model and ACCURATE for the 10 test simulations pertaining to the vertical domain case study.
        Figure \ref{fig:SimulationTime_scatter} shows the resulting average simulation time per time-step (seconds per time-step) for the XFEM model (shown in green) and the ACCURATE model (shown in red).
        Figure \ref{fig:Simulation_time_bar} shows the average across all 10 simulations from Figure \ref{fig:SimulationTime_scatter}, for XFEM (shown in light blue) versus ACCURATE (shown in light orange).
        The XFEM-based model was executed using a personal laptop with an Intel Core i9-12900H CPU of 2.50GHz and 16.0GB RAM, and ACCURATE using the same personal PC with laptop-grade GPU NVIDIA GeForce RTX 3070 Ti. 
        We computed the simulation time of ACCURATE starting from the generation of the initial graph representation at $t_{0}$, to the predictions of the prior GNNs, \textit{K}-GNN and \textit{C}-GNN, and the final GNN's prediction, \textit{P}-GNN.
        We used the optimized and open-source PyTorch Geometric (PyG) library for the development of ACCURATE, resulting in a significant difference in time improvement of up to 200x faster (2 orders of magnitude) compared to XFEM.
        The resulting significant speedup is shown in Figure \ref{fig:Simulation_Time}.
        The XFEM fracture model requires more than 10 seconds for each time-step in the simulation, while ACCURATE requires approximately 0.1 seconds per time-step. 
        This means that the XFEM model would require approximately 28 hours to generate a total of 100 simulations, while the ACCURATE framework requires approximately 17 minutes.
        On the other hand, we note that the XFEM model was not parallelized, restricting our performance comparison.
    
\section{Conclusion}\label{sec:Conclusion}

    To conclude, in this work, we developed an accelerated and universal fracture mechanics framework capable of emulating fracture due to multiple cracks' interaction, propagation, and coalescence in brittle materials with high accuracy across various problem configurations. 
    ACCURATE is capable of emulating stress evolution by predicting the Mode-I and Mode-II stress intensity factors for each crack-tip, and the crack growth by predicting future crack-tip positions.
    By leveraging TL on the trained MicroCrack-GNN, we were able to study new problem configurations while requring only 20 training simulations (as opposed to 960 for Microcrack-GNN).
    Also, by implementing a sequence of 5 TL updates involving cases with (i) arbitrary crack lengths, (ii) arbitrary crack orientations, (iii) square domains (i.e., $2500mm \times 2500mm$), (iv) horizontal domains (i.e., $3000mm \times 2000mm$), and (v) shear loadings, ACCURATE was able to learn generalized knowledge of the fracture mechanics.
    To show this, we demonstrate ACCURATE's ability to emulate crack propagation and stress evolution with good accuracy for new unseen cases involving the combination of new domain dimensions with both arbitrary crack lengths and crack orientations, when subjected to tension loads and shear loads.
    We believe these key features allow ACCURATE's use towards exploring a variety of additional problem configurations other than those presented in this work.
    
    Another key contribution of the ACCURATE framework is its significantly accelerated simulation time. 
    Compared to the XFEM fracture model, ACCURATE showed a speedup of approximately 200x.
    An ideally parallelized XFEM code would therefore require approximately 200 CPU cores to achieve the same level of performance.
    Lastly, we believe the developed framework demonstrates the benefits of using ML techniques such as GNNs and TL towards the development of fast reduced-order computational ML-based fracture models that can be trained with very small datasets.
    ACCURATE can be explored in future work to include dynamic effects, ductile material failure, and crack bifurcation.
    The framework can be extended in future work to include ductile material properties, dynamic effects, and crack bifurcation with very small training datasets and accelerated simulation time.

\section{Acknowledgements}
The authors are grateful for the financial support provided by the U.S. Department of Defense in conjunction with the Naval Air Warfare Center/Weapons Division through the SMART scholarship Program (SMART ID: $2021-17978$). 

\section{Supplementary data}
Supplementary animations for cases mentioned in the manuscript can be found at the following GitHub page. \url{https://github.com/rperera12/ACCURATE-GNN-Animations}

\bibliographystyle{ieeetr}
\bibliography{library}

\end{document}